\documentclass{tlp}

\usepackage{latexsym}
\usepackage{amssymb}
\usepackage{stmaryrd}

\submitted{10 February 2005}
\revised{28 August 2006} 
\accepted{11 December 2006}

\def\parfun{\mbox{$\succ\!\!\rightarrow$}}
\def\fbigx{X^\flat}

\def\tvals{{\sf TSub}}
\def\tvaltop{\top}
\def\tvalbot{\bot}
\def\tvalglb{\curlywedge}
\def\bigtvalglb{\bigcurlywedge}
\def\tvallub{\curlyvee}
\def\bigtvallub{\bigcurlyvee}

\def\rules{{\Delta}}
\newcommand\sem[2]{{[\!\![#2]\!\!]}_{#1}}
\newcommand\semb[2]{{\langle\!\langle#2\rangle\!\rangle}_{#1}}
\newcommand {\definedas}{=}
\newcommand {\first}[1]{{#1}}

\def\true{{\it true} }

\def\lor{\vee}
\def\land{\wedge}
\def\lequiv{\leftrightarrow}

\def\kcal{{\cal K}}
\def\gcal{\vec{T}}
\def\fcal{{\bf E}}

\def\ttop{{\bf 1} }
\def\tbot{{\bf 0} }
\def\tand{{\sf and} }
\def\tor{{\sf or} }
\def\tleq{\sqsubseteq}
\def\teq{\equiv}
\def\teqb{\doteq}

\def\tneg{{\sim} }

\def\vi{{V}_{P}}
\def\viplus{{V}_{P}'}

\def\emptybody{\rule{1ex}{1.2ex}}
\def\emptyhead{\Box}

\def\sub{{\it Sub}}

\newcommand{\circled}[1] {\mbox{$\bigcirc$\hspace{-1em}
      \raisebox{0.2ex}{$\scriptstyle #1$}\hspace{.4ex}}}

\newcommand{\point}[1] {\circled{#1}}

\def\scal{{\cal S}}

\def\dom{{\it dom}}

\def\term{{\sf Term}}
\def\atom{{\sf Atom}}
\def\cons{{\sf Cons}}
\def\pars{{\sf Para}}

\def\tnat{{\it nat}}
\def\teven{{\it even}}
\def\todd{{\it odd}}
\def\tlist{{\it list}}
\def\tint{{\it integer}}
\def\tfloat{{\it float}}
\def\tatom{{\it atom}}
\def\ttree{{\it tree}}

\def\sys{{\it sys}}

\newcommand {\Vector}[2]{{#1_1,\cdots,#1_#2}}
\newcommand {\values}{\mapsto}
\newcommand {\comments} [1]{{}}

\def\eqs{{\sf Eqn} }
\def\unify{{\it uf}}
\def\lfp{{\it lfp}}
\def\fail{{\it fail}}
\def\mgu{{\it mgu}}

\def\func{\Sigma}
\def\pred{\Pi}

\def\allvars{{\sf Var} }
\def\restrict{\mbox{$\mid\hspace{-0.45em} \raise 4pt
        \hbox{$\scriptscriptstyle \setminus$}$}}

\def\lor{\vee}
\def\land{\wedge}
\def\lequiv{\leftrightarrow}

\def\ctop{{\sub}}
\def\club{{\cup}}

\def\cleq{{\subseteq}}
\def\csub{{\wp(\sub)}}
\def\cunify{{\it Uf}}
\def\csys{{\it Sys}}

\def\alub{{\sqcup}^\flat}
\def\aglb{{\sqcap}^\flat}
\def\aleq{{\sqsubseteq}^\flat}
\def\asub{{\it ASub}^\flat}
\def\aunify{{\cunify}^\flat}
\def\aid{{\it Id}^\flat}
\def\asys{{\csys}^\flat}

\def\rtype{{\sf Type}}   
\def\gtype{{\sf Schm}}   
\def\ertype{{\sf eType}}   

\def\rpreceq{\preccurlyeq}  

\def\req{\thickapprox} 

\def\rsub{\asub}      
\def\rleq{\aleq}      
\def\rlub{\alub}         
\def\rglb{\aglb}         
\def\rcon{\gamma}         
\def\vtcon{\gamma_{{\sf VT}}}         

\def\munify{\aunify}
\def\mup{{\it up}}
\def\mdown{{\it down}}
\def\msolve{{\it solve}}
\def\mrestrict{{\it rest}}

\def\tdef{{\rightarrowtriangle}}

\def\type{{\it type}}
\def\vts{{\it vts}}
\def\tvs{{\it cover}}

\def\rt{{\it R}}
\def\rtvec{\vec{R}}
\def\tauvec{\vec{\tau}}
\def\at{{\it A}}

\newcommand{\atomat}[1]{{\mathbb{A}(#1)}}

\newcommand{\headat}[1]{{\mathbb{H}(#1)}}
\def\edges{{\sf Edge}}
\def\points{{\sf Pt}}
\newcommand{\edge}[2]{{#1\!\leftharpoonup\! #2}}

\def\endprf{~\hspace*{\fill}$\rule{1ex}{1.5ex}$\par}

\newtheorem {definition} {Definition} [section]
\newtheorem {corollary}[definition]{Corollary}
\newtheorem {lemma}[definition]{Lemma}
\newtheorem {theorem}[definition]{Theorem}
\newtheorem {algorithm}[definition]{Algorithm}
\newtheorem {example}[definition]{Example}

\newcommand{\lemmaproof}[3] {{{\noindent \bf
      Lemma}~\ref{#1}}.\/ #2  \begin{proof}#3\end{proof}\vspace{1pc}}

\newcommand{\coproof}[3] {{{\noindent \bf
      Corollary}~\ref{#1}}.\/ #2 \begin{proof}#3\end{proof}}

\newcommand{\theoremproof}[3] {{{\noindent \bf
      Theorem}~\ref{#1}}.\/ #2 \begin{proof}#3\end{proof}\vspace{1pc}}

\title[Improving Precision of Type Analysis Using Non-Discriminative Union]
{\begin{tabular}{c}
Improving Precision of  Type Analysis\\  Using Non-Discriminative Union
\end{tabular}}
\author[Lunjin Lu]
{
LUNJIN LU\\
Oakland University, Rochester, MI~48309, USA. \\
\email{lunjin@acm.org} }

\begin{document}

\maketitle

\input rotate.tex
\newbox\rotbox

\bibliographystyle{acmtrans}

\begin{abstract}

This paper presents a new type analysis for logic programs. The
analysis is performed with {\it a priori} type definitions; and type
expressions are formed from a fixed alphabet of type constructors.
Non-discriminative union is used to join type information from
different sources without loss of precision.  An operation that is
performed repeatedly during an analysis is to detect if a fixpoint has
been reached. This is reduced to checking the emptiness of types. Due
to the use of non-discriminative union, the fundamental problem of
checking the emptiness of types is more complex in the proposed type
analysis than in other type analyses with {\it a priori} type
definitions. The experimental results, however, show that use of
tabling reduces the effect to a small fraction of analysis time on a
set of benchmarks.

\end{abstract}


\section{Introduction} \label{sec:int}

Types play an important role in programming. They make programs easier
to understand and help detect errors. There has been much research
into types in logic programming. A type checker requires the
programmer to declare types for each predicate in the program and
verifies if the program is consistent with \comments {well-typed with
respect to} the declared
types~\cite{AikenL94,Dart:Zobel:JLP92,FagesC01,FruhwirthSVY:LICS91,Mycroft:OKeefe:84,Reddy:NACLP90,YardeniFS:ICLP91,Yardeni:Shapiro:91}.
\comments
{A call to a predicate is said to be well-typed if its arguments have
the declared type for the predicate.  Otherwise, it is ill-typed. The
execution of a well-typed call with a well-typed program never
encounters any ill-typed sub-call.}  A type analysis derives types for
the predicates or literals in the program from the text of the
program~\cite{GallagherP02,CharatonikP98,aci_types_full,GallagherW94,HeintzeJ90,HeintzeJ92,LuJLP98,Mishra:84,1995-saglam,Zobel:87}.

\comments {
Some type analyses infer types with respect to which the
program is well-typed. These analyses often allow the programmer to
declare types for some predicates in the program.  Other type analyses
derive types to describe the success set of the program or the set of
possible memory stores at each program point.

A type denotes a set of terms. In some type systems, types are sets of
ground terms~\cite{}.  Some other type systems use types to denote
sets of possibly non-ground terms that are closed under
instantiation~\cite{}.  There are also type systems in which types
denote sets of possibly non-ground terms that are not necessarily
closed under instantiation~\cite{}. These type systems may need to use
information about variable independence to ensure its soundness and/or
improve its precision since types capture also information about
freeness. Types are regular tree languages in most type systems.
Types are deterministic (recognizable by deterministic top-down tree
automata) in some type systems and non-deterministic in others. }

This paper presents a new type analysis that infers types with {\it a
priori} type definitions which determine possible types and their
meanings.  Types are formed from type constructors from a fixed
alphabet. This is in contrast to those type analyses that generate
type definitions during analysis. Both kinds of type analysis are
useful. An analysis that generates type definitions may be favored in
compile-time optimizations and program transformations whilst an
analysis with {\it a priori} type definitions may be preferred in
interactive programming tools such as debuggers because inferred types
are easier for the programmer to understand.

A number of factors compromise the precision of previous type analyses
with {\it a priori} type definitions. Firstly, they only allow
deterministic type definitions. A function symbol cannot occur more
than once in the definition of the same type. A type then denotes a
tree language recognized by a deterministic top-down tree
automaton~\cite{tata02} and hence called a deterministic type.  The
restriction to deterministic type definitions allows fast propagation
of type information.  However, it causes loss of precision because of
the limited power of deterministic types. The same restriction also
prevents many natural typings. For instance, these two type rules
$float\tdef +(integer,float)$ and $float\tdef +(float,float)$ violate
the restriction. Some previous work even disallows function
overloading~\cite{HoriuchiK87,Kanamori:Horiuchi:85,KanamoriJLP93},
which makes it hard to support built-in types. For instance, Prolog has
built-in type $\tatom$ that denotes the set of atoms. Without function
overloading, atoms such as $[~]$ cannot be a member of another type,
say $\tlist$.  Secondly, the type languages in previous type analyses
with {\it a priori} type definitions do not include set union as a
type constructor.  The denotation of the join of two types can be
larger than the set union of their denotations. For instance, the join
of $\tlist(integer)$ and $\tlist(float)$ is $\tlist(number)$. Let
$\tor$ be a type constructor that is interpreted as set union. Then
$\tlist(number)$ is a super-type of
$\tor(\tlist(integer),\tlist(float))$ since the list [1, 2.5] belongs
to the former but not the latter. Should non-deterministic type
definitions be allowed, there is also a need to use set intersection
as a type constructor as explained in Section~\ref{sec:mot}.  Finally,
previous type analyses with {\it a priori} type definitions describe a
set of substitutions by a single variable typing which maps variables
of interest into types. The least upper bound of two variable typings
is performed point-wise, effectively severing type dependency between
variables.

Our type analysis aims to improve precision by eliminating the above
mentioned factors. It supports non-deterministic type definitions,
uses a type language that includes set union and intersection as type
constructors and describes a set of substitutions by a set of variable
typings. All these help improve analysis precision. On the other hand,
they all incur performance penalty. However, experimental results with
a prototype implementation show that tabling~\cite{Warren92} reduces
the time increase to a small fraction on a suite of benchmark
programs. Our type analysis is presented as an abstract domain
together with a few primitive operations on the domain. The domain is
presented for an abstract semantics that is Nilsson's abstract
semantics~\cite{Nilsson:88} extended to deal with negation and built-in
predicates. The primitive operations on the domain can be easily
adapted to work with other abstract semantics such
as~\cite{Bruynooghe91}.

The remainder of the paper is organized as
follows. Section~\ref{sec:mot} provides motivation behind our work
with some examples and Section~\ref{sec:pre} briefly presents the
abstract semantics along with basic concepts and notations used in the
remainder of the paper. Section~\ref{sec:typ} is devoted to types ---
their definitions and denotations. Section~\ref{sec:dom} presents the
abstract domain and Section~\ref{sec:ops} the abstract operations. In
Section~\ref{sec:imp}, we present a prototype implementation of our
type analysis and some experimental results. Section~\ref{sec:related}
compares our type analysis with others and Section~\ref{sec:con}
concludes. An appendix contains proofs.

\section{Motivation} \label{sec:mot}
This section provides motivation behind our type analysis via
examples. The primary operations for propagating type information are
informally illustrated; and the need for using set union and
intersection as type constructors is highlighted.

\begin{example} \label{ex:one} This example
demonstrates the use of set union as a type constructor.  Consider the
following program and type rules.
\begin{eqnarray*}
p(Z) &\leftarrow& ~\point{2}~ X=a~\point{3}, Y=2.5~\point{4}~, Z=cons(X,cons(Y,nil))~\point{5}.\\
&\leftarrow & ~\point{1}~ p(Z)~\point{6}. ~\mbox{\% query} \\\\
\tlist(\beta) &\tdef & nil\\
\tlist(\beta) &\tdef & cons(\beta,\tlist(\beta))
\end{eqnarray*}
 The two type rules define lists. They state that a term is of type
 $\tlist(\beta)$ iff it is either $nil$ or of the form $cons(X,Y)$
 such that $X$ is of type $\beta$ and $Y$ of type $\tlist(\beta)$.
 Type rules are formally introduced in Section~\ref{sec:typ}.  The
 program has been annotated with circled numbers to identify relevant
 program points for the purpose of exposition.

 The type analysis can be thought of as an abstract execution that
 mimics the concrete (normal) execution of the program. A program
 state in the concrete execution is replaced with an abstract one that
 describes the concrete state. The abstract states are type
 constraints.

Suppose that no type information is given at program point $\point{1}$
--- the start point of the execution. This is described by the type
constraint $\mu_1=\true$. The execution reaches program point
$\point{2}$ with the abstract state $\mu_2=\true$. The abstract state
at program point $\point{3}$ is $\mu_3=(X\in{\it atom})$ which states
that $X$ is of type ${\it atom}$. The abstract state at program point
$\point{4}$ is $\mu_4=(X\in{\it atom})\land (Y\in{\it float})$. The
abstract execution of $Z=cons(X,cons(Y,nil))$ in $\mu_4$ obtains the
abstract state $\mu_5$ at program point $\point{5}$. The computation
of $\mu_5$ needs some explanation. The two terms that are unified have
the same type after the unification.  Since $\mu_4$ does not constrain
$Z$, there is no type information propagated from $Z$ to either $X$ or
$Y$.  The type for $Z$ in $\mu_5$ equals the type of
$cons(X,cons(Y,nil))$ in $\mu_4$ which is computed in a bottom-up
manner. To compute the type for $nil$, we apply the type rule for
$nil/0$. The type rule states that $nil$ is of type $\tlist(\beta)$
for any $\beta$.  Thus, the most precise type for $nil$ is
$\tlist(\tbot)$ where the type $\tbot$ denotes the empty set of
terms. We omit the process of computing the type $\tlist({\it float})$
for $cons(Y,nil)$ in $\mu_4$ since it is similar to the following.  To
compute the type for $cons(X, cons(Y,nil))$, we apply the type rule
for $cons/2$. The right hand side of the type rule is
$cons(\beta,\tlist(\beta))$. We first find the smallest value for
$\beta$ such that $\beta$ is greater than or equal to ${\it atom}$ ---
the type for $X$ in $\mu_4$ and the smallest value for $\beta$ such
that $\tlist(\beta)$ is greater than or equal to $\tlist({\it float})$
--- the type for $cons(Y,nil)$ in $\mu_4$. Those two values are
respectively ${\it atom}$ and ${\it float}$ and their least upper
bound is $\tor({\it atom},{\it float})$. Replacing $\beta$ with
$\tor({\it atom},{\it float})$ in the left hand side of the type rule
gives the most precise type $\tlist(\tor({\it atom},{\it float}))$ for
$cons(X,cons(Y,nil))$ in $\mu_4$. Conjoining $Z\in\tlist(\tor({\it
  atom},{\it float}))$ with $\mu_4$ results in $\mu_5=((X\in{\it
  atom})\land(Y\in{\it float})\land (Z\in\tlist(\tor({\it atom},{\it
  float}))))$. The abstract state at program point $\point{6}$ is
$\mu_6=(Z\in\tlist(\tor({\it atom},{\it float})))$ which is obtained
from $\mu_5$ by projecting out type constraints on $X$ and $Y$.

 The existence of the type constructor $\tor$ helps avoid
 approximations. Without it, the least upper bound of ${\it atom}$ and
 ${\it float}$ is $\ttop$ which denotes the set of all terms. Note
 that the collection of type rules is fixed during analysis. \endprf
\end{example}

When two or more type rules are associated with a single function
symbol, there is also a need to use set intersection as a type
constructor. The following example illustrates this point.
\begin{example} Suppose that types are defined by the following four type rules.
\begin{eqnarray*}
\tlist(\beta) &\tdef & nil\\ \tlist(\beta) &\tdef &
cons(\beta,\tlist(\beta))\\ \ttree(\beta) &\tdef & nil\\ \ttree(\beta)
&\tdef & node(\ttree(\beta),\beta,\ttree(\beta))
\end{eqnarray*}
Consider the problem of computing the type for $cons(X,nil)$ in the
abstract state $\mu=(X\in{\it integer})$.

There are two type rules for $nil/0$. The type rule
$\tlist(\beta)\tdef nil$ states that $nil$ belongs to $\tlist(\beta)$
for any $\beta$. The most precise type for $nil$ that can be inferred
from this rule is $\tlist(\tbot)$. Similarly, the most precise type
for $nil$ that can be inferred from the type rule $\ttree(\beta)\tdef
nil$ is $\ttree(\tbot)$. Thus, the most precise type for $nil$ is
$\tand(\tlist(\tbot),\ttree(\tbot))$ where $\tand$ is a type
constructor that denotes set intersection.

 To compute the type for $cons(X, nil)$, we apply the type rule for
 $cons/2$. Its right hand side is $cons(\beta,\tlist(\beta))$.  We
 first find the smallest value for $\beta$ such that $\beta$ is
 greater than or equal to ${\it integer}$ --- the type for $X$ in
 $\mu$.  The value is ${\it integer}$. We then find the smallest value
 for $\beta$ such that $\tlist(\beta)$ is greater than or equal to
 $\tand(\tlist(\tbot),\ttree(\tbot))$ --- the type for $nil$ in
 $\mu$. This is done by matching $\tlist(\beta)$ with $\tlist(\tbot)$
 and with $\ttree(\tbot)$ and intersecting values for $\beta$ obtained
 from these two matches.  The first match results in $\tbot$. The
 second match is unsuccessful and produces $\ttop$ since we are
 computing an upper approximation. The intersection of these two types
 is $\tand(\tbot,\ttop)$ which is equivalent to $\tbot$. The join of
 the two smallest values ${\it integer}$ and $\tbot$ for $\beta$ is
 $\tor({\it integer},\tbot)$ which is equivalent to ${\it
 integer}$. Finally, the type $\tlist({\it integer})$ for
 $cons(X,nil)$ is obtained by substituting ${\it integer}$ for $\beta$
 in the left hand side of the type rule.

Without $\tand$ in the type language, a choice must be made between
$\tlist(\tbot)$ and $\ttree(\tbot)$ as the type for $nil$. Though
these types are equivalent to $\tand(\tlist(\tbot),\ttree(\tbot))$,
the choice made could complicate the ensuing computation.  Should
$\ttree(\tbot)$ be chosen, we would need to find the smallest value
for $\beta$ such that $\tlist(\beta)$ is greater than or equal to
$\ttree(\tbot)$. This could only be solved by applying an algorithm
for solving type inclusion constraints. The presence of $\tand$ allows
us to avoid that. \endprf
\end{example}

For the purpose of improving the precision of analysis, there is also
a need for disjunction at the level of abstract states. The following
example illustrates this point.
\begin{example} Consider the following program
\begin{eqnarray*}
p(X) &\leftarrow & q(X,Y),~\point{1}~ ...\\
q(1,2). \\
q(a,b). \\
&\leftarrow &  p(X). ~\mbox{\% query}
\end{eqnarray*}
When the execution reaches program point \point{1}, X and Y are both
of type {\it integer} or they are both of type {\it atom}. This is
described by a type constraint $((X\in{\it integer})\land (Y\in{\it
integer}))\lor ((X\in{\it atom})\land (Y\in{\it atom}))$. Without
disjunction at the level of abstract states, we would have to replace
the type constraint with a less precise one: $((X\in\tor({\it
integer},{\it atom}))\land (Y\in\tor({\it integer},{\it
atom})))$. \endprf
\end{example}

\section{Preliminaries}\label{sec:pre}
The reader is assumed to be familiar with the terminology of logic
programming \cite{Lloyd:87} and that of abstract
interpretation~\cite{Cousot:Cousot:77}.  We consider a subset of
Prolog which contains definite logic programs extended with negation
as failure and some built-in predicates.

\subsection{Basic Concepts}

           We sometimes use Church's lambda notation for
functions, so that a function $f$ defined $f(x)=e$ will be denoted
$\lambda x.e$. Let $A$ and $B$ be sets. Then $A\mapsto B$ is the set
of total functions from $A$ to $B$ and $A\parfun B$ is the set of
partial functions from $A$ to $B$. The function composition $\circ$ is
defined $f \circ g=\lambda x. f(g(x))$. Let $D$ be a set. A sequence over
$D$ is either $\epsilon$ or $d\bullet\vec{d}$ where $d\in{D}$ and
$\vec{d}$ is a sequence over $D$. The infix operator $\bullet$
associates to the right and prepends an element to a sequence to form
a longer sequence. The set of all sequences over $D$ is denoted
$D^{*}$. Let $\vec{d}=d_1\bullet d_2 \bullet\cdots\bullet
d_n\bullet\epsilon$. We will sometimes write $\vec{d}$ as
$d_1,d_2,\cdots,d_n$. The dimension $\|\vec{d}\|$ of $\vec{d}$ is
$n$. Let $E\subseteq D$ and $S\subseteq D^{*}$. The set extension of
$\bullet$ is defined as $E\bullet S=\{d\bullet \vec{d}\mid
d\in{E}\land \vec{d}\in{S}\}$.

\subsection{Abstract Interpretation}
A semantics of a program is given by an interpretation $\langle
(C,\sqsubseteq_C),{\cal C}\rangle$ where $(C,\sqsubseteq_C)$ is a
complete lattice and ${\cal C}$ is a monotone function on $
(C,\sqsubseteq_C)$. The semantics is defined as the least fixed point
$\lfp~{\cal C}$ of ${\cal C}$. The concrete semantics of the program is
given by the concrete interpretation $\langle (C,\sqsubseteq_C),{\cal
C}\rangle$ while an abstract semantics is given by an abstract
interpretation $\langle (A,\sqsubseteq_A),{\cal A}\rangle$.  The
correspondence between the concrete and the abstract domains is
formalized by a Galois connection $(\alpha,\gamma)$ between
$(C,\sqsubseteq_C)$ and $(A,\sqsubseteq_A)$.  A Galois connection
between $A$ and $C$ is a pair of monotone functions
$\alpha:{C}\mapsto{A}$ and $\gamma:{A}\mapsto{C}$ satisfying $\forall
c\in{C}.  (c\sqsubseteq_C \gamma\circ\alpha(c))$ and $\forall
{a}\in{A}.  (\alpha\circ\gamma({a})\sqsubseteq_A{a})$. The function
$\alpha$ is called an abstraction function and the function $\gamma$ a
concretization function.  A sufficient condition for $\lfp{{\cal A}}$
to be a safe abstraction of $\lfp~{{\cal C}}$ is
$\forall{a}\in{A}.(\alpha\circ{\cal
C}\circ\gamma({a})~\sqsubseteq_A~{\cal A}({a}))$ or equivalently
$\forall{a}\in{A}.({\cal
C}\circ\gamma({a})~\sqsubseteq_C~\gamma\circ{\cal A}({a}))$, according
to propositions 24 and 25 in~\cite{Cousot:JLP92}. The abstraction and
concretization functions in a Galois connection uniquely determine
each other; and a complete meet-morphism $\gamma:{A}\mapsto{C}$
induces a Galois connection $(\alpha,\gamma)$ with
$\alpha(c)=\sqcap_{A} \{{a}~|~c\sqsubseteq_C\gamma({a})\}$. A function
$\gamma:{A}\mapsto{C}$ is a complete meet-morphism iff
$\gamma(\sqcap_{A} X)=\sqcap_{C}\{\gamma(x)\in{X}\}$ for any
$X\subseteq{A}$. Thus, an analysis can be formalized as a tuple
\( (\langle (C,\sqsubseteq_C),{\cal C}\rangle,\gamma,\langle (A,\sqsubseteq_A),{\cal A}\rangle)\) such that $\langle (C,\sqsubseteq_C),{\cal C}\rangle$ and
$\langle (A,\sqsubseteq_A),{\cal A}\rangle$ are interpretations, $\gamma$ is a complete meet-morphism from
$({C},\sqsubseteq_C)$ to $({A},\sqsubseteq_A)$, and
$\forall{a}\in{A}.({\cal
C}\circ\gamma({a})~\sqsubseteq_C~\gamma\circ{\cal A}({a}))$.

\subsection{Logic Programs}
Let $\func$ be a set of \first{function symbols}, $\pred$ a set of
\first{predicate symbols} and $\allvars$ a denumerable set of
variables. Each function or predicate symbol has an arity which is a
non-negative integer. We write $f/n\in\func$ for an $n$-ary function
symbol $f$ in $\func$ and $p/n\in\pred$ for an $n$-ary predicate
symbol $p$ in $\pred$.  Let $V\subseteq\allvars$. The set of all
\first{terms} over $\func$ and $V$, denoted $\term(\func,V)$, is the
smallest set satisfying: (i) $V\subseteq\term(\func,V)$; and (ii) if
$\{\Vector{t}{n}\}\subseteq\term(\func,V)$ and $f/n\in\func$ then
$f(\Vector{t}{n})\in\term(\func,V)$.  The set of all \first{atoms}
that can be constructed from $\pred$ and $\term(\func,V)$ is denoted
$\atom(\pred,\func,V)$;
$\atom(\pred,\func,V)=\{p(\Vector{t}{n})\mid(p/n\in\pred)\land(\{\Vector{t}{n}\}\subseteq\term(\pred,\func,V))\}$. Let
$\term=\term(\func,\allvars)$ and $\atom=\atom(\pred,\func,\allvars)$
for abbreviation. The set $\term$ contains all terms and the set
$\atom$ all atoms.  The negation of an atom $p(\Vector{t}{n})$ is
written $\neg p(\Vector{t}{n})$. A literal is either an atom or the
negation of an atom. The set of all literals is denoted ${\sf
Literal}$.  Let ${\sf Bip}$ denote the set of calls to built-in
predicates. Note that ${\sf Bip}\subseteq\atom$.

A \first{clause} $C$ is a formula of the form
$H\leftarrow\Vector{L}{n}\emptybody$ where
$H\in\atom\cup\{\emptyhead\}$ and $L_{i}\in{\sf Literal}$ for $1\leq
i\leq n$. $H$ is called the \first{head} of the clause and
$\Vector{L}{n}\emptybody$ the \first{body} of the clause. Note that
$\emptyhead$ denotes the empty head and $\emptybody$ denotes the empty
body. A query is a clause whose head is $\emptyhead$. A
\first{program} is a set of clauses of which one is a query. The query
initiates the execution of the program.

Program states which exist during the execution of a logic program are
called substitutions.  A substitution $\theta$ is a mapping from
$\allvars$ to $\term$ such that
$dom(\theta)\definedas\{x~|~(x\in\allvars)\land(\theta(x)\neq{x})\}$
is finite.  The set $dom(\theta)$ is called the domain of
$\theta$. Let $dom(\theta)=\{x_1,\cdots,x_n\}$. Then $\theta$ is
written as $\{x_1\values\theta(x_1),\cdots,x_n\values\theta(x_n)\}$. A
substitution $\theta$ is idempotent if $\theta\circ\theta=\theta$.
The set of idempotent substitutions is denoted ${\sub}$; and the
identity substitution is denoted $\epsilon$.  Let
${\sub}_{\fail}\definedas \sub\cup\{\fail\}$ and extend $\circ$ by
$\theta\circ\fail\definedas\fail$ and
$\fail\circ\theta\definedas\fail$ for any $\theta\in
{\sub}_{\fail}$.
Substitutions are not distinguished from their homomorphic extensions
to various syntactic categories.

An \first{equation} is a formula of the form $l=r$ where either
$l,r\in \term$ or $l,r\in\atom$.  The set of all equations is denoted
$\eqs$.  For a set of equations $E$,
$mgu:\wp(\eqs)\mapsto{\sub}_{\fail}$ returns either a most general
unifier for $E$ if $E$ is unifiable or $\fail$ otherwise. Let
$mgu(l,r)$ stand for $mgu(\{l=r\})$.  Define
$eq(\theta)=\{x=\theta(x)~|x\in\dom(\theta)\}$ for $\theta\in\sub$ and
$eq(\fail)=\fail$.

The set of variables in a syntactic object $o$ is denoted ${\it
vars}(o)$.  A renaming substitution $\rho$ is a substitution such that
$\{\rho(x)\mid x\in\allvars\}$ is a permutation of $\allvars$. The set
of all renaming substitutions is denoted ${\sf Ren}$. Define ${\sf
Ren}(o_1,o_2)\definedas\{\rho\in{\sf Ren}\mid {\it
vars}(\rho(o_1))\cap {\it vars}(o_2)=\emptyset\}$.

We assume that there is a function $\sys:{\sf
Bip}\times\sub\mapsto\wp(\sub)$ that models the behavior of built-in
predicates. The set $\sys(p(\Vector{t}{n}),\theta)$ consists of all
those substitutions $\sigma\circ\theta$ such that $\sigma$ is a
computed answer to $\theta(p(\Vector{t}{n}))$.

Let $\vi$ be the set of variables in the program and
$\atom_{P}=\atom(\pred,\func,\vi)$. Define
$\unify:\atom_{P}\times{\sub}\times\atom_{P}\times{\sub}\mapsto{\sub}_{\fail}$
by
\[
\unify(a_1,\theta,a_2,\omega) =
    let~\rho\in{\sf Ren}(\theta(a_1),\omega(a_2))~ in~
    mgu(\rho(\theta(a_1)),\omega(a_2))\circ\omega
\]
The operation $\unify(a_1,\theta,a_2,\omega)$ models both
procedure-call and procedure-exit operations. In a procedure-call
operation, $a_1$ and $\theta$ are the call and the program state
before the call, $a_2$ is the head of the clause that is used to
resolve with the call and $\omega$ the identity substitution
$\epsilon$. In a procedure-exit operation, $a_2$ and $\omega$ are the
call and the program state before the call, $a_1$ is the head of the
clause that was used to resolve with the call and $\theta$ is the
program state after the execution of the body of the clause. A
renaming is applied to the call in a procedure-call operation whilst
in a procedure-exit operation it is the head of the clause that is
renamed.

\subsection{Abstract Semantics} \label{sec:asem}
The new type analysis is presented as an abstract domain with four
abstract operations. The domain and the operations are designed for
an abstract semantics in~\cite{Nilsson:88} extended with supports for
negation-as-failure and built-in predicates. The extended abstract
semantics is a special case of an abstract semantics
in~\cite{Lu03path} where a formal presentation can be found.  The
adaptation of the analysis to other abstract semantics such
as~\cite{Bruynooghe91} is straightforward since they require abstract
operations with similar functionalities.

The abstract semantics is parameterized by an abstract domain $\langle
\asub, \aleq\rangle$\comments{ and four abstract operations:
$\alub:\asub\times\asub\mapsto\asub$, $\aid :\emptyset\mapsto\asub$,
$\asys:{\sf Bip}\times\asub\mapsto\asub$ and
$\aunify:\atom_{P}\times\asub\times\atom_{P}
\times\asub\mapsto\asub$}. The elements in $\asub$ are called abstract
substitutions since they are properties of substitutions.  The
abstract domain is related to the collecting domain $\langle
\csub,\cleq\rangle$ via a concretization function $\gamma:\asub\mapsto\csub$. We say that an abstract substitution
$\pi$ describes a set of substitutions $\Theta$ iff
$\Theta\subseteq\gamma(\pi)$. As usual, the abstract domain and the
concretization function are required to satisfy the following
conditions.
\begin{itemize}
\item [C1:] $<\asub,\aleq>$ is a complete lattice with  least upper bound operation $\alub$;
\item [C2:] $\gamma(\asub)$ is a Moore family where $\gamma(X)=\bigcup\{\gamma(x)\mid x\in X\}$.
\end{itemize}

We informally present the abstract semantics using the following
program as a running example.
 \[ \begin{array}{lll}
  \texttt{diff}(X, L, K )  &\leftarrow&
          ~\point{1}~ \texttt{member}(X, L),~\point{2}~ \neg
                        \texttt{member}(X, K) ~\point{3}\\
 \texttt{diff}(X, L, K) &\leftarrow& ~\point{4}~  \texttt{member}(X, K),~\point{5}~\neg
                         \texttt{member}(X, L)~\point{6}\\
 \texttt{member}(X,[X|L]) &\leftarrow&~\point{7}\\
 \texttt{member}(X,[H|L]) &\leftarrow& ~\point{8}~\texttt{member}(X, L)~\point{9}\\
 &\leftarrow& ~\point{{\hspace{-.5ex}10}}~Y=[a,b]~~\point{{\hspace{-.5ex}11}}~Z=[1,2] ~\point{{\hspace{-.5ex}12}}~ \texttt{diff}(X,Y,Z)~\point{{\hspace{-.5ex}13}}
   \end{array}
    \]
The intended interpretation for $\texttt{member}(X,L)$ is that $X$ is
a member of list $L$. The intended interpretation for
$\texttt{diff}(X,L,K)$ is that $X$ is in $L$ or $K$ but not in both.
For brevity of exposition, let $A=\texttt{member}(X,L)$;
$B=\texttt{member}(X,K)$; $C=\texttt{member}(X,[X|L])$;
$D=\texttt{member}(X,[H|L])$; $E=\texttt{diff}(X,L,K)$ and
$F=\texttt{diff}(X,Y,Z)$.  The atom in the literal to the right of a
program point $p$ is denoted $\atomat{p}$. For instance,
$\atomat{2}=\atomat{4}=B$.  Let $\headat{p}$ denote the head of the
clause with which $p$ is associated. For instance,
$\headat{1}=\headat{2}=E$. Let $p^{\_}$ be the point to the left of
$p$ if $p^{\_}$ exists. For instance, $2^{-}=1$ whilst $1^{-}$ is
undefined.

The abstract semantics associates each textual program point with an
abstract substitution. The abstract substitution describes all the
substitutions that may be obtained when the execution reaches the
program point. The abstract semantics is the least solution to a
system of data flow equations - one for each program point. The system
is derived from the control flow graph of the program whose vertices
are the textual program points. Let $\points$ be the set of the
textual program points.  An edge from vertex $p$ to vertex $q$ in the
graph is denoted $\edge{q}{p}$; and it indicates that the execution
may reach $q$ immediately after it reaches $p$.

Consider the example program. We have $\points=\{1,\cdots,13\}$. The
program point $\iota=10$ is called the initial program point since it
is where the execution of the program is initiated. The abstract
substitution at $\iota=10$ is an analysis input, denoted $\pi_\iota$,
and it does not change during analysis. Thus, the data flow equation
for program point $10$ is $\fbigx(10)=\pi_{\iota}$ where $\fbigx$ is a
mapping from program points to abstract substitutions. The data flow
equations for other program points are derived by considering four
kinds of control flow that may arise during program execution. The
first kind models the execution of built-in calls. For instance, the
control may flow from program point $10$ to program point $11$ by
executing $Y=[a,b]$.  The data flow equation for program point $11$ is
$\fbigx(11)=\asys(Y=[a,b],\fbigx(10))$ where the transfer function
$\asys: {\sf Bip}\times\asub\mapsto\asub$ emulates the execution of a
built-in call. Let $\points^{bip}$ be the set of all the program
points that follow the built-in calls in the program. We have
$\points^{bip}=\{11,12\}$ for the example program. Another kind of
control flow models negation-as-failure. The transfer function for
this kind of control flow is the identity function.  For instance, the
control may flow from program point $2$ to program point $3$ since
\texttt{member}(X,K) may fail, which yields this data flow equation
$\fbigx(3)=\fbigx(2)$. Denote by $\points^{\it nf}$ the set of all the
program points that follow negative literals. We have
$\points^{\it nf}=\{3,6\}$ for the example program.

The third kind of control flow arises when a procedure-call is
performed. For instance, the control may flow from program point $1$
to program point $8$. The description of data that flow from program
point $1$ to program point $8$ is expressed as $\aunify(A,\fbigx(1),
D, \aid)$ where $\aid$ is an abstract substitution that describes
$\{\epsilon\}$. Note that $A$ is the call and $D$ the head of the
clause to which program point $8$ belongs. The control may also flow
to program point $8$ from program points 4, 8, 2 and 5. The control
flows from program point $5$ to program point $8$ when the negated
sub-goal \texttt{member}(K,L) is executed.  The descriptions of data
that flow to program point $8$ from those five source program points
are merged together using the least upper bound operation $\alub$ on
$\asub$, yielding the following data flow equation.
\begin{eqnarray*}
\fbigx(8) &\!\!=\!\!& \aunify(A,\fbigx(1),D,\aid)~\alub~
              \aunify(B,\fbigx(4),D,\aid)~\alub~
          \aunify(A,\fbigx(8),D,\aid)\\
 && ~\alub~\aunify(B,\fbigx(2),D,\aid)~\alub~\aunify(A,\fbigx(5),D,\aid)
\end{eqnarray*}
The transfer function
$\aunify:\atom_{P}\times\asub\times\atom_{P}
\times\asub\mapsto\asub$ approximates $\cunify: \atom_{P}\times \csub\times\atom_{P}\times\csub\mapsto
\csub$ defined
\[\cunify(a_1,\Theta_1,a_2,\Theta_2)
 \definedas
\{\unify(a_1,\theta_1,a_2,\theta_2)\neq\fail~|~\theta_1\in\Theta_1
               \land\theta_2\in\Theta_2\}
\] which is the set extension of $\unify$.
Denote by $\points^{call}$ the set of program points that are reached
via procedure-calls. We have $\points^{call}=\{1,4,7,8\}$ for the
example program.

The fourth kind of control flow arises when a procedure exits. For
instance, the control may flow from program point $3$ to program point
$13$. The description of data that flow from program point $3$ to
program point $13$ is expressed by $\aunify(E,\fbigx(3),F,\fbigx(12))$
where $E$ is the head of the clause to which program point $3$ belongs
and $F$ the call that invoked the clause. The only other control flow
to program point $13$ is from program point $6$. Thus, the data flow
equation for program point $13$ is \( \fbigx(13) =
\aunify(E,\fbigx(3),F,\fbigx(12)) ~\alub~
\aunify(E,\fbigx(6),F,\fbigx(12)) \). Let $\points^{ret}$ be the set
of program points that are reached via procedure-exits. For the
example program, we have $\points^{ret}=\{2,5,9,13\}$.

Let $\edges^{\jmath}=\{\edge{q}{p}\mid q\in\points^{\jmath}\}$ where
$\jmath\in\{call,ret,nf,bip\}$. Note that $\edges^{\jmath}$ is the set
of control flows that sink in $\points^{\jmath}$.  The data flow
equation has the following general form.
\[
\fbigx(q) =
\left\{ \begin{array}{ll}
\pi_{\iota} & \mbox{if $q = \iota$}\\
\alub \{ \aunify(\atomat{p},\fbigx(p),\headat{q},\aid) \mid \edge{q}{p}\in\edges\}
      & \mbox{if $q\in\points^{call}$}\\
\alub \{\aunify(\headat{q},\fbigx(q),\atomat{p^{\_}},\fbigx(p^{\_}))
    \mid   \edge{q}{p}\in\edges\} & \mbox{if $q\in\points^{ret}$}\\
\fbigx(q^{\_})  & \mbox{if $q\in\points^{\it nf}$}\\
\asys(\atomat{q^{\_}},\fbigx(q^{\_})) &  \mbox{if $q\in\points^{bip}$}
\end{array}\right.
\]
where $\pi_\iota$ is the input abstract substitution.  The least
solution to the system of data flow equations is a correct analysis
if, in addition to C1 and C2, the following local
safety requirements are met.
\begin{itemize}
   \item [C3:] \label{scheme:c3} $\{\epsilon\}\subseteq \gamma(\aid)$;
\item [C4:] $\csys(a,\gamma(\pi))\subseteq\gamma(\asys(a,\pi))$ for
any $a\in{\sf Bip}$ with ${\it vars}(a)\subseteq\vi$ and $\pi\in\asub$; and
\item
[C5:] \(\cunify(a_1,\gamma(\pi_1), a_2,\gamma(\pi_2)) \subseteq
\gamma\circ \aunify(a_1,\pi_1,a_2,\pi_2)\) for any $\pi_1,\pi_2\in
\asub$, any $a_1,a_2\in\atom_{P}$. \label{scheme:c4}
\end{itemize}
Note that the condition C2 implies that $\alub$ safely abstracts
$\club$ with respect to $\gamma$. The operation $\aunify$ is called
abstract unification since it mimics the normal unification operation
whilst $\asys$ is called abstract built-in execution operation.

The complete system of
data flow equations for the example program is as follows.

\begin{eqnarray*}
\fbigx(1) &=& \aunify(F,\fbigx(12),E,\aid)\\
\fbigx(2) &=& \aunify(C,\fbigx(7),A,\fbigx(1))~\alub~ \aunify(D,\fbigx(9),A,\fbigx(1))\\
\fbigx(3) &=& \fbigx(2)\\
\fbigx(4) &=& \aunify(F,\fbigx(12),E,\aid)\\
\fbigx(5) &=& \aunify(C,\fbigx(7),B,\fbigx(4))~\alub~ \aunify(D,\fbigx(9),B,\fbigx(4))\\
\fbigx(6) &=& \fbigx(5)\\
\fbigx(7) &=& \aunify(A,\fbigx(1),C,\aid)~\alub~
              \aunify(B,\fbigx(4),C,\aid)~\alub~\\
         && \aunify(A,\fbigx(8),C,\aid)~\alub~\aunify(B,\fbigx(2),C,\aid)~\alub~\\
	 && \aunify(A,\fbigx(5),C,\aid)\\
\fbigx(8) &=& \aunify(A,\fbigx(1),D,\aid)~\alub~
              \aunify(B,\fbigx(4),D,\aid)~\alub~\\ &&\aunify(A,\fbigx(8),D,\aid)
           ~\alub~\aunify(B,\fbigx(2),D,\aid)~\alub~\\ &&\aunify(A,\fbigx(5),D,\aid)\\
\fbigx(9) &=& \aunify(C,\fbigx(7),A,\fbigx(8))~\alub~ \aunify(D,\fbigx(9),A,\fbigx(8))\\
\fbigx(10) &=& \pi_{\iota}\\
\fbigx(11) &=& \asys(Y=[a,b],\fbigx(10))\\
\fbigx(12) &=& \asys(Z=[1,2],\fbigx(11))\\
\fbigx(13) &=& \aunify(E,\fbigx(3),F,\fbigx(12))
              ~\alub~ \aunify(E,\fbigx(6),F,\fbigx(12))
\end{eqnarray*}

 The remainder of the paper presents our type analysis
 as an abstract domain and four abstract operations as required by the
 above abstract semantics. We begin with the type language and type
 definitions.

\section{Types}\label{sec:typ}

The type language in a type system decides which sets of terms are
types.  A type is syntactically a ground term constructed from a
ranked alphabet $\cons$ and $\{\tand,\tor,\ttop,\tbot\}$ where $\tand$
and $\tor$ are binary and $\ttop$ and $\tbot$ are nullary.  Elements
of $\cons\cup\{\tand,\tor,\ttop,\tbot\}$ are called type
constructors. It is assumed that
$(\cons\cup\{\tand,\tor,\ttop,\tbot\})\cap\func=\emptyset$.  The set
of types is
$\rtype=\term(\cons\cup\{\tand,\tor,\ttop,\tbot\},\emptyset)$.  The
denotations of type constructors in $\cons$ are determined by type
definitions whilst $\tand,\tor,\ttop$ and $\tbot$ have fixed
denotations.

\subsection{Type Rules}

Types are defined by type rules.  A type parameter is a variable
 from $\pars$. A type scheme is either a type parameter or of
 the form $c(\Vector{\beta}{m})$ where $c\in\cons$ and
 $\Vector{\beta}{m}$ are different parameters.  Let $\gtype$ be the
 set of all type schemes.  A type rule is of the form
 $c(\beta_1,\cdots,\beta_{m})\tdef f(\tau_1,\cdots,\tau_{n})$ where
 $c\in\cons$, $f/n\in\func$, $\beta_1,\cdots,\beta_{m}$ are different
 type parameters, and $\tau_j$ is a type scheme with type parameters
 from $\{\Vector{\beta}{m}\}$.  Note that every type parameter in the
 right-hand side of a type rule must occur in the left-hand side.
 Overloading of function symbols is permitted since a function symbol
 can appear in the right-hand sides of two or more type rules.  Let
 $\rules$ be the set of all type rules.  We assume that each function
 symbol occurs in at least one type rule and that each type
 constructor occurs in at least one type rule. Type rules are similar
 to type definitions used in typed logic programming languages
 Mercury~\cite{Mercury} and G\"{o}del~\cite{Godel}.

\begin{example} \label{ex1}  Let
$\func=\{0,s(),[~],[~|~], void, tr(,,)\}$ and
$\cons=\{\tnat,\teven,\todd, \tlist(),\ttree()\}$. The following set of type
rules will be used in examples throughout the paper.
\[ \rules= \left\{\begin{array}{ll}
                   \tnat \tdef 0, & \tnat \tdef s(\tnat), \\
                   \teven \tdef 0,& \teven \tdef s(\todd),\\
                   \todd\tdef s(\teven), &\\
                   \tlist(\beta) \tdef [~], &
                \tlist(\beta) \tdef [\beta|\tlist(\beta)] \\
                   \ttree(\beta) \tdef void, &
                \ttree(\beta) \tdef tr(\beta,\ttree(\beta),\ttree(\beta))
                  \end{array}
           \right\}
\]
Type rules in $\rules$ define natural numbers, even numbers, odd
numbers, lists and trees.  \endprf
\end{example}

\subsection{Denotations of Types}\label{sec:meaning}

A (ground) type substitution is a member of $\tvals\definedas
(\pars\parfun\rtype)\cup\{\tvaltop,\tvalbot\}$.  The application of a
type substitution to a type scheme is defined as follows.
$\tvaltop(\tau)\definedas\ttop$ and $\tvalbot(\tau)\definedas\tbot$
for any type scheme $\tau$. Let $\Bbbk \in
(\pars\parfun\rtype)$. Define $\Bbbk(\beta)\definedas\tbot$ for each
$\beta\not\in\dom(\Bbbk)$ where $\dom(\Bbbk)$ is the domain of
$\Bbbk$.  Then $\Bbbk(\tau)$ is obtained by replacing each $\beta$ in
$\tau$ with $\Bbbk(\beta)$.  For instance, $\{\beta_1\values
\tlist(\tnat),\beta_2\values \tnat\}(\tlist(\beta_1)) =
\tlist(\tlist(\tnat))$.
\begin{definition}\label{def:sem}
The meaning of a type is defined by a function
$\sem{\rules}{\cdot}:\rtype\mapsto\wp(\term)$.
\[\begin{array}{rcl}
\sem{\rules}{\ttop} &\definedas & \term\\
\sem{\rules}{\tbot} &\definedas & \emptyset\\
\sem{\rules}{\tand(\rt_1,\rt_2)} &\definedas & \sem{\rules}{\rt_1}\cap\sem{\rules}{\rt_2}\\
\sem{\rules}{\tor(\rt_1,\rt_2)} &\definedas & \sem{\rules}{\rt_1}\cup\sem{\rules}{\rt_2}\\
\sem{\rules}{c(\Vector{\rt}{m})} &\definedas &\\
\multicolumn{3}{r}{~\hspace{2pc}
     \bigcup_{(c(\Vector{\beta}{m})\tdef f(\Vector{\tau}{n}))\in\rules}
     \left(\begin{array}{l}
       let~\Bbbk=\{\beta_j\values\rt_j~|~1\leq{j}\leq{m}\}\\
       in~\\
       \{f(\Vector{t}{n})\mid \forall 1\leq{i}\leq{n}.
       t_{i}\in \sem{\rules}{\Bbbk(\tau_{i})}
       \}
     \end{array}\right)
     }
\end{array}
\]\endprf
\end{definition}

The function $\sem{\rules}{\cdot}$ gives fixed denotations to
$\tand,\tor,\ttop$ and $\tbot$. Type constructors $\tand$ and $\tor$
are interpreted as set intersection and set union respectively. The
type constructor $\ttop$ denotes $\term$ and $\tbot$ the empty set. We
say that a term $t$ is in a type $\rt$ iff $t\in\sem{\rules}{\rt}$.
Set inclusion and $\sem{\rules}{\cdot}$ induce a pre-order $\tleq$ on
types: $(\rt_1\tleq\rt_2)\definedas
(\sem{\rules}{\rt_1}\subseteq\sem{\rules}{\rt_2})$ and an equivalence
relation $\teq$ on types: $(\rt_1\teq\rt_2)\definedas
(\rt_1\tleq\rt_2)\land (\rt_2\tleq\rt_1)$.

\begin{example} Continuing with Example~\ref{ex1}, we have
\[ \sem{\rules}{\tnat} = \{0,s(0),s(s(0)),\cdots\}\]
\[\sem{\rules}{\tlist(\tbot)} = \{[~]\}\]
\[ \sem{\rules}{\tlist(\ttop)} =\{[~],[x|[~]],\cdots\} \]
where $x\in\allvars$. Observe that
\mbox{$\tor (\tlist(\teven),\tlist(\todd))\not\teq\tlist(\tnat)$} since 
$[0,s(0)]\in\sem{\rules}{\tlist(\tnat)}$ and 
$[0,s(0)]\not\in\sem{\rules}{\tlist(\teven)}$ and 
$[0,s(0)]\not\in\sem{\rules}{\tlist(\todd)}$.
\endprf
\end{example}

The type constructors $\tand$ and $\tor$ will sometimes be written as
infix operators, i.e., $\tand(\rt_1,\rt_2)$ is written as
$(\rt_1~\tand~\rt_2)$ and $\tor(\rt_1,\rt_2)$ as $(\rt_1~\tor~\rt_2)$.
A type is atomic if its main constructor is neither $\tand$ nor
$\tor$.  A type is conjunctive if it is of the form
$\tand_{1\leq{i}\leq{k}} \at_i$ where each $\at_i$ is atomic. By an
obvious analogy to propositional logic, for any type $\rt$, there is a
type of the form $\tor_{1\leq{i}\leq{m}}C_i$ such that each $C_i$ is
conjunctive and $\rt~\teq~\tor_{1\leq{i}\leq{m}}C_i$. We call
$\tor_{1\leq{i}\leq{m}}C_i$ a disjunctive normal form of $\rt$.

A term in a type may contain variables. This lemma states that types
are closed under instantiation.

\begin{lemma} \label{lm:closed}
Let $\rt\in\rtype$ and $t\in\term$. If $t\in\sem{\rules}{\rt}$ then
$\sigma(t)\in\sem{\rules}{\rt}$ for any $\sigma\in\sub$. \endprf
\end{lemma}

Type rules in $\rules$ are production rules for a context-free tree
grammar~\cite{tata02,GecsegS84}. The complement of the denotation of a
type is not necessarily closed under instantiation. For an instance,
let $\rules$ be defined as in Example~\ref{ex1}, $x\in\allvars$ and
$\sigma=\{x\mapsto s(0)\}$. Observe that $x\not\in\sem{\rules}{\tnat}$
and $\sigma(x)\in\sem{\rules}{\tnat}$. Since $x\in\term\setminus
\sem{\rules}{\tnat}$ and $\sigma(x)\not\in\term\setminus
\sem{\rules}{\tnat}$, $\term\setminus \sem{\rules}{\tnat}$ is not
closed under instantiation and cannot be denoted by a type in
$\rtype$. The example shows that the family of types is not closed
under complement. This explains why set complement is not a type
constructor.

Types have also been defined using tree
automata~\cite{GecsegS84,tata02}, regular term
grammars~\cite{DartZ92,Smaus01,LS01}, and regular unary logic
programs~\cite{Yardeni:Shapiro:91}. A type defined in such a
formalism denotes a regular set of ground terms. The meaning
function $\sem{\rules}{\cdot}$ interprets a type as a set of
possible non-ground terms; in particular, it interprets $\ttop$ as
the set of all terms.  Type rules are used to propagate type
information during analysis. Let $x$ be of type $\tnat$ and $y$ of
type $\tlist(\tatom)$. Then the type rule $\tlist(\beta)\tdef
[\beta|\tlist(\beta)]$ is used to infer that $[x|y]$ is of type
$\tlist(\tor(\tnat,\tatom))$. The type parameter $\beta$ is not
only used as a placeholder but also used in folding heterogeneous
types precisely via non-discriminated union operator.

\subsection{Type Sequences}
During propagation of type information, it is necessary to work with
type sequences.  A type sequence expression is an expression
consisting of type sequences of the same dimension and constructors
$\tand$ and $\tor$. Note that constructors $\tand$ and $\tor$ are
overloaded.  The dimension of the type sequence expression is defined
to be that of a type sequence in it.  Let $\rt\in\rtype$,
$\rtvec\in\rtype^{*}$ and $\fcal_1$ and $\fcal_2$ be type sequence
expressions.  We extend $\sem{\rules}{\cdot}$ to type sequence
expressions as follows.
\begin{eqnarray*}
\sem{\rules}{\epsilon} & \definedas &\{\epsilon\}\\
\sem{\rules}{\rt\bullet\rtvec} & \definedas &
\sem{\rules}{\rt}\bullet\sem{\rules}{\rtvec}\\
\sem{\rules}{\fcal_1~\tand~\fcal_2} &\definedas&
                   \sem{\rules}{\fcal_1}\cap\sem{\rules}{\fcal_2}\\
\sem{\rules}{\fcal_1~\tor~\fcal_2} &\definedas&
                   \sem{\rules}{\fcal_1}\cup\sem{\rules}{\fcal_2}
\end{eqnarray*}

The relations $\tleq$ and $\teq$ on types carry over naturally to type
sequence expressions.  An occurrence of $\vec{\tbot}$ (respectively
$\vec{\ttop}$) stands for the type sequence of $\tbot$'s (respectively
$\ttop$'s) with a dimension appropriate for the occurrence.

\section{Abstract Domain} \label{sec:dom}
Abstract substitutions in our type analysis are type constraints
represented as a set of variable typings which are mappings from
variables to types. A variable typing represents the conjunction of
primitive type constraints of the form $x\in\rt$. For instance, the
variable typing $\{x\mapsto\tnat, y\mapsto\teven\}$ represents the
type constraint $(x\in\tnat)\land(y\in\teven)$.  The restriction of a
variable typing $\mu$ to a set $V$ of variables is defined as
\[ \mu\uparrow V\definedas \lambda x. (\mbox{if $x\in V$ then $\mu(x)$ else $\ttop$})
\]
 The denotation of a variable typing is given by
$\vtcon:(\vi\mapsto\rtype)\mapsto\csub$ defined
\[ \vtcon(\mu)\definedas\{\theta~|~\forall x\in\vi.(\theta(x)\in\sem{\rules}{\mu(x)})\}
\]
 For instance,
$\vtcon(\{x\values\tnat,y\values\tlist(\tnat)\})=\{\theta~|~\theta(x)\in\sem{\rules}{\tnat}\land\theta(y)\in\sem{\rules}{\tlist(\tnat)}\}$.
The denotation of a set of variable typings is the set union of the
denotations of its elements.
\begin{example} \label{ex:scal} For instance, letting
$\scal=\{\{x\values\tnat,y\values\tlist(\tnat)\},
\{x\values\tlist(\tnat),y\values\tnat\}\}$, $\scal$ denotes
$\{\theta~|~\theta(x)\in\sem{\rules}{\tnat}\land\theta(y)\in\sem{\rules}{\tlist(\tnat)}\}
\cup\{\theta~|~\theta(x)\in\sem{\rules}{\tlist(\tnat)}\land\theta(y)\in\sem{\rules}{\tnat}\}$. \endprf
\end{example}

There may be many sets of variable typings that denote the same set of
substitutions.  Firstly, two different type expressions in $\rtype$
may denote the same set of terms. For instance,
$\sem{\rules}{\tnat~\tand~\tlist(\ttop)}=\sem{\rules}{\tbot}$ using
$\rules$ in Example~\ref{ex1}. Secondly, an element of a set of
variable typings may have a smaller denotation than another. For an
example, let
$\scal=\{\{x\mapsto\tlist(\ttop)\},\{x\mapsto\tlist(\tnat)\}\}$. Then
$\scal$ has the same denotation as one of its proper subset
$\scal'=\{\{x\mapsto\tlist(\ttop)\}\}$.  Those abstract elements that
have the same denotation are identified.  Let $\rpreceq$ on
$\wp(\vi\mapsto\rtype)$ be defined as
$\scal_1\rpreceq\scal_2\definedas
(\bigcup_{\mu\in\scal_1}\vtcon(\mu))\subseteq(\bigcup_{\nu\in\scal_2}\vtcon(\nu))$.
It is a pre-order and induces an equivalence relation $\req$ on
$\wp(\vi\mapsto\rtype)$:
$(\scal_1\req\scal_2)\definedas(\scal_1\rpreceq\scal_2)\land(\scal_2\rpreceq\scal_1)$.
The equivalence classes with respect to $\req$ are abstract
substitutions. Thus, the abstract domain is $\langle\rsub,\rleq
\rangle$ where
\begin{eqnarray*}
\rsub &\definedas& {\wp(\vi\mapsto\rtype)}_{/\req} \\
\rleq &\definedas& {\rpreceq}_{/\req}
\end{eqnarray*}
$\langle\rsub,\rleq \rangle$ is a complete lattice. Its join and meet
operators are respectively \( {[\scal_1]}_\req \rlub {[\scal_2]}_\req
= {[\scal_1\cup\scal_2]}_\req \) and \( {[\scal_1]}_\req \rglb
{[\scal_2]}_\req =
{[\scal_1^{\downarrow}\cap\scal_2^{\downarrow}]}_\req \) where
$\scal_{i}^{\downarrow}\definedas
\{\mu\in(\vi\mapsto\rtype)\mid
\exists\nu\in\scal_{i}.(\vtcon(\mu)\subseteq\vtcon(\nu))\}$. The
infimum is ${[\emptyset]}_\req$ and the supremum
${[\{x\values\ttop~|~x\in\vi\}]}_\req$.  The concretization function
$\rcon:\rsub\mapsto\csub$ is defined
\[\rcon({[\scal]}_\req)\definedas\bigcup_{\mu\in\scal}\vtcon(\mu)\]
The following lemma states that $\rcon$ satisfies the safety
requirement C2.

\begin{lemma} $\rcon(\rsub)$ is a Moore family. \label{lm:A} \endprf
\end{lemma}

The definition of $\rglb$ is not constructive since the downward
closure of a set of variable typings $\scal$ can be infinite. For
instance, letting $\scal=\{\{x\values\tlist(\ttop)\}\}$,
$\{x\values{\tlist}^k(nat)\}$ is in $\scal^{\downarrow}$ for any
$k\geq 1$. The following operator
$\otimes:\wp(\vi\mapsto\rtype)\times\wp(\vi\mapsto\rtype)\mapsto\wp(\vi\mapsto\rtype)$
computes effectively the meet of abstract substitutions.
\[\scal_1\otimes\scal_2 \definedas
   \{ \{x\values(\mu(x)~\tand~\nu(x))~|~x\in\vi\}
      ~|~\mu\in\scal_1\land\nu\in\scal_2\}
\]

If $\scal_1$ and $\scal_2$ are finite representatives of two abstract
substitutions then $\scal_1\otimes\scal_2$ is a finite representative
of the meet of the abstract substitutions, which is stated in this
lemma.

\begin{lemma} \label{lm:B}   $\rcon({[\scal_1\otimes\scal_2]}_\req)
=\rcon({[\scal_1]}_\req\rglb{[\scal_2]}_\req)$. \endprf
\end{lemma}

We will use a fixed renaming substitution $\Psi$ such that
$\vi\cap\Psi(\vi)=\emptyset$ and define
$\viplus\definedas\vi\cup\Psi(\vi)$.  The relation $\req$, the
functions $\vtcon$ and $\rcon$ and the operator $\otimes$ extend
naturally to sets of variable typings over $\viplus$. Let $\mu$ be a
variable typing, $\scal$ a set of variable typings and $\theta$ a
substitution. We say that $\theta$ satisfies $\mu$ if
$\theta\in\vtcon(\mu)$; and we say that $\theta$ satisfies $\scal$ if
$\theta\in\rcon({[\scal]}_\req)$.

The conditions C1 and C2 are satisfied. C1 holds because
$\langle\rsub,\rleq,\rlub\rangle$ is a complete lattice. C2 is implied
by Lemma~\ref{lm:A}.

\section{Abstract Operations} \label{sec:ops}
The design of our type analysis is completed with four abstract
operations required by the abstract semantics given in
Section~\ref{sec:pre}. One operation is $\alub$ which is the least
upper bound on $\langle\asub,\aleq\rangle$. Let $\aid={[\{\lambda
x\in\vi.\ttop\}]}_\req$. The operation $\aid$ obviously satisfies the
condition $C3$ and thus safely abstracts $\{\epsilon\}$ with respect to
$\rcon$. Since abstract built-in execution operation $\asys$ makes use of
ancillary operations for abstract unification operation $\aunify$, we
present $\aunify$ before $\asys$.

\subsection{Outline of Abstract Unification}

The abstract unification operator $\aunify$ takes two atoms and two
abstract substitutions and computes an abstract substitution.  The
computation is reduced to solving a constraint that consists of a set
of equations in solved form $E$ and a set of variable typings
$\scal_i$.  The solution to the constraint is a set of variable
typings $\scal_o$. In order to ensure that $\aunify$ safely abstracts
$\cunify$, $\scal_o$ is required to describe the set of all those
substitutions that satisfy both $E$ and $\scal_i$. Let $E=\{x_1=
t_1,\cdots,x_n= t_n\}$. The set $\scal_o$ is computed in two steps. In
the first step, type information about $x_i$ is used to derive more
type information about the variables in $t_i$. This is a downward
propagation since type information is propagated from a term to its
sub-terms.  The second step propagates type information in the opposite
direction. It derives more type information about $x_i$ from type
information about the variables in $t_i$.

For an illustration, let $E=\{x=[w],y=[w]\}$ and $S_i=\{\mu\}$ where
$\mu=\{w\values\ttop, x\values\tlist(\tatom~\tor~\tfloat),y\values
\tlist(\tatom~\tor~\tint)\}$. During the downward propagation step,
more type information for $w$ is derived from type information for
both $x$ and $y$. Since $\mu(x)=\tlist(\tatom~\tor~\tfloat)$ and
$x=[w]$, $[w]$ is of type $\tlist(\tatom~\tor~\tfloat)$.  Since there
is only one type rule for $[\cdot|\cdot]$: $\tlist(\beta)\tdef
[\beta|\tlist(\beta)]$, we deduce that $w$ is of type
$(\tatom~\tor~\tfloat)$.  Similarly, we deduce that $w$ is of type
$(\tatom~\tor~\tint)$ since $\mu(y)=\tlist(\tatom~\tor~\tint)$ and
$y=[w]$. So, $w$ is of type
$((\tatom~\tor~\tfloat)~\tand~(\tatom~\tor~\tint))$ that is equivalent
to $\tatom$. The derived type $\tatom$ for $w$ is used to strengthen
$\mu$ into $\nu=\{w\values\tatom,
x\values\tlist(\tatom~\tor~\tfloat),y\values
\tlist(\tatom~\tor~\tint)\}$. During the upward propagation step, more
type information for both $x$ and $y$ is derived type information for
$w$. Note that $[w]$ is an abbreviation for $[w|[~]]$. By applying the
type rule $\tlist(\beta)\tdef [~]$, we infer that $[~]$ is of type
$\tlist(\tbot)$.
Since $\nu(w)=\tatom$, we derive that $[w]$ is of type
$\tlist(\tatom)$ by applying the type rule $\tlist(\beta)\tdef
[\beta|\tlist(\beta)]$.  We deduce that both $x$ and $y$ are of type
$\tlist(\tatom)$ since $x=[w]$ and $y=[w]$. The derived type
$\tlist(\tatom)$ for $x$ and $y$ is used to strengthen $\nu$,
resulting in this singleton set of variable typing
$S_o=\{\{w\values\tatom, x\values\tlist(\tatom),
y\values\tlist(\tatom)\}\}$. Both the downward and upward propagation
steps in the preceding example produce a single output variable typing
from an input variable typing. In more general cases, both steps may
yield multiple output variable typings from an input variable
typing. We now present in details these two steps.

\subsection{Downward Propagation} \label{sec:down}
Downward propagation requires propagating a type $\rt$ downwards (the
structure of) a term $t\in\term(\func,\viplus)$. Let
\(\Theta=\{\theta\mid \theta(t)\in\sem{\rules}{\rt}\}\). Propagation
of $\rt$ downwards $t$ calculates a set of variable typings $\scal$
(computed as $\vts(\rt,t)$) such that
$\Theta\subseteq\rcon({[\scal]}_\req)$, that is, $\scal$ describes the
set of all those substitutions that instantiate $t$ to a term of type
$\rt$. This is done by a case analysis. If $\rt=\ttop$ then
$\Theta=\sub$ since $\theta(t)$ is in $\rt$ for any
$\theta\in\sub$. Put $\scal=\{\lambda y\in\viplus. \ttop\}$. Then
$\scal$ satisfies the condition that
$\Theta\subseteq\rcon({[\scal]}_\req)$.  If $t\in\viplus$ then
$\scal=\{\lambda y\in\viplus.(\mbox{if $y=t$ then $\rt$ else
$\ttop$})\}$ satisfies the condition that
$\Theta\subseteq\rcon({[\scal]}_\req)$. Consider the case
$\rt=(\rt_1~\tor~\rt_2)$. We have $\Theta=\Theta_1\cup\Theta_2$ where
\(\Theta_1=\{\theta_1\mid \theta_1(t)\in\sem{\rules}{\rt_}\}\) and
\(\Theta_2=\{\theta_2\mid \theta_2(t)\in\sem{\rules}{\rt_2}\}\). We
propagate the types $\rt_1$ and $\rt_2$ downwards $t$ separately,
obtaining two sets of variable typings $\scal_1$ and $\scal_2$ such
that $\Theta_1\subseteq\rcon({[\scal_1]}_\req)$ and
$\Theta_2\subseteq\rcon({[\scal_2]}_\req)$. Put
$\scal=\scal_1\cup\scal_2$. Then the condition that
$\Theta\subseteq\rcon({[\scal]}_\req)$ is satisfied. For the case
$\rt=\rt_1~\tand~\rt_2$, $\scal=\scal_1\otimes\scal_2$ satisfies the
condition that $\Theta\subseteq\rcon({[\scal]}_\req)$ where $\scal_1$
and $\scal_2$ are obtained as above. Consider the remaining case
$\rt=c(\rt_1,\cdots,\rt_2)$ and $t=f(t_1,\cdots,t_n)$. Assume that
there are $k$ type rules $\Upsilon^1,\cdots,\Upsilon^k$ for $c/m$ and
$f/n$ and $\Upsilon^j$ is $c(\beta_1^j,\cdots,\beta_m^j)\tdef
f(\tau_1^j,\cdots,\tau_n^j)$. By the definition of
$\sem{\rules}{\cdot}$, $\Theta=\bigcup_{1\leq j\leq k} \Theta_j$ where
\begin{eqnarray*}
\Theta_j &=& \{\theta\mid \theta(f(t_1,\cdots,t_n))\in
\{f(s_1,\cdots,s_n)\mid \forall 1\leq i\leq
n. (s_i\in\sem{\rules}{\kappa^j(\tau_i^j))} \}\}\\
&=& \{\theta\mid f(\theta(t_1),\cdots,\theta(t_n))\in
\{f(s_1,\cdots,s_n)\mid \forall 1\leq i\leq
n. (s_i\in\sem{\rules}{\kappa^j(\tau_i^j))} \}\}\\
&=& \{\theta\mid
\forall 1\leq i\leq
n. (\theta(t_i)\in\sem{\rules}{\kappa^j(\tau_i^j))} \\
&=& \Theta_1^j\cap\Theta_2^j\cap\cdots\cap\Theta_n^j
\end{eqnarray*}
and $\kappa^j=\{\beta_1^j\mapsto\rt_1,\cdots,\beta_m^j\mapsto\rt_m\}$
and
$\Theta_i^j=\{\theta\mid\theta(t_i)\in\sem{\rules}{\kappa^j(\tau_i^j)}\}$.
We obtain $\scal$ as follows. We first propagate type
$\kappa^j(\tau_i^j)$ downwards term $t_i$, obtaining a set of variable
typings $\scal_i^j$. We have that
$\Theta_i^j\subseteq\rcon({[\scal_i^j]}_\req)$. We then calculate
$\scal^j=\scal_1^j\otimes\cdots\otimes\scal_n^j$ for the type rule
$\Upsilon^j$. The set $\scal^j$ satisfies the condition that
$\Theta^j\subseteq \rcon({[\scal^j]}_\req)$.  Finally, we compute
$\scal=\scal^1\cup\cdots\cup\scal^k$. Since $\Theta^j\subseteq
\rcon({[\scal^j]}_\req)$ and $\Theta=\bigcup_{1\leq j\leq k}\Theta^j$,
$\scal$ satisfies the condition that
$\Theta\subseteq\rcon({[\scal]}_\req)$. In summary,
$\scal=\vts(\rt,t)$ where
$\vts:\rtype\times\term(\Sigma,\viplus)\mapsto\wp(\viplus\mapsto\rtype)$
is defined
\[
\begin{array}{rcl}
\vts(\ttop,t) &\definedas& \{\lambda y\in\viplus. \ttop\}\\
\vts(\rt,x) &\definedas& \{\lambda y\in\viplus.(\mbox{if $y=x$ then $\rt$ else $\ttop$}) \}\\
\vts((\rt_1~\tand~\rt_2),t) &\definedas&
    \vts(\rt_1,t)\otimes\vts(\rt_2,t)\\
\vts((\rt_1~\tor~\rt_2),t) &\definedas&
    \vts(\rt_1,t)\cup\vts(\rt_2,t)\\
\vts(c(\Vector{\rt}{m}),f(\Vector{t}{n})) &\definedas & \\
\multicolumn{3}{r}{
 \bigcup_{(c(\Vector{\beta}{m})\tdef f(\Vector{\tau}{n}))\in\rules}
  \left( \begin{array}{l} let~\Bbbk=\{\beta_j\values\rt_j~|~1\leq{j}\leq{m}\}\\
   in\\
   \bigotimes_{1\leq{i}\leq{n}} \vts(\Bbbk(\tau_i),t_i)
   \end{array}\right)
}
\end{array}
\]
where $x\in\viplus$, $f/n\in\func$ and $c/m\in\cons$. The first one
applies when there are multiple applicable alternatives.

The following lemma states that $\vts(\rt,t)$ describes all the
substitutions that instantiate $t$ to a term of type $\rt$.
\begin{lemma}\label{lm:C}
For any $\rt\in\rtype$ and $t\in\term(\func,\viplus)$,
$\{\theta~|~\theta(t)\in\sem{\rules}{\rt}\}\subseteq\rcon({[\vts(\rt,t)]}_\req)$.
 \endprf
\end{lemma}

We now consider the overall downward propagation given a set of
variable typings $\scal$ and a set of equations in solved form
$E=\{x_1=t_1,\cdots,x_n=t_n\}$. Each variable typing $\mu$ in $\scal$
is processed separately as follows.  We first propagate the type
$\mu(x_i)$ downwards $t_i$. This results in a set of variable typings
$\vts(\mu(x_i),t_i)$ which describes all the substitutions that
instantiate $t_i$ to a term of type $\mu(x_i)$. We then calculate
$\scal_\mu=\vts(\mu(x_1),t_1)\otimes\cdots\otimes\vts(\mu(x_n),t_n)$. The
set $\scal_\mu$ describes all the substitutions that instantiate $t_i$
to a term of type $\mu(x_i)$ for all $1\leq i\leq n$. We finally
conjoin $\scal_\mu$ with $\{\mu\}$, obtaining
$\{\mu\}\otimes\scal_\mu$ which describes all the substitutions that
satisfy both $\mu$ and $E$.  After each variable typing in $\scal$
is processed, results from different variable typings are joined
together using set union.  The overall downward propagation function
$\mdown:\wp(\eqs)\times\wp(\viplus\mapsto\rtype) \mapsto
\wp(\viplus\mapsto\rtype)$ is defined

\begin{equation}
\mdown(E,\scal)\definedas
\bigcup_{\mu\in\scal}(\{\mu\}\otimes\bigotimes_{(x=t)\in{E}}\vts(\mu(x),t))
\end{equation}

\begin{example} \label{ex:mdown} Let $\viplus=\{x,y\}$,
 $\scal=\{\{x\values\ttop,y\values(\tlist(\tnat)~\tor~\tnat)\}\}$ and
 $\rules$ be that in Example~\ref{ex1}. We have
 $\vts(\tlist(\tnat),[x|[~]]) = \{\{x\values\tnat,y\values\ttop\}\}$
 and $\vts(\tnat,[x|[~]])=\emptyset$. So,
\[\vts(\tlist(\tnat)~\tor~\tnat,[x|[~]]) =
 \{\{x\values\tnat,y\values\ttop\}\}\]
and
\begin{eqnarray*}
\lefteqn{\mdown(\{y=[x|[~]]\},\scal)}\\
& = &
 \{\{ x\values\ttop, y\values(\tlist(\tnat)~\tor~\tnat)\}\}\otimes
 \{\{x\values\tnat,y\values\ttop\}\}\\
& = & \{\mu\}
\end{eqnarray*}
 where
 $\mu=\{x\values\tnat,y\values(\tlist(\tnat)~\tor~\tnat)\}$.
 \endprf
\end{example}

The following lemma states the correctness of downward propagation.

\begin{lemma}\label{lm:G}
Let $\scal'=\mdown(E,\scal)$. Then
$\mgu(\theta(E))\circ\theta\in\rcon({[\scal']}_\req)$ for all
$\theta\in\rcon({[\scal]}_\req)$. \endprf
\end{lemma}

\subsection{Upward Propagation}
We now consider upward propagation of type information. The key step
in upward propagation is to compute a type for a term from those of
its variables.  We first consider how a type rule $\tau\tdef
f(\Vector{\tau}{n})$ can be applied to compute a type of
$f(\Vector{t}{n})$ from types of its top-level sub-terms
$\Vector{t}{n}$.  Let $\rt_i$ be the type of $t_i$.  A simplistic
approach would compute a type substitution $\Bbbk$ such that
${\langle\Vector{\rt}{n}\rangle}\tleq{\Bbbk(\langle\Vector{\tau}{n}\rangle)}$
and then return $\Bbbk(\tau)$ as the type of $t$. However, this leads
to loss of precision. Consider the term $[x|y]$ and the type rule
$list(\beta)\tdef [\beta|\tlist(\beta)]$. Let the types of $x$ and $y$
be $(\teven~\tor~\todd)$ and $\tlist(\tbot)$. Then the minimal type
substitution $\Bbbk$ such that ${\langle \teven~\tor~\todd,
\tlist(\tbot)\rangle}\tleq{\Bbbk(\langle\beta,\tlist(\beta)\rangle)}$
is $\Bbbk=\{\beta\mapsto(\teven~\tor~\todd)\}$. We would obtain
$\Bbbk(\tlist(\beta))=\tlist(\teven~\tor~\todd)$ as a type of
$[x|y]$. A more precise type of $[x|y]$ is
$(\tlist(\teven)~\tor~\tlist(\todd))$. We first compute a set of type
substitutions $\kcal$ such that
${\langle\Vector{\rt}{n}\rangle}\tleq\tor_{\Bbbk\in\kcal}
{\Bbbk(\langle\Vector{\tau}{n}\rangle)}$ and then return
$\tor_{\Bbbk\in\kcal}\Bbbk(\tau)$ as a type of
$f(\Vector{t}{n})$. Continue with the above example.  Let
$\kcal=\{\{\beta\mapsto\teven\},\{\beta\mapsto\todd\}\}$.  Then
${\langle \teven~\tor~\todd, \tlist(\tbot)\rangle}\tleq
\tor_{\Bbbk\in\kcal}{\Bbbk(\langle\beta,\tlist(\beta)\rangle)}$. We
obtain $(\tlist(\teven)~\tor~\tlist(\todd))$ as a type of $[x|y]$.

\begin{definition}
Let $\tau\in\gtype$, $\vec{\tau}\in\gtype^*$, $\rt\in\rtype$,
$\vec{\rt}\in\rtype^*$ and $\kcal\in\wp(\tvals)$.  We say that $\kcal$
is a cover for $\rt$ and $\tau$ iff
$\rt\tleq\tor_{\Bbbk\in\kcal}\Bbbk(\tau)$. We say that $\kcal$ is a
cover for $\vec{\rt}$ and $\vec{\tau}$ iff
$\vec{\rt}\tleq\tor_{\Bbbk\in\kcal}\Bbbk(\vec{\tau})$.
\endprf
\end{definition}

Calculating a cover for a type and a type scheme is a key task in
upward propagation of type information. Before defining a function
that does the computation, we need some operations on type
substitutions.

\subsubsection{Operations on Type Substitutions}
  We first introduce an operation for calculating an upper bound of
  two type substitutions. It is the point-wise extension of $\tor$
  when both of its operands are mappings from type parameters to
  types.  Define $\tvallub:\tvals\times\tvals\mapsto\tvals$ as
  follows.
\[
{\Bbbk_1\tvallub\Bbbk_2}
    \definedas
       \left\{\begin{array}{lr}
         \tvaltop, & \mathtt{if}~(\Bbbk_1=\tvaltop)\lor(\Bbbk_2=\tvaltop);\\
         \Bbbk_2, & \mathtt{else~if}~(\Bbbk_1=\tvalbot);\\
         \Bbbk_1, & \mathtt{else~if}~(\Bbbk_2=\tvalbot);\\
         \multicolumn{2}{l}{\{\beta\values(\Bbbk_1(\beta)~\tor~\Bbbk_2(\beta)) \mid \beta\in dom(\Bbbk_1)\cup dom(\Bbbk_2)\},~~~~~~\mathtt{otherwise.}}
       \end{array}\right.
\]
An operation $\tvalglb:\tvals\times\tvals\mapsto\tvals$ that
calculates a lower bound of  type substitutions is defined dually:
\[ {\Bbbk_1\tvalglb\Bbbk_2}
   \definedas
       \left\{\begin{array}{lr}
         \tvalbot, & \mathtt{if}~(\Bbbk_1=\tvalbot)\lor(\Bbbk_2=\tvalbot);\\
         \Bbbk_2, & \mathtt{else~if}~(\Bbbk_1=\tvaltop);\\
         \Bbbk_1, & \mathtt{else~if}~(\Bbbk_2=\tvaltop);\\
         \multicolumn{2}{l}{\{\beta\values (\Bbbk_1(\beta)~\tand~\Bbbk_2(\beta))\mid\beta\in dom(\Bbbk_1)\cap dom(\Bbbk_2)\},~~~~~~\mathtt{otherwise}.}
       \end{array}\right.
\]

The following lemma states that the operations $\tvallub$ and
$\tvalglb$ indeed compute upper and lower bounds of two type
substitutions respectively.
\begin{lemma}
\label{P:1} For any $\tau\in\gtype$ and any $\Bbbk_1,\Bbbk_2\in\tvals$,
\begin{itemize}
\item [(a)] $(\Bbbk_1(\tau)~\tor~\Bbbk_2(\tau)) \tleq {(\Bbbk_1\tvallub\Bbbk_2)(\tau)}$; and
\item [(b)] $(\Bbbk_1(\tau)~\tand~\Bbbk_2(\tau)) \teq {(\Bbbk_1\tvalglb\Bbbk_2)(\tau)}$.
\end{itemize} \endprf
\end{lemma}

While the type substitution operation is a meet homomorphism according
to Lemma~\ref{P:1}.(b), it is not a join homomorphism. For an
instance, let $\tau=\tlist(\beta)$, $\Bbbk_1=\{\beta\mapsto \tnat\}$
and $\Bbbk_2=\{\beta\mapsto\tlist(\tnat)\}$. Then
$\Bbbk_1\tvallub\Bbbk_2=\{\beta\mapsto (\tnat~\tor~\tlist(\tnat))\}$,
$(\Bbbk_1\tvallub\Bbbk_2)(\tau)=\tlist(\tnat~\tor~\tlist(\tnat))$, and
$\Bbbk_1(\tau)~\tor~\Bbbk_2(\tau) =
\tlist(\tnat)~\tor~\tlist(\tlist(\tnat))$. Observe that
$(\Bbbk_1(\tau)~\tor~\Bbbk_2(\tau))\not\teq
(\Bbbk_1\tvallub\Bbbk_2)(\tau)$ since the term $[0,[0]]$ has type
\tlist(\tnat~\tor~\tlist(\tnat)) but it does not have type
$(\tlist(\tnat)~\tor~\tlist(\tlist(\tnat)))$.

Let $\kcal_1$ and $\kcal_2$ be sets of type substitutions. We say that
 $\kcal_1$ and $\kcal_2$ are equivalent, denoted as
 $\kcal_1\cong\kcal_2$, iff $(\tor_{\Bbbk\in\kcal_1}\Bbbk(\tau))\teq
 (\tor_{\Bbbk\in\kcal_2}\Bbbk(\tau))$ for any type scheme $\tau$.
 Define
 $\bigtvallub,\bigtvalglb:\wp(\tvals)\times\wp(\tvals)\mapsto\wp(\tvals)$
 as the set extensions of $\tvallub$ and $\tvalglb$ respectively:

\begin{eqnarray*}
\kcal_1 \bigtvallub \kcal_2 &\definedas& \{\Bbbk_1\tvallub\Bbbk_2~|~ \Bbbk_1\in\kcal_1\land\Bbbk_2\in\kcal_2\}\\
\kcal_1 \bigtvalglb \kcal_2 &\definedas& \{\Bbbk_1\tvalglb\Bbbk_2~|~ \Bbbk_1\in\kcal_1\land\Bbbk_2\in\kcal_2\}
\end{eqnarray*}

\begin{example}
Let $\kcal_1=\{\{\beta_1\values\ttree(\tnat),\beta_2\values\tnat\},
     \{\beta_1\values\tlist(\tnat),\beta_2\values\tnat\}\}$ and
     $\kcal_2=\{\{\beta_1\values\tlist(\teven),\beta_2\values\teven\}\}$.
     Since $\teven\tleq\tnat$ and $\tlist(\teven)\tleq\tlist(\tnat)$,
     we have
\begin{eqnarray*}
   \kcal_1 \bigtvallub \kcal_2 &=&
    \left\{\begin{array}{c}
          \{\beta_1\values\ttree(\tnat)~\tor~\tlist(\teven),\beta_2\values\tnat~\tor~\teven\},\\
      \{\beta_1\values\tlist(\tnat)~\tor~\tlist(\teven),\beta_2\values\tnat~\tor~\teven\}
      \end{array}
    \right\} \\
    &\cong&
    \left\{\begin{array}{c}
      \{\beta_1\values\ttree(\tnat)~\tor~\tlist(\teven),\beta_2\values\tnat\},\\
      \{\beta_1\values\tlist(\tnat),\beta_2\values\tnat\}
    \end{array}\right\}
\end{eqnarray*} We also have
\begin{eqnarray*}
   \kcal_1 \bigtvalglb \kcal_2 &=&
    \left\{\begin{array}{c}
          \{\beta_1\values(\ttree(\tnat)~\tand~\tlist(\teven)),
        \beta_2\values(\tnat~\tand~\teven)\},\\
      \{\beta_1\values(\tlist(\tnat)~\tand~\tlist(\teven)),
        \beta_2\values(\tnat~\tand~\teven)\}
      \end{array}
    \right\} \\
    &\cong&
    \left\{\begin{array}{c}
      \{\beta_1\values(\ttree(\tnat)~\tand~\tlist(\teven)),\beta_2\values\teven\},\\
      \{\beta_1\values\tlist(\teven),\beta_2\values\teven\}
    \end{array}\right\} \\
    &\cong&
    \{ \{\beta_1\values\tlist(\teven),\beta_2\values\teven\}
        \}
\end{eqnarray*} since $(\ttree(\tnat)~\tand~\tlist(\teven))\teq\tbot$. \endprf
\end{example}

A cover for a type sequence and a type scheme sequence can be computed
compositionally according to the following lemma.
\begin{lemma} \label{P:2}
 Let $\kcal_1,\kcal_2\in\wp(\tvals)$,
$\rt\in\rtype$, $\tau\in\gtype$, $\rtvec\in\rtype^{*}$ and
$\tauvec\in\gtype^{*}$ such that $\|\rtvec\|=\|\tauvec\|$. If
${\rt}\tleq\tor_{\Bbbk_1\in\kcal_1}{\Bbbk_1(\tau)}$ and ${\rtvec}
\tleq \tor_{\Bbbk_2\in\kcal_2}{\Bbbk_2(\tauvec)}$ then
${\rt\bullet\rtvec} \tleq
\tor_{\Bbbk\in(\kcal_1\bigtvallub\kcal_2)}{\Bbbk(\tau\bullet\tauvec)}$. \endprf
\end{lemma}

\subsubsection{Calculating a Cover}
We now consider how to compute a cover $\kcal$ for a type $\rt$ and a
type scheme $\tau$. In the case $\rt=\ttop$, $\kcal=\{\tvaltop\}$ is a
cover since $\tvaltop(\tau)=\ttop$; and $\kcal=\{\tvalbot\}$ is a
cover in the case $\rt=\tbot$ since $\tvalbot(\tau)=\tbot$.  Consider
the case $\rt=(\rt_1~\tor~\rt_2)$, a cover $\kcal_j$ can be
recursively computed for $\rt_j$ and $\tau$ for $j=1,2$. We have that
$\rt_j\tleq\tor_{\Bbbk\in\kcal_j}\Bbbk(\tau)$ and hence that
$(\rt_1~\tor~\rt_2)\tleq\tor_{\Bbbk\in(\kcal_1\cup\kcal_2)}\Bbbk(\tau)$. So,
the union of $\kcal_1$ and $\kcal_2$ is a cover for $\rt$ and
$\tau$. Consider the case $\rt=(\rt_1~\tand~\rt_2)$. A cover $\kcal_j$
can be recursively computed for $\rt_j$ and $\tau$ for $j=1,2$. Let
$\kcal=\kcal_1\bigtvalglb\kcal_2 =\{\Bbbk_1\tvalglb\Bbbk_2\mid
\Bbbk_1\in\kcal_1\land \Bbbk_2\in\kcal_2 \}$. Then
$\tor_{\Bbbk\in\kcal}(\tau)=\tor_{\Bbbk_1\in\kcal_1\land
\Bbbk_2\in\kcal_2}(\Bbbk_1\tvalglb\Bbbk_2)(\tau)$. By
Lemma~\ref{P:1}.(b),
$\tor_{\Bbbk\in\kcal}(\tau)=\tor_{\Bbbk_1\in\kcal_1\land
\Bbbk_2\in\kcal_2} (\Bbbk_1(\tau)~\tand~\Bbbk_2(\tau))$ and hence
$\tor_{\Bbbk\in\kcal}(\tau)=(\tor_{\Bbbk_1\in\kcal_1}\Bbbk_1(\tau))~\tand~
(\tor_{\Bbbk_2\in\kcal_2}\Bbbk_2(\tau))$. So,
$\kcal=\kcal_1\bigtvalglb\kcal_2$ is a cover for $\rt$ and $\tau$. In
the case $\rt$ is atomic and $\tau$ is a type parameter,
$\kcal=\{\{\tau\mapsto\rt\}\}$ is a cover for $\rt$ and $\tau$. In the
remaining case, $\rt=c(\rt_1,\cdots,\rt_m)$ and
$\tau=d(\Vector{\beta}{k}))$. If $c/m=d/k$ then $\{\{\beta_j\mapsto
\rt_j~|~1\leq{j}\leq{m}\}\}$ is a cover. Otherwise $\{\tvaltop\}$ is a
cover. In summary, the function that computes a cover is
$\tvs:\rtype\times\gtype\mapsto\wp(\tvals)$ defined

\[\begin{array}{rcl}
\tvs(\ttop,\tau) &\definedas & \{\tvaltop\}\\
\tvs(\tbot,\tau) &\definedas & \{\tvalbot\}\\
\tvs((\rt_1~\tor~\rt_2),\tau) &\definedas&
         \tvs(\rt_1,\tau)\cup\tvs(\rt_2,\tau)\\
\tvs((\rt_1~\tand~\rt_2),\tau) &\definedas&
         \tvs(\rt_1,\tau)\bigtvalglb\tvs(\rt_2,\tau)\\
\tvs(\rt,\beta)  &\definedas & \{\{\beta\mapsto\rt\}\}\\
\tvs(c(\Vector{\rt}{m}),d(\Vector{\beta}{k}))  & \definedas &\\
\multicolumn{3}{r}{ \left\{
         \begin{array}{l}
           \mathit{if}~(c/m)=(d/k)~then~
          \{\{\beta_j\mapsto \rt_j~|~1\leq{j}\leq{m}\}\} \\
       else~\{\tvaltop\}
         \end{array}
                  \right.}
\end{array}
\]

\begin{example} Let $\cons$ be given in Example~\ref{ex1}. Then,
\begin{eqnarray*}
\lefteqn{\tvs((\tlist(\tnat)~\tand~\ttree(\teven)),\tlist(\beta))}\\
   &=& \tvs(\tlist(\tnat),\tlist(\beta)) \bigtvalglb
     \tvs(\ttree(\teven),\tlist(\beta)) \\
   &\cong&
   \{\{\beta\values\tnat\}\} \bigtvalglb \{\tvaltop\}\\
   & = &
   \{\{\beta\values\tnat\}\}
\end{eqnarray*} \endprf
\end{example}

The following lemma states that $\tvs(\rt,\tau)$ is a cover for $\rt$
and $\tau$.

\begin{lemma}\label{lm:D}
Let $\tau\in\gtype$, $\rt\in\rtype$ and $\kcal=\tvs(\rt,\tau)$. Then
${\rt} \tleq \tor_{\Bbbk\in\kcal}{\Bbbk(\tau)}$. \endprf
\end{lemma}

\subsubsection{Computing a Type}

The type of a term $t$ is computed from those of its variables in a
bottom-up manner. The types of the variables are given by a variable
typing $\mu$.  For a compound term $t=f(\Vector{t}{n})$, a type
$\rt_i$ is first computed from $t_i$ and $\mu$ for each
$1\leq{i}\leq{n}$. Each type rule for $f/n$ is applied to compute a
type of $t$. Types resulting from all type rules for $f/n$ are
conjoined using $\tand$. The result is a type of $t$ since
conjunctions of two or more types of $t$ is also a type of $t$.  For a
type rule $\tau\tdef f(\Vector{\tau}{n})$, a cover $\kcal_i$ for
$\rt_i$ and $\tau_i$ is computed for each $1\leq{i}\leq{n}$. Joining
covers for $1\leq i \leq n$ obtains a
cover $\kcal$ for ${\langle\Vector{\rt}{n}\rangle}$ and
$\langle\Vector{\tau}{n}\rangle$. The type that is computed
from the type rule is $\tor_{\Bbbk\in\kcal}{\Bbbk(\tau)}$. Define
$\type:\term(\func,\viplus)\times(\viplus\mapsto\rtype)\mapsto\rtype$
by
\[
\begin{array}{rcl}
\type(x,\mu)&\definedas&\mu(x)\\
\type(f(\Vector{t}{n}),\mu) &\definedas &
{ \tand_{\tau\tdef f(\Vector{\tau}{n})\in\rules}
    (\tor_{\Bbbk\in(\bigtvallub_{1\leq{i}\leq{n}}
    \tvs(\type(t_i,\mu),\tau_i))}\Bbbk(\tau))}
\end{array}
\]

\begin{example}\label{ex:mup:0}
 Let $\mu=\{x\values\tnat,y\values(\tlist(\tnat)~\tor~\tnat)\}$,
$\Bbbk_1=\{\beta\values\tnat\}$ and $\Bbbk_2=\{\beta\values\tbot\}$.
By the definition of $\tvs$, $\tvs(\tnat,\beta)=\{\Bbbk_1\}$ and
$\tvs(\tlist(\tbot),\tlist(\beta))=\{\Bbbk_2\}$.  By
the definition of  $\type$,  $\type(x,\mu)=\tnat$
and $\type([~],\mu)=\tlist(\tbot)$. So,
\({\type([x|[~]],\mu)}
=(\Bbbk_1\tvallub\Bbbk_2)(\tlist(\beta))
=\tlist(\tnat)\).\endprf
\end{example}

The following lemma says that $\type(t,\mu)$ is a type that contains all the
instances of $t$ under the substitutions described by $\mu$.

\begin{lemma}\label{lm:F}
Let $t\in\term(\func,\viplus)$ and
$\mu\in(\viplus\mapsto\rtype)$. Then
$\theta(t)\in\sem{\rules}{\type(t,\mu)}$ for all $\theta\in\vtcon(\mu)$.
\endprf
\end{lemma}

\subsubsection{Upward Propagation}
We are now ready to present the overall upward propagation.  For a set
$\scal$ of variable typings and a set $E$ of equations in solved form,
upward propagation strengthens each variable typing $\mu$ in $\scal$
as follows. For each equation $x=t$ in $E$, $\type(t,\mu)$ is a type
of $x$ if variables occurring in $t$ satisfy $\mu$.  The overall
upward propagation is performed by a function
$\mup:\wp(\eqs)\times\wp(\viplus\mapsto\rtype)
\mapsto\wp(\viplus\mapsto\rtype)$  defined
\[{\mup(E,\scal) \definedas}
\bigcup_{\mu\in\scal}
   \left\{\lambda x\in\viplus.\left(
    \begin{array}{l}
    \mathit{if}~\exists t.(x=t)\in E\\
    then~(\mu(x)~\tand~\type(t,\mu))\\
    else~\mu(x)
    \end{array}\right)\right\}
\]

\begin{example} \label{ex:mup}
Continue with Example~\ref{ex:mup:0}. We have
\begin{eqnarray*}
\mup(\{y=[x|[~]]\},\{\mu\}) &=& \{\mu[y\values ((\tlist(\tnat)~\tor~\tnat)~\tand~\tlist(\tnat))]\}\\
  &\req& \{\{x\values\tnat,y\values\tlist(\tnat)\}\} 
\end{eqnarray*}
 \endprf
\end{example}

The correctness of upward propagation is ensured by this lemma.

\begin{lemma} \label{lm:H} Let $\scal\in\wp(\viplus\mapsto\rtype)$ and
                           $E\in\wp(\eqs)$. Then
$\mgu(\theta(E))\circ\theta\in\rcon({[\mup(E,\scal)]}_\req)$ for all
$\theta\in\rcon({[\scal]}_\req)$. \endprf
\end{lemma}

\subsection{Abstract Unification}

Algorithm~\ref{algo:aunify} defines the abstract unification operation
$\aunify$.  Given two atoms $a_1,a_2\in\atom_{P}$ and two
abstract substitutions
${[\scal_{1}]}_\req,{[\scal_{2}]}_\req\in\rsub$, it first applies the
renaming substitution $\Psi$ to $a_1$ and $\scal_{1}$ and computes
$E_{0}=eq\circ \mgu(\Psi(a_1),a_2)$. If $E_{0}=\fail$, it returns
${[\emptyset]}_\req$ -- the smallest abstract substitution which
describes the empty set of substitutions.  Otherwise, it calculates
$\scal'_{0}=\Psi(\scal_{1})\biguplus\scal_{2}$ where
$\biguplus:(\Psi(\vi)\mapsto\rtype)\times(\vi\mapsto\rtype)\mapsto(\viplus\mapsto\rtype)$. A
variable typing represents a conjunctive type constraint. If $\mu$ and
$\nu$ have disjoint domains then $\mu\cup\nu$ represents the
conjunction of $\mu$ and $\nu$. The first operand of $\biguplus$ is a
set of variable typings over $\Psi(\vi)$ and the second operand a set
of variable typings over $\vi$. The result of $\biguplus$ describes
the set of all the substitutions that satisfy both of its two
operands.  Thus $\scal'_0$ describes the set of all the substitutions
that satisfy both $\Psi(\scal_{1})$ and $\scal_{2}$. Note that
$\scal'_0\in \wp(\viplus\mapsto\rtype)$. The abstract unification
operation then calls a function
$\msolve:\eqs\times\wp(\viplus\mapsto\rtype)\mapsto\wp(\viplus\mapsto\rtype)$
to perform downward and upward propagations. The result $\scal'_1$ is
$\msolve(E_0,\scal'_0)$ which describes the set of all the
substitutions that satisfy both $E_0$ and $\scal'_0$.  Finally, it
calls a function
$\mrestrict:\wp(\viplus\mapsto\rtype)\mapsto\wp(\vi\mapsto\rtype)$ to
restrict each variable typing in $\scal'_1$ to $\vi$.

\begin{algorithm} \label{algo:aunify}
\[\begin{array}{rcl}
  {\munify(a_1,{[\scal_{1}]}_\req,a_2,{[\scal_{2}]}_\req)}
 &  \definedas &
  \left\{\begin{array}{l}
        let~~~E_{0}=eq\circ\mgu(\Psi(a_1),a_2)~in\\
        \mathit{if}~~~~E_{0}\neq\fail\\
        then~{[\mrestrict\circ
     \msolve(E_{0},\Psi(\scal_{1})\biguplus\scal_{2})]}_\req\\
        else~~{[\emptyset]}_\req
     \end{array}\right. \\
  \scal_1\biguplus\scal_2 &\definedas& \{\mu\cup\nu~|~\mu\in\scal_1\land\nu\in\scal_2\}\\
  \mrestrict(\scal)&\definedas&
    \{\mu\uparrow\vi~|~\mu\in\scal\land\forall x\in\viplus.({\mu(x)}\not\teq\tbot)\}\\
  \msolve(E,\scal) &\definedas & \mup(E,\mdown(E,\scal))
\end{array}
\]
\end{algorithm}
The function $\mrestrict$ removes those variable typings that denote
the empty set of substitutions and projects the remaining variable
typings onto $\vi$.

\begin{example} Let $\vi=\{x\}$, $\Psi(x)=y$, $a_1=p(x)$,
$a_2=p([x|[~]])$,
$\scal_1=\{\{x\values(\tlist(\tnat)~\tor~\tnat)\}\}$, and
$\scal_2=\{\{x\values\ttop\}\}$. Then $E_0=\{y=[x|[~]]\}$ and
$\Psi(\scal_1)\biguplus\scal_2=\scal$ with $\scal$ being that in
Example~\ref{ex:mdown}. By Examples~\ref{ex:mdown} and~\ref{ex:mup},
\begin{eqnarray*}
\msolve(E_0,\scal)&=&\mup(E_0,\mdown(E_0,\scal))=\mup(E_0,\{\mu\})\\
&=&\{\{x\values\tnat,y\values\tlist(\tnat)\}\}
\end{eqnarray*}
 with $\mu$ given in
Example~\ref{ex:mup}.  \endprf
\end{example}

The following theorem states that $\munify$ safely abstracts $\cunify$
with respect to $\rcon$.

\begin{theorem} \label{monosafeness}  \label{lm:I}
For any
  ${[\scal_{1}]}_\req,{[\scal_{2}]}_\req\in
  \rsub$ and any $a_1,a_2\in\atom_{P}$,
  \begin{eqnarray*}
  {\cunify(a_1,\rcon({[\scal_{1}]}_\req),
  a_2,\rcon({[\scal_{2}]}_\req))}
 & \subseteq & \rcon
  (\munify(a_1,{[\scal_{1}]}_\req,a_2,{[\scal_{2}]}_\req))
  \end{eqnarray*}  \endprf
\end{theorem}

\subsection{Abstract Built-in Execution Operation}
For each built-in, it is necessary to specify an operation that
transforms an input abstract substitution to an output abstract
substitution. These operations are given in Table~\ref{table:builtin}
where abstract substitutions are displayed as sets of variable
typings.  The primitive types $integer$, $float$, $number$, $string$,
$atom$ and $atomic$ have their usual denotations in Prolog. Observe
that $number = (integer~\tor~float)$ and $atomic =
(number~\tor~atom)$.

Unification $t_1 \texttt{=} t_2$ is modeled by
$\lambda\scal.\msolve(\mgu(t_1,t_2),\scal)$. Let $\theta$ be the
program state before the execution of $t_1 \texttt{=} t_2$ and assume
that $\theta$ satisfies $\scal$. The program state after the execution
of $t_1 \texttt{=} t_2$ is $\mgu(\theta(t_1),\theta(t_2))\circ\theta$
and satisfies $\msolve(\mgu(t_1,t_2),\scal)$.  Built-ins such as
$\texttt{</2}$ succeed only if their arguments satisfy certain type
constraints. Such type constraints are conjoined with the input
abstract substitution to obtain the output abstract substitution. For instance,
the execution of $t_1\texttt{<} t_2$ in an input program state
$\theta$ succeeds only if $\theta$ instantiates both $t_1$ and $t_2$
to numbers. So, the abstract operation for $t_1\texttt{<}t_2$ is
$f_3=\lambda\scal.(\scal\otimes \vts(number, t_1) \otimes
\vts(number,t_2))$ where $\vts$ defined in Section~\ref{sec:down} is
extended to deal with built-in types. The extension is straightforward
and omitted. For another instance, $format(t_1)$ succeeds only if $t$
is an atom, or a list of character codes or a string in its input
program state. The above type constraint is obtained as
$\vts(atom~\tor~\tlist(integer)~\tor~string, t_1)$. The type
$\tlist(integer)$ describes lists of character codes since character
codes are integers.  The type checking built-ins such as $atom/1$ are
modeled in the same way.  Built-ins such as \texttt{@</2} do not
instantiate their arguments or check types of their arguments. They
are modeled by the identity function $\lambda\scal.\scal$. The
built-in $fail/0$ never succeeds and hence is modeled by the constant
function that always returns $\emptyset$.

Consider a built-in to which a call $p(t_1,\cdots,t_n)$ will definitely
instantiate $t_i$ to a term of type $\rt_i$ upon success. The type
$\rt_i$ can be propagated downwards $t_i$, resulting in a set of
variable typings. The input abstract substitution can be strengthened
by this set of variable typings to give the output abstract
substitution.  For an instance, consider $name(t_1,t_2)$. Upon
success, $t_1$ is either an atom or an integer and $t_2$ is a
string. So, $name(t_1,t_2)$ is modeled by $\lambda\scal.(\scal\otimes
\vts(atom~\tor~integer, t_1) \otimes \vts(string,t_2))$. The built-ins
$length(t_1,t_2)$ and $compare(t_1,t_2,t_3)$ fall into this category.

Consider the built-in $var(t)$. The execution of $var(t)$ succeeds in a
program state $\theta$ iff $\theta(t)$ is a variable. All types that
contains variables are equivalent to $\ttop$. Thus, the built-in
$var(t)$ is modeled by $\lambda\scal.\{\mu\mid \mu\in\scal \land
\type(t,\mu)\teq\ttop\}$. The output abstract substitution contains
only those variable typings in which $t$ has no type smaller than
$\ttop$. The built-in $nonvar(t)$ is modeled by the identity function
$\lambda\scal.\scal$ since a term being a non-variable does not
provide any information about its type unless non-freeness is defined
as a type. So is the built-in $ground(t)$ since a term being ground
says nothing about its type unless groundness is defined as a
type. The operation for the built-in $compound(t)$ makes use of the
property that a compound term is not atomic. It removes from the input
abstract substitution any variable typing in which $t$ is atomic.

\begin{table}\centering
\begin{tabular}{@{}p{2.5in}|c}
\textit{Predicate} & \textit{Operation} \\
\cline{1-2}
\cline{1-2}
$abort$, $fail$, $false$ & $\lambda\scal.\emptyset$\\
\cline{1-2}
$!$,
$t_1 \texttt{@<} t_2$, $t_1 \texttt{@>} t_2$,  $t_1 \texttt{=<@} t_2$,  $t_1 \texttt{@>=} t_2$,
$t_1 \texttt{$\backslash$==} t_2$,
$t_1 \texttt{$\backslash$=} t_2$,
$display(t_1)$,
$ground(t_1)$,
$listing$,
$listing(t_1)$,
$nl$,
$nonvar(t_1)$,
$portray\_clause(t_1)$,
$print(t_1)$,
$read(t_1)$,
$repeat$,
$true$,
$write(t_1)$,
$writeq(t_1)$
             &  $\lambda\scal.\scal$\\

\cline{1-2}
$compound(t)$ & $\lambda\scal.(\{\mu\mid \mu\in\scal \land \type(t,\mu)\not\tleq atomic  \})$\\

\cline{1-2}
$atom(t)$ & $\lambda\scal. (\scal\otimes\vts(atom,t))$\\
\cline{1-2}
$atomic(t)$ & $\lambda\scal. (\scal\otimes\vts(atomic,t))$\\
\cline{1-2}
$float(t)$ & $\lambda\scal. (\scal\otimes\vts(float,t))$\\
\cline{1-2}
$erase(t)$, $integer(t)$, $tab(t)$ & $\lambda\scal. (\scal\otimes\vts(integer,t))$\\
\cline{1-2}
$number(t)$ & $\lambda\scal. (\scal\otimes\vts(number,t))$\\
\cline{1-2}
$put(t)$ & $\lambda\scal. (\scal\otimes\vts(atom~\tor~integer,t))$\\
\cline{1-2}
$string(t)$ & $\lambda\scal. (\scal\otimes\vts(string,t))$\\

\cline{1-2}
$var(t)$ & $\lambda\scal.\{\mu\mid \mu\in\scal \land \type(t,\mu)\teq\ttop\}$ \\

\cline{1-2}
$t_1 \texttt{=} t_2$, $t_1 \texttt{==} t_2$ &  $\lambda\scal.\msolve(\mgu(t_1,t_2),\scal)$\\

\cline{1-2}

$format(t_1)$,
$format(t_1,t_2)$,
$format(t_0,t_1,t_2)$ & $f_4$ \\

\cline{1-2}
$t_1 {<} t_2$,
$t_1 > t_2$,
$t_1 \texttt{=$<$} t_2$,
$t_1 \texttt{$>$=} t_2$,
$t_1 \texttt{=:=} t_2$,
$t_1 \texttt{=$\backslash$=} t_2$, $is(t_1,t_2)$   & $f_3$ \\

\cline{1-2} $length(t_1,t_2)$ & ${f}_1$\\
\cline{1-2} $compare(t_1,t_2,t_3)$ & $\lambda\scal.(\scal\otimes\vts(atom,t_1))$\\
\cline{1-2} $name(t_1,t_2)$ & ${f}_2$\\
\cline{1-2}
\end{tabular}
\caption{\label{table:builtin} Abstract operations for built-ins where
${f}_1 = \lambda\scal.(\scal\otimes \vts(\tlist(\ttop),t_1)\otimes\vts(integer,t_2))$,
${f}_2 = \lambda\scal.(\scal\otimes \vts(atom~\tor~integer, t_1) \otimes \vts(string,t_2))$,
${f}_3 = \lambda\scal.(\scal\otimes \vts(number, t_1) \otimes \vts(number,t_2))$,
 and
$f_4=\lambda\scal.(\scal\otimes\vts(atom~\tor~\tlist(integer)~\tor~string,t_1))$.
}
\end{table}

\section{Implementation} \label{sec:imp}

We have implemented a prototype of our type analysis in
SWI-Prolog. The prototype is a meta-interpreter using ground
representations for program variables. The prototype supports the
primitive types $integer$, $float$, $number$, $string$, $atom$ and
$atomic$ with their usual denotations in Prolog.

\subsection{Examples}
\begin{example} \label{ex:intersect}
The following is the {\sf intersect} program that computes the
intersection of two lists and its analysis result. Lists are defined
in Example~\ref{ex1}.  Abstract substitutions are displayed as
comments. The abstract substitution associated with the entry point of
the query is an analysis input whilst all other abstract substitutions
are analysis outputs.  Sets are displayed as lists. A binding $V\mapsto T$ is
written as $V/T$, $\tor$ as ${\it or}$ and $\tand$ as ${\it and}$.
The code for the predicate {\sf member}/2 is omitted.

{

\begin{verbatim}
:-  %[[X/list(atom or float),Y/list(atom or integer)]]
  intersect(X,Y,Z).
    %[[X/list(atom or float),Y/list(atom or integer),Z/list(atom)]]

intersect([],L,[]).
    %[[L/list(atom or integer)]]
intersect([X|Xs],Ys,[X|Zs]) :-
    %[[X/atom,Xs/list(atom or float),Ys/list(atom or integer)],
    % [X/float,Xs/list(atom or float),Ys/list(atom or integer)]]
  member(X,Ys),
    %[[X/atom,Xs/list(atom or float),Ys/list(atom or integer)]]
  intersect(Xs,Ys,Zs).
    %[[X/atom,Xs/list(atom or float),Ys/list(atom or integer),
    %  Zs/list(atom)]]
intersect([X|Xs],Ys,Zs) :-
    %[[X/atom,Xs/list(atom or float),Ys/list(atom or integer)],
    % [X/float,Xs/list(atom or float),Ys/list(atom or integer)]]
  \+ member(X,Ys),
    %[[X/float,Xs/list(atom or float),Ys/list(atom or integer)],
    % [X/atom,Xs/list(atom or float),Ys/list(atom or integer)]]
  intersect(Xs,Ys,Zs).
    %[[X/float,Xs/list(atom or float),Ys/list(atom or integer),
    %  Zs/list(atom)],
    % [X/atom,Xs/list(atom or float),Ys/list(atom or integer),
    %  Zs/list(atom)]]
\end{verbatim}
}

The result shows that the intersection of a list containing atoms and
float numbers and another list containing atoms and integer numbers is
a list of atoms. This is precise because the type
$((atom~\tor~float)~\tand~(atom~\tor~integer))$ is equivalent to the
type $atom$. Without the set operators $\tand$ and $\tor$ in their
type languages, previous type analyses with {\it a priori} type
definitions cannot produce a result as precise as the above.  \endprf
\end{example}

\begin{example}
The following is a program $p/1$. The analysis result is displayed
with the typing binding $x\mapsto\ttop$ omitted for any variable $x$.

\begin{verbatim}
p([]). % [[]]

p([X|Y]) :-
    % [[]]
  integer(X),
    % [[X/integer]]
  p(Y).
    % [[X/integer, Y/list(or(atom, integer))]]

p([X|Y]) :-
    % [[]]
  atom(X),
    % [[X/atom]]
  p(Y).
    % [[X/atom, Y/list(or(atom, integer))]]

:-  % [[]]
   p(U).
    % [[U/list(or(atom, integer))]]
\end{verbatim}
The result captures precisely type information in the success set of
the program, that is, $U$ is a list consisting of integers and atoms
upon success of p(U).\footnote{This example was provided by an
anonymous referee of a previous version of this paper.} \endprf
\end{example}

During analysis of a program, the analyzer repeatedly checks if two
sets of variable typings are equivalent and if a set of variable
typings contains redundant elements. Both of these decision problems
are reduced to checking if a given type denotes the empty set of
terms.

\comments { We then show the effect of tabling the emptiness checks of
types and evaluate the effect of non-deterministic type definitions
and set operators in terms of accuracy and time performance. We
finally address the issue of the termination of our type analysis.  }

\subsection{Emptiness of Types}
Type rules in $\rules$ are production rules for a context-free tree
grammar in restricted form~\cite{GecsegS84}. According
to~\cite{LuC:empty}, if $\ttop$ denotes the set of all ground terms
instead of all terms then each type denotes a regular tree language.
We now show how an algorithm in~\cite{LuC:empty} can be used for
checking the emptiness of types.  We first extend the type language
with the complement operator $\tneg$ and define
$\ertype=\term(\cons\cup\{\tneg,\tand,\tor,\ttop,\tbot\},\emptyset)$. Observe
that $\rtype\subset \ertype$ and that $\sem{\rules}{\cdot}$ is not
defined for elements in $\ertype\setminus\rtype$.  Since the algorithm
in~\cite{LuC:empty} was developed for checking the emptiness of types
that denote sets of ground terms, we need to justify its application
by closing the gap between the two different semantics of types. This
is achieved by extending the signature $\func$ with an extra constant
$\varrho$ ($\varrho\in\func$) that is used to encode variables in
terms. Use of extended signatures in analysis of logic programs can be
traced to~\cite{GallagherBS95} where extra constants are used to
encode non-ground terms. In fact, by introducing an infinite set of
extra constants one can obtain an isomorphism between the set of all
terms in the original signature and the set of the ground terms in the
extended signature.

\begin{definition}
The meaning of a type in $\ertype$ is given by a function
$\semb{\rules}{\cdot}:\ertype\mapsto\wp(\term(\func\cup\{\varrho\},\emptyset)$.
\[\begin{array}{rcl}
\semb{\rules}{\ttop} &\definedas & \term(\func\cup\{\varrho\},\emptyset)\\
\semb{\rules}{\tbot} &\definedas & \emptyset\\
\semb{\rules}{\tneg\rt} &\definedas & \semb{\rules}{\ttop}
                   \setminus \semb{\rules}{\rt}\\
\semb{\rules}{\tand(\rt_1,\rt_2)} &\definedas & \semb{\rules}{\rt_1}\cap\semb{\rules}{\rt_2}\\
\semb{\rules}{\tor(\rt_1,\rt_2)} &\definedas & \semb{\rules}{\rt_1}\cup\semb{\rules}{\rt_2}\\
\semb{\rules}{c(\Vector{\rt}{m})} &\definedas &\\
\multicolumn{3}{r}{~\hspace{1pc}
     \bigcup_{(c(\Vector{\beta}{m})\tdef f(\Vector{\tau}{n}))\in\rules}
     \left(\begin{array}{l}
       let~\Bbbk=\{\beta_j\values\rt_j~|~1\leq{j}\leq{m}\}\\
       in~\\
       \{f(\Vector{t}{n})\mid \forall 1\leq{i}\leq{n}.
       t_{i}\in \semb{\rules}{\Bbbk(\tau_{i})}
       \}
     \end{array}\right)
     }
\end{array}
\]\endprf
\end{definition}

There are two differences between $\semb{\rules}{\cdot}$ and
$\sem{\rules}{\cdot}$.  Firstly, $\tneg$ is interpreted as set
complement under $\semb{\rules}{\cdot}$ whilst it has no denotation
under $\sem{\rules}{\cdot}$. Type constructor $\tneg$ can be
interpreted as set complement by $\semb{\rules}{\cdot}$ because
$\semb{\rules}{\rt}$ is a regular tree language for any
$\rt\in\ertype$~\cite{LuC:empty}. It cannot be interpreted as set
complement by $\sem{\rules}{\cdot}$ because the complement of
$\sem{\rules}{\rt}$ is not closed under instantiation.  Secondly, the
universal type $\ttop$ denotes $\term(\func,\allvars)$ in
$\sem{\rules}{\cdot}$ whilst it denotes
$\term(\func\cup\{\varrho\},\emptyset)$ in $\semb{\rules}{\cdot}$. An
implication is that a type denotes a set of terms closed under
instantiation under $\sem{\rules}{\cdot}$ whilst it denotes a set of
ground terms under $\semb{\rules}{\cdot}$.

Let $\chi:\term(\func,\allvars)\mapsto
\term(\func\cup\{\varrho\},\emptyset)$ be defined $\chi(x)=\varrho$
for all $x\in\allvars$ and
$\chi(f(t_1,\cdots,t_n))=f(\chi(t_1),\cdots,\chi(t_n))$. The function
$\chi(\cdot)$ transforms a term into a ground term by replacing all
variables in the term with the same constant $\varrho$. The following
theorem states that, given a term $t$ and a type $\rt$, the membership
of $t$ in $\sem{\rules}{\rt}$ is equivalent to that of $\chi(t)$ in
$\semb{\rules}{\rt}$.
\begin{theorem} \label{th:eq} For any term $t$ in $\term(\func,\allvars)$ and any type
$\rt$ in $\rtype$, $t\in\sem{\rules}{\rt}$ iff
$\chi(t)\in\semb{\rules}{\rt}$. \endprf
\end{theorem}
As a consequence, checking the emptiness of a type under
$\sem{\rules}{\cdot}$ can be reduced to checking the emptiness of the
type under $\semb{\rules}{\cdot}$, and vice versa. Therefore, whether
$\sem{\rules}{\rt}=\emptyset$ can be decided by employing the
algorithm developed in~\cite{LuC:empty} that checks if
$\semb{\rules}{\rt}=\emptyset$. The following corollary of the theorem
allows us to reduce a type inclusion test under $\sem{\rules}{\cdot}$
to a type inclusion test under $\semb{\rules}{\cdot}$.

\begin{corollary} \label{co:eq} For any $\rt_1,\rt_2\in\rtype$,
$\sem{\rules}{\rt_1}\subseteq\sem{\rules}{\rt_2}$ iff
$\semb{\rules}{\rt_1}\subseteq\semb{\rules}{\rt_2}$.\endprf
\end{corollary}

In order to reduce the decision problems to the emptiness of types, we
need to extend the syntax for type sequence expressions with the
operator $\tneg$ and $\semb{\rules}{\cdot}$ to type sequences. The
expression $\tneg \fcal$ is a type sequence expression whenever
$\fcal$ is a type sequence expression. Let $\rt$ be a type in
$\ertype$, $\rtvec$ a type sequence in $\ertype^{*}$, $\fcal_1$ and
$\fcal_2$ be type sequence expressions. Define
$\semb{\rules}{\epsilon}\definedas\{\epsilon\}$,
$\semb{\rules}{\rt\bullet\rtvec}\definedas\semb{\rules}{\rt}\bullet\semb{\rules}{\rtvec}$,
$\semb{\rules}{\fcal_1~\tand~\fcal_2} \definedas
\semb{\rules}{\fcal_1}\cap\semb{\rules}{\fcal_2}$,
$\semb{\rules}{\fcal_1~\tor~\fcal_2} \definedas
\semb{\rules}{\fcal_1}\cup\semb{\rules}{\fcal_2}$ and
$\semb{\rules}{\tneg~\fcal} \definedas
\semb{\rules}{\vec{\ttop}}-\semb{\rules}{\fcal}$. It can be shown that
both Theorem~\ref{th:eq} and Corollary~\ref{co:eq} carry over to type
sequence expressions that do not contain $\tneg$.

Set inclusion and $\semb{\rules}{\cdot}$ induces an equivalence
between types and type sequence expressions. Let $\rt_1\teqb\rt_2$ iff
$\semb{\rules}{\rt_1}=\semb{\rules}{\rt_2}$ and $\fcal_1\teqb\fcal_2$
iff $\semb{\rules}{\fcal_1}=\semb{\rules}{\fcal_2}$.  The following
function eliminates the complement operator $\tneg$ over type sequence
expressions.
\begin{eqnarray*}
push(\tneg (\tor_{i\in{I}}\fcal_{i})) &\definedas&
               \tand_{i\in{I}}push(\tneg\fcal_{i})\\
push(\tneg (\tand_{i\in{I}}\fcal_{i})) &\definedas&
               \tor_{i\in{I}}push(\tneg\fcal_{i})\\
push(\tneg(\rt_1,\rt_2,\cdots,\rt_{k})) &\definedas &
    \tor_{1\leq{l}\leq{k}}
(\underbrace{\ttop,\cdots,\ttop}_{l-1},\tneg
\rt_{l},\underbrace{\ttop,\cdots,\ttop}_{k-l})~~~~\mbox{for $k\geq 1$}
\end{eqnarray*}
It follows from De Morgan's law and the definition of
$\semb{\rules}{\cdot}$ that $push(\tneg\fcal)\teqb\tneg\fcal$. Note
that the complement operator $\tneg$ does not apply to any type
sequence expression in $push(\tneg\fcal)$; it only applies to type
expressions.  Let $\rt\in\ertype$ and define $etype(\rt)\definedas
(\semb{\rules}{\rt}=\emptyset)$. The formula $etype(\rt)$ is true iff
$\rt\teqb\tbot$ is true. By Theorem~\ref{th:eq}, if $\rt\in\rtype$
then $etype(\rt)$ is true iff $\rt\teq\tbot$ is true.

\subsection{Equivalence between Sets of Variable Typings} \label{sec:equiv}
An indispensable operation in a static analyzer is to check if a
fixpoint has been reached. This operation reduces to checking if two
sets of variable typings denote the same set of concrete
substitutions. This equivalence test is reduced to checking emptiness
of types as follows.  Let $\vi=\{x_1,\cdots,x_k\}$ and
$\scal_1,\scal_2\in\wp(\vi\mapsto\rtype)$.  By definition,
$\scal_1\req\scal_2$ iff
$\bigcup_{\mu\in\scal_1}\vtcon(\mu)\subseteq\bigcup_{\nu\in\scal_2}\vtcon(\nu)$
and
$\bigcup_{\nu\in\scal_2}\vtcon(\nu)\subseteq\bigcup_{\mu\in\scal_1}\vtcon(\mu)$.
Suppose $\scal_1=\{\mu_1,\mu_2,\cdots,\mu_m\}$ and
$\scal_2=\{\nu_1,\nu_2,\cdots,\nu_n\}$. We construct
$\rtvec_1,\rtvec_2,\cdots,\rtvec_m$ and
$\gcal_1,\gcal_2,\cdots,\gcal_n$ as follows.
$\rtvec_i=\langle\mu_i(x_1),\mu_i(x_2),\cdots,\mu_i(x_k)\rangle$ and
$\gcal_j=\langle\nu_j(x_1),\nu_j(x_2),\cdots,\nu_j(x_k)\rangle$. Then
$\bigcup_{\mu\in\scal_1}\vtcon(\mu)\subseteq\bigcup_{\nu\in\scal_2}\vtcon(\nu)$
is true iff ${\tor_{1\leq{i}}\rtvec_i} \tleq {\tor_{1\leq{j}}\gcal_j}$
is true.  By Corollary~\ref{co:eq}, ${\tor_{1\leq{i}}\rtvec_i} \tleq
{\tor_{1\leq{j}}\gcal_j}$ is true iff \(
(\tor_{1\leq{i}}\rtvec_i)~\tand~\tneg(\tor_{1\leq{j}}\gcal_j)\teqb\vec{\tbot}
\) is true. The latter can be reduced to emptiness of types as shown
in~\cite{LuC:empty}.

\begin{example} \label{ex:eq}
Let $\rules$ be given as in Example~\ref{ex1}, $\vi=\{x,y\}$,
$\scal_1=\{\mu_1,\mu_2\}$ and $\scal_2=\{\mu_3\}$ where
\begin{eqnarray*}
\mu_1 &=&\{x\values\tlist(\teven),y\values\tlist(\tnat)\}\\
\mu_2 &=&\{x\values\tlist(\todd),y\values\tlist(\tnat)\}\\
\mu_3 &=&\{x\values\tlist(\teven)~\tor~\tlist(\todd),y\values\tlist(\tnat)\}
\end{eqnarray*}
The truth value of
$(\vtcon(\mu_1)\cup\vtcon(\mu_2))\subseteq\vtcon(\mu_3)$ is decided by
testing emptiness of  types as follows.  Let
\begin{eqnarray*} \rtvec_1&=&\langle\tlist(\teven),\tlist(\tnat)\rangle\\
    \rtvec_2&=&\langle\tlist(\todd),\tlist(\tnat)\rangle\\
    \gcal_1&=&\langle\tlist(\teven)~\tor~\tlist(\todd),\tlist(\tnat)\rangle
\end{eqnarray*}
Then $(\vtcon(\mu_1)\cup\vtcon(\mu_2))\subseteq\vtcon(\mu_3)$ iff
${(\rtvec_1~\tor~\rtvec_2)~\tand~\tneg\gcal_1}\teqb\vec{\tbot}$ which,
by replacing $\tneg\gcal_1$ with $push(\tneg\gcal_1)$ and distributing
$~\tand~$ over $\tor$, is equivalent to the conjunction of the
following formulas.
\begin{eqnarray*}
{\langle\tlist(\teven),\tlist(\tnat)\rangle~\tand~\langle\ttop,\tneg\tlist(\tnat)\rangle}&\teqb&\vec{\tbot}\\
{\langle\tlist(\todd),\tlist(\tnat)\rangle~\tand~\langle\ttop,\tneg\tlist(\tnat)\rangle}&\teqb&\vec{\tbot}\\
{\langle\tlist(\teven),\tlist(\tnat)\rangle~\tand~\langle\tneg\tlist(\teven)~\tand~\tneg\tlist(\todd),\ttop\rangle}&\teqb&\vec{\tbot}\\
{\langle\tlist(\todd),\tlist(\tnat)\rangle~\tand~\langle\tneg\tlist(\teven)~\tand~\tneg\tlist(\todd),\ttop\rangle}&\teqb&\vec{\tbot}\\
\end{eqnarray*}
The first of the above holds iff either ${\tlist(\teven)}\teqb\tbot$
or ${\tlist(\tnat)~\tand~\tneg\tlist(\tnat)}\teqb\tbot$, both of which
are emptiness tests on types. Since
${\tlist(\tnat)~\tand~\tneg\tlist(\tnat)}\teqb\tbot$, the first
formula is decided to be true.  The other three can be decided to be
true similarly. Therefore,
$(\vtcon(\mu_1)\cup\vtcon(\mu_2))\subseteq\vtcon(\mu_3)$ holds. In a
similar way, we can show that
$\vtcon(\mu_3)\subseteq(\vtcon(\mu_1)\cup\vtcon(\mu_2))$
holds. So, $\scal_1\req\scal_2$.  \endprf
\end{example}

\subsection{Redundancy Removal}

For the sake of an efficient implementation, an abstract substitution
${[\scal]}_\req$ should be represented by a set of variable typings
that does not contain redundancy.  A set of variable typings can be redundant in
two ways. Firstly, a variable typing $\mu$ in $\scal$ may denote the
empty set of substitutions i.e., $\mu(x)\teq\tbot$ for some $x\in\vi$.
Secondly, a variable typing $\mu$ in $\scal$ can be subsumed by other
variable typings in that $\vtcon(\mu)\subseteq
\bigcup_{\nu\in\scal\land\nu\neq\mu}\vtcon(\nu)$. In both cases,
$\scal\setminus\{\mu\}$ and $\scal$ denote the same set of
substitutions and $\mu$ can be removed from $\scal$. Suppose
$\vi=\{x_1,\cdots,x_k\}$.  The detection of $\vtcon(\mu)=\emptyset$
reduces to $etype(\mu(x_1))\lor\cdots\lor etype(\mu(x_k))$ while the
detection of $\vtcon(\mu)\subseteq
\bigcup_{\nu\in\scal\land\nu\neq\mu}\vtcon(\nu)$ can be reduced to
checking emptiness of types as in Section~\ref{sec:equiv}.

\begin{example} Let $\rules$ be given Example~\ref{ex1}, $\vi=\{x,y\}$,
$\scal=\{\mu_1,\mu_2,\mu_3\}$ where
\begin{eqnarray*}
\mu_1 &=&\{x\values\tlist(\teven),y\values\tlist(\tnat)\}\\
\mu_2 &=&\{x\values\tlist(\todd),y\values\tlist(\tnat)\}\\
\mu_3 &=&\{x\values\tlist(\tnat),y\values\tlist(\tnat)\}
\end{eqnarray*}
We now show how  $\mu_1$ is decided to be redundant in $\scal$. Let
\begin{eqnarray*}
\rtvec_1 &=&\langle\tlist(\teven),\tlist(\tnat)\rangle\\
\rtvec_2 &=&\langle\tlist(\todd),\tlist(\tnat)\rangle\\
\rtvec_3 &=&\langle\tlist(\tnat),\tlist(\tnat)\rangle
\end{eqnarray*}
Then $\vtcon(\mu_1)\subseteq\vtcon(\mu_2)\cup\vtcon(\mu_3)$ holds iff
${\rtvec_1~\tand~\tneg(\rtvec_2~\tor~\rtvec_3)}\teqb\vec{\tbot}$ holds
iff
${\rtvec_1~\tand~\tneg\rtvec_2~\tand~\tneg\rtvec_3}\teqb\vec{\tbot}$ holds. The
latter, after replacing $\tneg\rtvec_2$ and $\tneg\rtvec_3$ with
$push(\tneg\rtvec_2)$ and $push(\tneg\rtvec_3)$ respectively and
distributing $~\tand~$ over $\tor$, is equivalent to the conjunction
of the following formulas.
\begin{eqnarray*}
{\langle\tlist(\teven),\tlist(\tnat)\rangle~\tand~
         \langle\ttop,\tneg\tlist(\tnat)\rangle~\tand~
         \langle\ttop,\tneg\tlist(\tnat)\rangle} &\teqb&\vec{\tbot}\\
{\langle\tlist(\teven),\tlist(\tnat)\rangle~\tand~
         \langle\ttop,\tneg\tlist(\tnat)\rangle~\tand~
         \langle\tneg\tlist(\tnat),\ttop\rangle} &\teqb&\vec{\tbot}\\
{\langle\tlist(\teven),\tlist(\tnat)\rangle~\tand~
         \langle\tneg\tlist(\todd),\ttop\rangle~\tand~
         \langle\ttop,\tneg\tlist(\tnat)\rangle} &\teqb&\vec{\tbot}\\
{\langle\tlist(\teven),\tlist(\tnat)\rangle~\tand~
         \langle\tneg\tlist(\todd),\ttop\rangle~\tand~
         \langle\tneg\tlist(\tnat),\ttop\rangle} &\teqb&\vec{\tbot}\\
\end{eqnarray*} Each of the above  can be decided to be true by testing  emptiness of   types as in Example~\ref{ex:eq}. Therefore, $\mu_1$ is redundant in $\scal$. In a similar way,
 $\mu_2$ is decided to be redundant in $\scal\setminus\{\mu_1\}$. So,
$\scal\req\{\mu_3\}$. That $\mu_3$ is not redundant in $\scal$ is
decided similarly.  \endprf
\end{example}

\subsection{Tabling}
The operations in our type analysis are complex because of
non-deterministic type definitions and non-discriminative union at
both the level of types and the level of abstract substitutions.  The
equality of two abstract substitutions in an analysis without these
features can be done in linear
time~\cite{HoriuchiK87,Kanamori:Horiuchi:85,KanamoriJLP93,Lu95}. The
same operation is exponential in our type analysis because deciding
the emptiness of a type is exponential. This indicates that our type
analysis could be much more time consuming.

As shown later, there is a high degree of repetition in emptiness
 checks during the analysis of a program. Making use of this
 observation, we have reduced time increase to 15\% on average using a
 simple tabling technique.  We memoize each call to $etype(R)$ and its
 success or failure by asserting a fact
 $\$etype\_tabled(R,Ans)$. The fact $\$etype\_tabled(R,yes)$
 (resp. $\$etype\_tabled(R,no)$) indicates that $etype(R)$ has been
 called before and $etype(R)$ succeeded (resp. failed). The tabled
 version of $etype(R)$ is $etype\_tabled(R)$. It first checks if a
 fact $\$etype\_tabled(R,Ans)$ exists. If so, the call $etype\_tabled(R)$
 succeeds or fails immediately. Otherwise, it calls $etype(R)$ and
 memoizes its success or failure.

We now present some experimental results with the prototype analyzer.
The experiments were done with a Pentium (R) 4 CPU 2.26 GHz running
GNU/Linux and SWI-Prolog-5.2.13.

\subsubsection{Time Performance}
Table~2 shows analysis time on a suite of benchmark programs. Each row
except the last one corresponds to a test case.  The first three
columns contain the name, the size of the program in terms of the
number of program points and the top-level goal. The top abstract
substitution which contains no type information is used as the input
abstract substitution for each test case.  These test cases will be
used in subsequent tables where only the program names are given.  The
fourth column gives analysis time in milliseconds. The time is
obtained by running the analyzer ten times on the test case and
averaging analysis time from these runs. Timing data in other tables
are also obtained in this way.  The table shows that the analyzer
takes an average of 1.27 milliseconds per program point.

\begin{table}
\caption{\label{tab:time} Time Performance}
\begin{tabular}{|r|r|l|l|}\cline{1-4}\cline{1-4}
Program & Program Points &  Goal & Time\\\cline{1-4}\cline{1-4}
browse&103&q&110\\\cline{1-4}
cs\_r&277&pgenconfig(C)&661\\\cline{1-4}
disj\_r&132&top(K)&171\\\cline{1-4}
dnf&77&go&200\\\cline{1-4}
kalah&228&play(G, R)&590\\\cline{1-4}
life&100&life(MR, MC, LC, SFG)&89\\\cline{1-4}
meta&89&interpret(G)&50\\\cline{1-4}
neural&341&go&250\\\cline{1-4}
nbody&375&go(M, G)&281\\\cline{1-4}
press&318&test\_press(X, Y)&161\\\cline{1-4}
serialize&37&go(S)&80\\\cline{1-4}
zebra&43&zebra(E, S, J, U, N, Z, W)&40\\\cline{1-4}\cline{1-4}
& Sum = 2120&&Sum = 2683\\\cline{1-4}\cline{1-4}
\end{tabular}
\end{table}

\subsubsection{Repetition of Emptiness Checks}
Table~3 shows that there is a high degree of repetition in emptiness
 checks during analysis.  Each test case corresponds to a row of the
 table. The first column of the row is the name of the program, the
 second is the total number of emptiness checks that occur during
 analysis. The third column gives the number of different types that
 are checked for emptiness.  The fourth column gives the average
 repetition of emptiness checks, which is the ratio of the second and
 the third columns.  While the total number of emptiness checks can be
 very large for a test case, the number of different emptiness checks
 is small, exhibiting a high degree of repetition in emptiness checks.
 The repetition of the emptiness checks ranges from 36.00 to
 698.88. The weighted repetition average is about 192.23. This
 motivated the use of tabling to reduce the time spent on emptiness
 checks.

\begin{table}
\caption{\label{tab:repeat} Repetition of Emptiness Checks}
\begin{tabular}{|l|r|r|r|}\cline{1-4}\cline{1-4}
Program & Total    & Different & Degree of \\
        & Checks   & Checks    & Repetition\\\cline{1-4}\cline{1-4}
browse&3050&64&47.65\\\cline{1-4}
cs\_r&23846&53&449.92\\\cline{1-4}
disj\_r&4500&37&121.62\\\cline{1-4}
dnf&6290&9&698.88\\\cline{1-4}
kalah&31182&86&362.58\\\cline{1-4}
life&3277&24&136.54\\\cline{1-4}
meta&468&13&36.00\\\cline{1-4}
neural&7985&131&60.95\\\cline{1-4}
nbody&8567&39&219.66\\\cline{1-4}
press&1734&23&75.39\\\cline{1-4}
serialize&2019&37&54.56\\\cline{1-4}
zebra&947&22&43.04\\\cline{1-4}
\cline{1-4}
&&&Ave.=192.23\\\cline{1-4}\cline{1-4}
\end{tabular}

\end{table}

\subsubsection{Effect of Tabulation}
Table~\ref{tab:tabling} illustrates the effect of tabling.  Statistics are obtained by
running the analyzer with and without tabling. For both experiments,
we measured analysis time and time spent on emptiness checks.  The
table shows that tabling reduces \comments{emptiness checking time to
$\frac{1}{93}$ and } analysis time to $\frac{1}{2.6}$. The
table also gives the proportion of analysis time that is spent on
emptiness checks. An average of 53\% of analysis time is spent on
emptiness checks without tabling while only a negligible portion of
analysis time is spent on emptiness checks with tabling.

\begin{table}
\caption{\label{tab:tabling} Effect of Tabulation}

\begin{tabular}{|l|r|r|r|r|r|r|}\cline{1-7}\cline{1-7}
     & \multicolumn{3}{c|}{\underline{\hspace{8ex}With Tabling\hspace{8ex}}}
     & \multicolumn{3}{c|}{\underline{\hspace{8ex}Without Tabling\hspace{8ex}}}
         \\
Program & Analysis    & Check  &  &
                    Analysis    & Check  &
           \\
        & Time        & Time      & Proportion &
                    Time        & Time      & Proportion
           \\\cline{1-7}\cline{1-7}
browse&110&10&0.09&269&129&0.47\\\cline{1-7}
cs\_r&661&0&0&1700&891&0.52\\\cline{1-7}
disj\_r&171&0&0&359&157&0.43\\\cline{1-7}
dnf&200&0&0&581&363&0.62\\\cline{1-7}
kalah&590&0&0&1939&1124&0.57\\\cline{1-7}
life&89&0&0&210&117&0.55\\\cline{1-7}
meta&50&0&0&60&21&0.35\\\cline{1-7}
neural&250&30&0.12&731&469&0.64\\\cline{1-7}
nbody&281&0&0&620&314&0.50\\\cline{1-7}
press&161&0&0&250&29&0.11\\\cline{1-7}
serialize&80&0&0&179&82&0.45\\\cline{1-7}
zebra&40&0&0&70&31&0.44\\\cline{1-7}
\cline{1-7}
&Sum=2683& &Ave.=0.01&Sum=6968& &Ave.=0.53\\\cline{1-7}\cline{1-7}
\end{tabular}
\end{table}

\begin{table}
\caption{\label{tab:compare} Cost and Effect of Precision Improvement Features}
\begin{tabular}{|l|r|r|r|r|}\cline{1-5}
Program & Simplified & Full-fledged  & Time & Precision  \\
        & Analysis      & Analysis & Ratio  & Ratio   \\
        & Time          & Time     &        &   \\\cline{1-5}\cline{1-5}
browse&100.00&110.00&0.90 & 0.74\\\cline{1-5}
cs\_r&589.00&661.00&0.89 & 0.77 \\\cline{1-5}
disj\_r&151.00&171.00&0.88 & 0.73 \\\cline{1-5}
dnf&190.00&200.00&0.95& 0.00 \\\cline{1-5}
kalah&420.00&590.00&0.71& 0.49\\\cline{1-5}
life&89.00&89.00&1.00& 0.35 \\\cline{1-5}
meta&40.00&50.00&0.80& 0.05 \\\cline{1-5}
neural&200.00&250.00&0.80& 0.49\\\cline{1-5}
nbody&231.00&281.00&0.82& 0.23\\\cline{1-5}
press&159.00&161.00&0.98& 0.95\\\cline{1-5}
serialize&80.00&80.00&1.00& 0.83 \\\cline{1-5}
zebra&40.00&40.00&1.00& 0.35\\\cline{1-5}
\cline{1-5}
&&&Ave. =0.85 & Ave. =0.54\\\cline{1-5}\cline{1-5}
\end{tabular}
\end{table}

\subsection{Cost and Effect of Precision Improvement Features}
The precision improvement features in our type analysis all incur some
performance penalty. In order to evaluate the effect of these
features, we also implemented a simplified type analysis. The
simplified analysis is obtained by removing the precision improvement
features from the full-fledged analysis. In the simplified analysis,
type expressions do not contain the constructors $\tor$ or $\tand$; an
abstract substitution is simply a variable typing; and
non-deterministic type definitions are disallowed. Function
overloading is still allowed. Abstract operations are simplified
accordingly. For instance, since
$(\tlist(\tlist(\tnat))~\tor~\tlist(\tnat))$ is not in the type
language of the simplified analysis, the least upper bound of
$\tlist(\tlist(\tnat))$ and $\tlist(\tnat)$ is $\tlist(\ttop)$. The
least upper bound operation on abstract substitutions is the
point-wise extension of the least upper bound operation on types.

Table~\ref{tab:compare} compares two type analyses.  The two analyses
are performed on each test case with the same input type
information. The input abstract substitution for the full-fledged
analysis is a singleton set of a variable typing. The corresponding
abstract substitution for the simplified analysis is the variable
typing.  For each test case, the table gives analysis times by the two
analyzers and their ratio. The relative performance of the two
analyzers varies with the test case. On average, the simplified
analysis takes 85 percent of the analysis time of the full-fledged
type analysis. This illustrates that the precision improvement
features does not substantially increase analysis time. 

The fifth column in Table~\ref{tab:compare} gives information about
the effect of the precision improvement features. For each program, it
lists the ratio of the number of the program points at which the
full-fledged analysis derives more precise type information than the
simplified analysis over the number of all program points.  Whether or
not these features improve analysis precision depends on the program
that is analysed. For some programs like {\sf dnf} and {\sf meta},
there is little or no improvement. For some other programs like {\sf
press} and {\sf serialize}, there is a substantial improvement. On
average, the full-fledged analysis derives more precise type
information at 54\% of the program points in a program. This indicates
that the precision improvement features is cost effective.

\subsection{Termination}

The abstract domain of our type analysis contains chains of
infinite length, which may lead to non-termination of the analysis of
a program.
\begin{example} \label{ex:ter} Let the program consist of a single clause
\texttt{p(x) :- $\circledast$ p([x])} where $\circledast$ is a
label of a program point. Let the query be of the form \texttt{:-
p(u)} with $u$ being of type $\tnat$. Then $x$ is a
term of type $\tlist^{i}(\tnat)$ at the $i^{th}$ time the execution
reaches the program point $\circledast$. Thus, the chain of the
abstract substitutions at the program point $\circledast$ is
\[
\begin{array}{l}
\{\{x\values\tlist(\tnat)\}\}\\
\{\{x\values\tlist(\tnat)\},
    \{x\values\tlist(\tlist(\tnat))\}\}\\
\vdots\\
\{\{x\values\tlist^{j}(\tnat)\} \mid 0<j\leq k\}\\
\vdots\\
\end{array}
\]
which is infinite. The  program is an instance of polymorphic
recursion~\cite{Kahrs} which is prohibited in ML. \endprf
\end{example}

The analyzer uses a canonical representation of types and a depth
abstraction to ensure termination. A conjunctive type is compact if it
contains no duplicated type atoms.  A type in disjunctive normal form
is compact if it contains no duplicated conjuncts and all of its
conjuncts are compact.  A type is canonical if it is in disjunctive
normal form, it is compact and all arguments of its type atoms are
canonical.  For every type $\rt$, a canonical equivalent of $\rt$ -- a
canonical type $\rt_c$ such that $\rt_c\teq\rt$ -- can be obtained as
follows. A disjunctive normal form $\rt'$ of $\rt$ is first computed.
Each argument of each type atom in $\rt'$ is then replaced with its
canonical equivalent, resulting in a type $\rt''$. Finally, $\rt_c$ is
obtained by deleting duplicate type atoms in each conjunct of $\rt''$
and then deleting duplicate conjuncts.  Let $cn(\rt)$ denote the
canonical equivalent of $\rt$ obtained by the above procedure. For
instance, $cn(\ttree(\ttree(\tlist(\ttop)~\tor~\tlist(\ttop)))) =
\ttree(\ttree(\tlist(\ttop)))$.

Let $\rt$ be a type. An atomic sub-term $\at$ of $\rt$ is both a
sub-term of $\rt$ and a type atom.  The depth of $\at$ in $\rt$ is the
number of the occurrences of type constructors in $\cons$ on the path
from the root of $\rt$ to but excluding the root of $\at$.  Thus, the
depth of the only occurrence of $\tlist(\tnat)$ in
$\ttree(\ttree(\tlist(\teven)~\tor~\tlist(\tlist(\tnat))))$ is 3 and
the depth of the only occurrence of $\tlist(\teven)$ in the same type
is 2. Note that type constructors $\tand$ and $\tor$ are ignored in
determining the depth of $\at$ in $\rt$. If the depth of $\at$ in
$\rt$ is $k$ then $\at$ is called an atomic sub-term of $\rt$ at depth
$k$. The depth of $\rt$ is defined as the maximum of the depths of all
its atomic sub-terms.

\begin{definition}
Let $\rt$ be a type and $k$ a positive integer. The depth $k$
abstraction of $\rt$, denoted as $d_{k}(\rt)$, is the result of
replacing each argument of any atomic sub-term of $\rt$ at depth $k$
by $\ttop$. \endprf
\end{definition}
For instance,
\begin{eqnarray*}
\lefteqn{d_{2}(\ttree(\ttree(\tlist(\teven)~\tor~\tlist(\tlist(\tnat)))))}\\ &&
= \ttree(\ttree(\tlist(\ttop)~\tor~\tlist(\ttop)))
\end{eqnarray*}

During analysis, the abstract substitution for a program point is
initialized to the empty set of variable typings. It is updated by
adding new variable typings and removing redundant ones. The analyzer
ensures termination as follows.  For each program point, the analyzer
determines a depth $k$ the first time a non-empty set of variable
typings $\scal_0$ is added. The depth $k$ is the maximum of the depths
of the types occurring in $\scal_0$ plus some fixed constant $k_0$
with $k_0\geq 0$.  After that, each time a set of variable typings
$\scal$ is added, each type $\rt$ occurring in $\scal$ is replaced by
$cn(d_{k}(cn(\rt)))$.  The above abstraction preserves analysis
correctness because $\rt'\tleq d_{k}(\rt')$ and $cn(\rt)\teq\rt$. The
number of depth $k$ abstractions of the canonical types occurring in
the abstract substitution is bounded and so is the number of variable
typings in the abstract substitution. This ensures termination.
\begin{example} Continue Example~\ref{ex:ter} and let $k_0=1$.
We have $\scal_0=\{\{x\values\tlist(\tnat)\}\}$ and hence $k=2$ since
the depth of the only type $\tlist(\tnat)$ in $\scal_0$ is 1. The
chain of the abstract substitutions for the program point
$\circledast$ is
\[
\begin{array}{l}
\{\{x\values\tlist(\tnat)\}\}\\
\{\{x\values\tlist(\tnat)\},
    \{x\values\tlist(\tlist(\tnat))\}\}\\
\{
    \{x\values\tlist(\tnat)\},
    \{x\values\tlist(\tlist(\tnat))\},
    \{x\values\tlist(\tlist(\tlist(\ttop)))\}
\}
\end{array}
\]
The last in the chain is the final abstract substitution for the
program point $\circledast$. \endprf
\end{example}

\section{Related Work} \label{sec:related}

There is a rich literature on type inference analysis for logic
programs.  Type analyses
in~\cite{FruhwirthSVY:LICS91,GallagherW94,GallagherP02,Mishra:84,Zobel:87} are
performed without {\it a priori} type definitions. They generate
regular tree grammars, or type
graphs~\cite{CortesiCH_JLP95,Janssens:JLP92} or set
constraints~\cite{HeintzeJ90,HeintzeJ92} as type definitions.  These
different formalisms for expressing type definitions are equivalent. A
type graph is equivalent to a regular tree grammar such that a
production rule in the grammar corresponds to a subgraph that is
composed of a node and its successors in the graph. For a system of
set constraints, there is a regular tree grammar that generates the
least solution to the system of set constraints, and vice
versa~\cite{CousotCousot95}. The production rules in a regular tree
grammar are similar to type rules used in our analysis but are not
parameterized. \comments{In most cases, analysis precision can be
adjusted by tuning the widening mechanism which are used both to
generate type definitions and to enforce termination.}  This kind of
analysis is useful for compiler-time optimizations and transformations
but inferred type definitions can be difficult for the programmer to
interpret. Like those
in~\cite{HoriuchiK87,Kanamori:Horiuchi:85,BarbutiG92,KanamoriJLP93,CodishD94,Lu95,1995-saglam,aci_types_full,LuJLP98,HillS02},
our type analysis is performed with {\it a priori} type definitions.
The type expressions it infers are formed of given type
constructors. Since the meaning of a type constructor is given by {\it
a priori} type definitions that are well understood to the programmer,
the inferred types are easier for the programmer to interpret and thus
they are more useful in an interactive programming environment.

\comments
{There are two general approaches to type analysis for logic programs.
A bottom-up type analysis computes an approximation of the success set
of the
program~\cite{BarbutiG92,FruhwirthSVY:LICS91,HeintzeJ90,HeintzeJ92,Mishra:84,Zobel:87}. It
is done by mimicking a bottom-up semantics of logic programs such
as~\cite{Emden:Kowalski:76}. A top-down type analysis simulates an
operational semantics for logic
programs~\cite{BruynoogheJCD87,HoriuchiK87,Janssens:JLP92,
Kanamori:Horiuchi:85,KanamoriJLP93,Lu95,LuJLP98}.
}

The type analyses with {\it a priori} type definitions
in~\cite{HoriuchiK87,Kanamori:Horiuchi:85,KanamoriJLP93,Lu95} are
based on top-down abstract interpretation frameworks.  They are
performed with a type description of possible queries as an input and
are thus goal-dependent.  They infer for each program point a type
description of all the program states that might be obtained when the
execution of the program reaches that program point. These are also
characteristics of our analysis.  However, these analyses do not
support non-deterministic type definitions or non-discriminative union
at the levels of types and abstract substitutions. The analysis
in~\cite{LuJLP98} traces non-discriminative union at the level of
abstract substitutions but not at the level of types. In addition, it
does not allow non-determinism in type definitions.  The above
mentioned top-down type analyses with {\it a priori} type definitions
approximate non-discriminative union of two types by their least upper
bound. The least upper bound may have a strictly larger denotation
than the set union of the denotations of the two types since set union
is not a type constructor.  Thus, our type analysis is strictly more
precise
than~\cite{HoriuchiK87,Kanamori:Horiuchi:85,KanamoriJLP93,Lu95,LuJLP98}.

The type analyses with {\it a priori} type definitions
in~\cite{BarbutiG92,CodishD94,1995-saglam,aci_types_full,HillS02} are
based on bottom-up abstract interpretaion frameworks. They infer a
type description of the success set of the program. The inferred type
description is a set of type atoms each of which is a predicate symbol
applied to a tuple of types. Some general remarks can be made about
the differences between our analysis and these analyses. Firstly, our
analysis is goal-dependent while these analyses are
goal-independent. Secondly, our analysis allows non-deterministic type
definitions that are disallowed by these analyses. Consequently, more
natual typings are allowed by our type analysis than by these
analyses. However, non-deterministic type definitions also make
abstract operations in our analysis more complex than in these
analyses. Thirdly, like our analysis, these analyses can express
non-discriminative union at the level of predicates. For example, the
two type atoms $p(\tlist(integer))$ and $p(\ttree(integer))$ express
the same information as $\langle
p(x),x\in(\tlist(integer)~\tor~\ttree(integer))\rangle$ in our type
analysis. However, these analyses except an informal proposal
in~\cite{BarbutiG92} cannot trace non-discriminative union at the
level of arguments, which leads to imprecise analysis results. For
instance, the inferred type for the concrete atom $p([1,[1]])$ is
$p(\tlist(\ttop))$ according
to~\cite{CodishD94,1995-saglam,aci_types_full,HillS02} and the main
proposal in~\cite{BarbutiG92}. The inferred type $p(\tlist(\ttop))$ is
less precise than $\langle
p(x),x\in\tlist(integer~\tor~\tlist(integer))\rangle$ which is
inferred by our type analysis. Lastly, as a minor note, set
intersection is not used as a type constructor in these type analyses
except~\cite{HillS02}. The two type clauses
$x(\tlist(\beta))\leftarrow$ and $x(\ttree(\beta))\leftarrow$ in an
abstract substitution of~\cite{HillS02} indicates that $x$ is both a
list and a tree. Some comparisons on other aspects between our type
analysis and these bottom-up analyses are in order.

Barbuti and Giacobazzi~\shortcite{BarbutiG92} infer polymorphic types
of Horn clause logic programs using a bottom-up abstract
interpretation framework~\cite{BarbutiGL93}.  The type description of
the success set of a Horn logic program is computed as the least
fixed-point of an abstract immediate consequence operator associated
with the program.  The abstract immediate consequence operator is
defined in terms of abstract unification and abstract application.
Abstract unification computes an abstract substitution given a term
and a type.  Abstract application computes a type given an abstract
substitution and a term. Both computations are derivations of a Prolog
program that is derived from {\it a priori} type definitions.  The
inferred type description describes only part of the success set of
the program though abstract operations can be modified so that the
type description approximates the whole success set. Ill-typed atoms
are not described by the type description. Nor are those well-typed
atoms that possess only ill-typed SLD resolutions. An SLD resolution
is ill-typed if any of its selected atoms is ill-typed.  Their type
definitions are slightly different form ours. For instance, they
define the type of the empty list $[~]$ as $[~] \rightarrow
\tlist(\bot)$ which is equivalent to $\tlist(\tbot)\tdef [~]$ in our
notation whilst the empty list $[~]$ is typed by $\tlist(\beta)\tdef
[~]$ in our analysis.  Barbuti and Giacobazzi also informally
introduced and exemplified an associative, commutative and idempotent
operator $\cup$ that expresses non-deterministic union at the level of
types. However, abstract unification and abstract application
operations for this modified domain of types are not given. In
addition, it requires changing type definitions, for instance, from
$cons(\beta,\tlist(\beta))\rightarrow \tlist(\beta)$ to
$cons(\alpha,\tlist(\beta))\rightarrow \tlist(\alpha\cup\beta)$.
Barbuti and Giacobazzi's analysis captures more type dependency than
ours. This is achieved through type parameters. For instance, the type
description for the program $\{p(X,[X])\leftarrow\}$ is
$\{p(\alpha,\tlist(\alpha))\}$. Abstract unification of the query
$p(X,Y)$ with the only type atom in the type description yields the
abstract substitution $\{X\values \alpha,Y\values\tlist(\alpha)\}$,
This kind of type dependency will be lost in our analysis.  The use of
type parameters and the use of non-discriminative union are orthogonal
to each other and it is an interesting topic for future research to
combine them for more analysis precision.

Codish and Demoen~\shortcite{CodishD94} apply abstract
compilation~\cite{HermenegildoWD92} to infer type dependencies by
associating each type with an incarnation of the abstract domain ${\sf
Prop}$~\cite{MarriottS89}.  The incarnations of ${\sf Prop}$ define
meanings of types and capture interactions between types. The type
dependencies of a logic program is similar to the type description of
the program inferred by the type analysis of Barbuti and
Giacobazzi~\shortcite{BarbutiG92} except that the type dependencies
describe the whole success set of the program. Codish and
Lagoon~\shortcite{aci_types_full} improve~\cite{CodishD94} by
augmenting abstract compilation with ACI-unification. An associative,
commutative and idempotent operator $\oplus$ is introduced to form the
type of a term from the types of its sub-terms.  It has the flavor of
set union. Nevertheless, it does not denote the set union. For an
example, the term $[1,[1]]$ has type
$\tlist(integer)\oplus\tlist(\tlist(integer))$ according
to~\cite{aci_types_full} while it has type
$\tlist(integer~\tor~\tlist(integer))$ in our type
analysis. Like~\cite{BarbutiG92}, type analyses
in~\cite{CodishD94,aci_types_full} capture type dependency via type
parameters. In addition, they have the desired property of condensing
which our analysis does not have.

Hill and Spoto~\shortcite{HillS02} provide a method that enriches an
abstract domain with type dependency information. The enriched domain
contain elements like $(x\in\tnat)\rightarrow (y\in\tlist(\tnat))$
meaning if $x$ has type $\tnat$ then $y$ has type
$\tlist(\tnat)$. Each element in the enriched domain is represented as
a logic program. Type analysis is performed by abstract compilation.
Their approach to improving precision of type analysis is different
from ours.  Their domain can express type dependencies that ours
cannot whilst our domain can express non-discriminative union at the
level of types but theirs cannot. Hill and Spoto do not take subtyping
into account in their design of abstract operations possibly because
subtyping is outside the focus of their work.

Gallagher and de Waal~\shortcite{GallagherW94} approximates the
success set of the program by a unary regular logic
program~\cite{Yardeni:Shapiro:91}.  This analysis infers both type
definitions and types and is incorporated into the Ciao
System~\cite{HermenegildoBPL99}. Saglam and
Gallagher~\shortcite{1995-saglam} extend~\cite{GallagherW94} by
allowing the programmer to supply deterministic type definitions for
some function symbols. The supplied type definitions are used to
transform the program and the transformed program is analyzed as
in~\cite{GallagherW94}. An interesting topic for further study is to
integrate non-deterministic type definitions and non-discriminative
union into~\cite{1995-saglam} and evaluate their impact on analysis
precision and analysis cost.

Finally, it is also worthy mentioning work on directional
types~\cite{BronsardJICSLP92,AikenL94,BoyeM96,CharatonikP98,RychlikowskiT01}.
Aiken and Lakshman~\shortcite{AikenL94} present an algorithm for
automatic checking directional types of logic programs. Directional
types describe both the structure of terms and the directionality of
predicates. A directional type for a predicate $p/n$ is of the form
${\tau}_I\rightarrow{\tau}_O$. Type $\tau_I$ is called an input type
and type $\tau_O$ an output type. They are type tuples of dimension
$n$. The directional type expresses two requirements. Firstly, if
$p/n$ is called with an argument of type ${\tau}_I$ then the argument
has type ${\tau}_O$ upon its success. Secondly, each predicate $q/m$
invoked by $p$ is called with an argument that has the input type of a
directional type for $q/m$. A program is well-typed with respect to a
collection of directional types if each directional type in the
collection is verified.  The type checking problem is reduced to a
decision problem on systems of inclusion constraints over set
expressions. The algorithm is sound and complete for discriminative
directional types. Charatonik and Podelski~\shortcite{CharatonikP98}
provide an algorithm for inferring directional types with respect to
which the program is well-typed.

\section{Conclusion} \label{sec:con}

We have presented a type analysis. The type analysis supports
non-deterministic type definitions, allows set operators in type
expressions, and uses a set of variable typings to describe type
information in a set of substitutions. The analysis is presented as an
abstract domain and four abstract operations for Nilsson's abstract
semantics~\cite{Nilsson:88} extended to deal with negation and
built-in predicates. These operations are defined in detail and their
local correctness proved. The abstract unification involves
propagation of type information downwards and upwards the structure of
a term. Given a set of equations in solved form and an abstract
substitution, abstract unification is accomplished in two steps. In
the first step, more type information for variables occurring on the
right-hand side of each equation is derived from type information for
the variable on the left-hand side. The second step derives more type
information for the variable on the left-hand side of each equation
from type information for the variables on the right-hand side. The
abstract built-in execution operation approximates the execution of built-in
predicates.  Each built-in is modeled as a function of abstract
substitutions.

Detection of the least fixpoint and elimination of redundancy in a set
of variable typings are both reduced to checking the emptiness of
types. Though types denote sets of possibly non-ground terms and are
not closed under set complement, checking the emptiness of types can
be done by using an algorithm that checks for the emptiness of the
types that denote sets of ground terms.  An experimental study shows
that due to a large repetition of emptiness checks, with tabling, the
precision improvement measures incurs only a small increase in
analysis time.

\paragraph{Acknowledgments:} The work is supported in part by the National Science
Foundation under grants CCR-0131862 and INT-0327760. A preliminary
version of this article appeared, under the title ``A Precise Type
Analysis of Logic Programs'', in {\it Proceedings of the Second
International ACM SIGPLAN Conference on Principles and Practice of
Declarative Programming}, Montreal, Canada, 2000. We would like to
thank anonymous reviewers for insightful comments on previous drafts
of this report.

\appendix

\section{Proofs}
Let ${\sf N}$ denote the set of natural numbers.  Define
  $h:\term(\func,\allvars)\mapsto{\sf N}$ as follows.  $h(x)\definedas
  0$ for all $x\in\allvars$ and $h(f(t_1,\cdots,t_n))\definedas 1+
  max\{h(t_i)~|~1\leq i\leq n\}$.  Define $h:\rtype\mapsto{\sf N}$ in
  the same way. Note $h(\rt)\geq 1$ for any $\rt\in\rtype$. Let
  $\langle x^1,y^1\rangle \prec \langle x^2,y^2\rangle\definedas ( x^1
  < x^2) \lor (( x^1= x^2)\land( y^1 < y^2))$.
\vspace{1pc}

\lemmaproof{lm:closed}
{Let $\rt\in\rtype$ and $t\in\term$. If $t\in\sem{\rules}{\rt}$ then
$\sigma(t)\in\sem{\rules}{\rt}$ for any $\sigma\in\sub$.
}
{
The proof is done by induction on $\langle h(t),h(\rt)\rangle$.  Let
$\sigma\in\sub$ be an arbitrary substitution.

Basis. We have that $h(t)=0$ and that $h(\rt)=1$. So, $t\in\allvars$
which implies $\rt=\ttop$ since $t\in\sem{\rules}{\rt}$ and
$h(\rt)=1$. Thus, $\sem{\rules}{\rt}=\term$ and
$\sigma(t)\in\sem{\rules}{\rt}$.

Induction.  Either that $h(t)=0$ or that $h(t)>0$.  Consider the case
where $h(t)=0$ first. Then $t\in\allvars$. Either (i) $\rt=\ttop$;
(ii) $\rt=\rt_1~\tor~\rt_2$; or (iii) $\rt=\rt_1~\tand~\rt_2$. The
case (i) is a special case of the base case. Consider the case (ii). We have
either that $t\in\sem{\rules}{\rt_1}$ or that
$t\in\sem{\rules}{\rt_2}$. If $t\in\sem{\rules}{\rt_j}$ then, by
induction hypothesis, $\sigma(t)\in\sem{\rules}{\rt_j}$ for $j=1,2$
since $h(\rt_j)<h(\rt)$. So, $\sigma(t)\in\sem{\rules}{\rt}$ by the
definition of $\sem{\rules}{\cdot}$ . The case (iii) is symmetric to
the case (ii). Thus, $\sigma(t)\in\sem{\rules}{\rt}$.

Now consider the case where $h(t)>0$. Then $t=f(t_1,\cdots,t_n)$.
Either (i) $\rt=\ttop$; (ii) $\rt=\rt_1~\tor~\rt_2$; (iii)
$\rt=\rt_1~\tand~\rt_2$; or (iv) $\rt=c(\rt_1,\cdots,\rt_m)$. The
proof for that $\sigma(t)\in\sem{\rules}{\rt}$ in the cases (i), (ii)
and (iii) is the same as in the previous paragraph. Consider the case
(iv).  Since $t\in\sem{\rules}{\rt}$, there is a type rule
$c(\beta_1,\cdots,\beta_m)\tdef f(\tau_1,\cdots,\tau_n)$ such that
$t_j\in\sem{\rules}{\Bbbk(\tau_j)}$ where
$\Bbbk=\{\beta_1\mapsto\rt_1,\cdots,\beta_m\mapsto\rt_m\}$. We have
that $h(\Bbbk(\tau_j))\leq h(\rt)$ and that $h(t_j)< h(t)$. By the
induction hypothesis, $\sigma(t_j)\in\sem{\rules}{\Bbbk(\tau_j)}$,
which together with the definition for $\sem{\rules}{\cdot}$, implies
that $\sigma(t)\in\sem{\rules}{\rt}$.}

\lemmaproof{lm:A} { $\rcon(\rsub)$ is a Moore family.  } { Since,
$\rcon({[\{x\values\ttop~|~x\in\vi\}]}_\req)=\ctop$ and $\ctop$ is the
supremum on $\csub$, $\rcon(\rsub)$ contains the supremum on $\csub$.
Let ${[\scal_1]}_\req,{[\scal_2]}_\req\in\rsub$.  Then
${[\scal_1]}_\req\rglb{[\scal_2]}_\req\in\rsub$.  Furthermore,
\begin{eqnarray*}
\rcon({[\scal_1]}_\req\rglb{[\scal_2]}_\req) &=&
   \rcon({[\scal_1^{\downarrow}\cap\scal_2^{\downarrow}]}_\req)\\
&=& (\bigcup_{\mu\in\scal_1^{\downarrow}}\vtcon(\mu))\cap
     (\bigcup_{\nu\in\scal_2^{\downarrow}}\vtcon(\nu))\\
&=& (\bigcup_{\mu\in\scal_1}\vtcon(\mu))\cap
     (\bigcup_{\nu\in\scal_2}\vtcon(\nu))\\
&=& \rcon({[\scal_1]}_\req)\cap\rcon({[\scal_2]}_\req)
\end{eqnarray*}
Thus, $\rcon(\rsub)$ is closed under $\cap$ -- the meet  on
$\csub$. So, $\rcon(\rsub)$ is a Moore family.  }

\lemmaproof{lm:B}
{  $\rcon({[\scal_1\otimes\scal_2]}_\req)
=\rcon({[\scal_1]}_\req\rglb{[\scal_2]}_\req)$.
}
 { We first prove that
$\rcon({[\scal_1\otimes\scal_2]}_\req)
\subseteq\rcon({[\scal_1]}_\req\rglb{[\scal_2]}_\req)$. Let
$\theta\in\rcon({[\scal_1\otimes\scal_2]}_\req)$. Then there is $\rho$
in $(\scal_1\otimes\scal_2)$ such that $\theta\in\vtcon(\rho)$.  This
implies that there are $\mu$ in $\scal_1$ and $\nu\in\scal_2$ such
that $\rho=\lambda x\in\vi.(\mu(x)~\tand~\nu(x))$.  We have
$\vtcon(\rho)\subseteq\vtcon(\mu)$ and
$\vtcon(\rho)\subseteq\vtcon(\nu)$, implying
$\rho\in\scal_1^{\downarrow}$ and
$\rho\in\scal_2^{\downarrow}$. Therefore,
$\rho\in(\scal_1^{\downarrow}\cap\scal_2^{\downarrow})$. Since
$\theta\in\vtcon(\rho)$, we have that
$\theta\in\rcon({[\scal_1^{\downarrow}\cap\scal_2^{\downarrow}]}_\req)$
and $\theta\in\rcon({[\scal_1]}_\req\rglb{[\scal_2]}_\req)$.

 We now prove that $\rcon({[\scal_1\otimes\scal_2]}_\req)
\supseteq\rcon({[\scal_1]}_\req\rglb{[\scal_2]}_\req)$. Let
$\theta\in\rcon({[\scal_1]}_\req\rglb{[\scal_2]}_\req)$. Then
$\theta\in\vtcon(\rho)$ for some
$\rho\in(\scal_1^{\downarrow}\cap\scal_2^{\downarrow})$ by the
definition of $\rglb$. There are  $\mu\in\scal_1$ and
$\nu\in\scal_2$ such that $\vtcon(\rho)\subseteq\vtcon(\mu)$ and
$\vtcon(\rho)\subseteq\vtcon(\nu)$, implying $\forall
x\in\vi.(\theta(x)\in\sem{\rules}{\mu(x)~\tand~\nu(x)})$. Thus,
$\theta\in\rcon({[\scal_1\otimes\scal_2]}_\req)$ by the definition of
$\otimes$.
}

\lemmaproof{lm:C} { For any $\rt\in\rtype$ and
$t\in\term(\func,\viplus)$,
$\{\theta~|~\theta(t)\in\sem{\rules}{\rt}\}\subseteq\rcon({[\vts(\rt,t)]}_\req)$.
} {The proof is done by induction on $\langle h(t),h(\rt)\rangle$.

Basis. $\langle h(t),h(\rt)\rangle =\langle 0,1\rangle$. Then
$t\in\viplus$ and
\[\vts(\rt,t)=\{\lambda x\in\viplus.(\mbox{if $x=t$ then $\rt$ else $\ttop$})\}
\]  The
lemma holds since
$\{\theta~|~\theta(t)\in\sem{\rules}{\rt}\}=\rcon({[\vts(\rt,t)]}_\req)$.

Induction. Assume that the lemma holds for all
  $\rt'\in\rtype$ and $t'\in\term(\func,\viplus)$ such
  that $\langle h(t'),h(\rt')\rangle \prec \langle
  h(t),h(\rt)\rangle$. Either (1) $h(\rt)>1$ or (2)
  $h(\rt)=1$.

Consider the case (1). Either (i) $\rt = \rt_1~\tand~\rt_2$ or (ii)
$\rt=\rt_1~\tor~\rt_2$ or (iii) $\rt=c(\rt_1,\cdots,\rt_m)$ for some
$m\geq 1$. The cases (i) and (ii) are immediate.  Consider the case
(iii). By the definition of $\sem{\rules}{\cdot}$,
$\theta(f(t_1,\cdots,t_n))\in \sem{\rules}{\rt}$ implies that there is
a type rule $c(\beta_1,\cdots,\beta_m)\tdef f(\tau_1,\cdots,\tau_n)$
in $\rules$ such that
$\theta(t_i)\in\sem{\rules}{\Bbbk(\tau_i)}$ where
$\Bbbk=\{\beta_j\mapsto\rt_j~|~1\leq{j}\leq{m}\}$. We have
$h(t_i)<h(f(t_1,\cdots,t_n))$. By the induction hypothesis,
\[\theta\in\rcon({[\vts(\Bbbk(\tau_i),t_i)]}_\req)\] for all $1\leq i\leq
n$.
 By Lemmas~\ref{lm:A} and~\ref{lm:B},
\[\theta\in\rcon({\left[\bigotimes_{1\leq{i}\leq{n}}\vts(\Bbbk(\tau_i),t_i)\right]}_\req)\]
By the definition of $\vts$, we have
\[\theta\in\rcon({[\vts(c(\rt_1,\cdots,\rt_m),f(t_1,\cdots,t_n))]}_\req)\]
 Thus, the lemma holds for the case (1).

Now consider the case (2). We have that $t=f(t_1,\cdots,t_n)$. The
proof is the same as that for the case (1).(iii).  }

 \lemmaproof{lm:G}
{Let $\scal'=\mdown(E,\scal)$. Then
$\mgu(\theta(E))\circ\theta\in\rcon({[\scal']}_\req)$ for all
$\theta\in\rcon({[\scal]}_\req)$.
}
{Let
    $\scal_{\mu}=\{\mu\}\otimes\bigotimes_{(x=t)\in{E}}\vts(\mu(x),t)$. It
    suffices to prove that
    $\mgu(\sigma(E))\circ\sigma\in\rcon({[\scal_{\mu}]}_\req)$ for all
    $\sigma\in\vtcon(\mu)$. $\mgu(\sigma(E))\circ\sigma\in\vtcon(\mu)$
    as the denotation of any type in $\rtype$ is closed under
    substitution. By Lemma~\ref{lm:C}, we have
    $\mgu(\sigma(E))\circ\sigma\in\rcon({[\vts(\mu(x),t)]}_\req)$ for any
    $(x=t)$ in $E$. So,
    $\mgu(\sigma(E))\circ\sigma\in\rcon({[\scal_{\mu}]}_\req)$.
 }

\lemmaproof{P:1}
{For any $\tau\in\gtype$ and any $\Bbbk_1,\Bbbk_2\in\tvals$,
\begin{itemize}
\item [(a)] $(\Bbbk_1(\tau)~\tor~\Bbbk_2(\tau)) \tleq {(\Bbbk_1\tvallub\Bbbk_2)(\tau)}$; and
\item [(b)] $(\Bbbk_1(\tau)~\tand~\Bbbk_2(\tau)) \teq {(\Bbbk_1\tvalglb\Bbbk_2)(\tau)}$.
\end{itemize}
} { We prove only (a) since the proof for (b) is similar to that for
(a). Let
$t\in\sem{\rules}{\Bbbk_1(\tau)~\tor~\Bbbk_2(\tau)}$. Either (1)
$t\in\sem{\rules}{\Bbbk_1(\tau)}$ or (2)
$t\in\sem{\rules}{\Bbbk_2(\tau)}$. Without loss of generality, we
assume (1).  We prove
$t\in\sem{\rules}{(\Bbbk_1\tvallub\Bbbk_2)(\tau)}$ by induction on
$\langle h(\tau),h(t)\rangle$.

Basis. $h(\tau)=0$. Then $\tau\in\pars$ and (a) holds since
$(\Bbbk_1(\tau)~\tor~\Bbbk_2(\tau)) \teq
{(\Bbbk_1\tvallub\Bbbk_2)(\tau)}$ by definition of $\tvallub$.

Induction. $h(\tau)\neq 0$ implies that $\tau=c(\Vector{\tau}{m})$.
 If $t\in\allvars$ then $\Bbbk_1(\tau)\teq\ttop$ and hence
 $(\Bbbk_1\tvallub\Bbbk_2)(\tau)\teq\ttop$ and
 $t\in\sem{\rules}{(\Bbbk_1\tvallub\Bbbk_2)(\tau)}$. Otherwise,
 $t=f(\Vector{t}{n})$. Since $t\in\sem{\rules}{\Bbbk_1(\tau)}$, there
 is a type rule $\tau\tdef f(\Vector{\tau}{n})$ such that
 $t_i\in\sem{\rules}{\Bbbk_1(\tau_i)}$ for $1\leq{i}\leq{n}$. We have
 $h(\tau_i)\leq h(\tau)$ and $h(t_i)<h(t)$. Thus,
 $t_i\in\sem{\rules}{(\Bbbk_1\tvallub\Bbbk_2)(\tau_i)}$ by the
 induction hypothesis and hence
 $t\in\sem{\rules}{(\Bbbk_1\tvallub\Bbbk_2)(\tau)}$ by the definition
 $\sem{\rules}{\cdot}$.
}

\lemmaproof{P:2} { Let $\kcal_1,\kcal_2\in\wp(\tvals)$,
$\rt\in\rtype$, $\tau\in\gtype$, $\rtvec\in\rtype^{*}$ and
$\tauvec\in\gtype^{*}$ such that $\|\rtvec\|=\|\tauvec\|$. If
${\rt}\tleq\tor_{\Bbbk_1\in\kcal_1}{\Bbbk_1(\tau)}$ and ${\rtvec}
\tleq \tor_{\Bbbk_2\in\kcal_2}{\Bbbk_2(\tauvec)}$ then
${\rt\bullet\rtvec} \tleq
\tor_{\Bbbk\in(\kcal_1\bigtvallub\kcal_2)}{\Bbbk(\tau\bullet\tauvec)}$.
} { Let $t\bullet\vec{t}\in\sem{\rules}{\rt\bullet\rtvec}$. Then
$t\in\sem{\rules}{\rt}$ and $\vec{t}\in\sem{\rules}{\rtvec}$. By
assumption, there are $\Bbbk_1\in\kcal_1$ such that
$t\in\sem{\rules}{\Bbbk_1(\tau)}$ and $\Bbbk_2\in\kcal_2$ such that
$\vec{t}\in\sem{\rules}{\Bbbk_2(\tauvec)}$. Let
$\Bbbk=\Bbbk_1\tvallub\Bbbk_2$. We have
$\Bbbk\in(\kcal_1\bigtvallub\kcal_2)$ by the definition of
$\bigtvallub$ and $t\in\sem{\rules}{\Bbbk(\tau)}$ and
$\vec{t}\in\sem{\rules}{\Bbbk(\tauvec)}$ by Lemma~\ref{P:1}. Thus,
$t\bullet\vec{t}\in\sem{\rules}{\Bbbk(\tau\bullet\tauvec)}$ by
the definition of $\sem{\rules}{\cdot}$. }

\lemmaproof{lm:D}
{Let $\tau\in\gtype$, $\rt\in\rtype$ and $\kcal=\tvs(\rt,\tau)$. Then
${\rt} \tleq \tor_{\Bbbk\in\kcal}{\Bbbk(\tau)}$.
}
{ The proof is done by induction on the structure of $\rt$.

   Basis. $\rt$ is atomic. $\rt=\ttop$ or $\rt=\tbot$
    or $\rt=c(\Vector{\rt}{m})$ for some
    $c/m\in\cons$ and
    $\Vector{\rt}{m}\in\rtype$. If $\rt=\ttop$ or
    $\rt=\tbot$ then the lemma holds by the
    definitions of $\tvaltop$, $\tvalbot$ and
    $\sem{\rules}{\cdot}$.  Let
    $\rt=c(\rt_1,\cdots,\rt_m)$. Either (a)
    $\tau\in\pars$ or (b)
    $\tau=d(\beta_1,\cdots,\beta_k)$ with
    $\beta_1,\cdots,\beta_k$ being different type
    parameters in $\pars$. In the case (a), we have
    $\kcal=\{\Bbbk\}$ with
    $\Bbbk=\{\tau\values\rt\}$. The lemma holds
    because $\Bbbk(\tau)=\rt$. Consider the case
    (b), if $c/m=d/k$ then $\kcal=\{\Bbbk\}$ with
    $\Bbbk=\{\beta_j\values\rt_j~|~1\leq{j}\leq{m}\}$
    by the definition of $\tvs$ and we have
    $\Bbbk(\tau)=\rt$. Otherwise, $\kcal=\{\Bbbk\}$
    with $\Bbbk=\tvaltop$ by the definition of
    $\tvs$ and $\Bbbk(\tau)=\ttop$ by the
    definition of $\tvaltop$. So, the lemma holds
    in the case (b).

    Induction. Either (1) $\rt=\rt_1~\tor~\rt_2$ or (2)
$\rt=\rt_1~\tand~\rt_2$.  In
the case (1), let $\kcal_1=\tvs(\rt_1,\tau)$ and
$\kcal_2=\tvs(\rt_2,\tau)$.  We have $\sem{\rules}{\rt_i}\subseteq
\bigcup_{\Bbbk\in\kcal_i}\sem{\rules}{\Bbbk(\tau)}$ for
$1\leq{i}\leq{2}$ by the induction hypothesis. Therefore,
\begin{eqnarray*}
   \sem{\rules}{\rt_1~\tor~\rt_2} &=&
   \sem{\rules}{\rt_1}\cup\sem{\rules}{\rt_2}\\
    &\subseteq &
   \bigcup_{\Bbbk\in\kcal_1}\sem{\rules}{\Bbbk(\tau)}\cup
   \bigcup_{\Bbbk\in\kcal_2}\sem{\rules}{\Bbbk(\tau)} \\
   &=&
    \bigcup_{\Bbbk\in(\kcal_1\cup\kcal_2)}\sem{\rules}{\Bbbk(\tau)}\\
   & =& \bigcup_{\Bbbk\in\tvs(\rt,\tau)}\sem{\rules}{\Bbbk(\tau)}
\end{eqnarray*}
So the lemma holds for the case (1). Consider the case  (2).
Let
$\kcal_1=\tvs(\rt_1,\tau)$ and
$\kcal_2=\tvs(\rt_2,\tau)$.  We have
$\sem{\rules}{\rt_i}\subseteq \bigcup_{\Bbbk\in\kcal_i}
\sem{\rules}{\Bbbk(\tau)}$ for
$1\leq{i}\leq{2}$ by the induction hypothesis. So,
\begin{eqnarray*}
   \sem{\rules}{\rt_1~\tand~\rt_2} &=&
   \sem{\rules}{\rt_1}\cap\sem{\rules}{\rt_2}\\
    &\subseteq &
   \bigcup_{\Bbbk\in\kcal_1}\sem{\rules}{\Bbbk(\tau)}\cap
   \bigcup_{\Bbbk\in\kcal_2}\sem{\rules}{\Bbbk(\tau)} \\
   &=&
    \bigcup_{\Bbbk\in(\kcal_1\bigtvalglb\kcal_2)}\sem{\rules}{\Bbbk(\tau)}~~\mbox{by Lemma~\ref{P:1}.(b)}\\
   & =& \bigcup_{\Bbbk\in\tvs(\rt,\tau)}\sem{\rules}{\Bbbk(\tau)}
\end{eqnarray*}
Thus, the lemma holds for the case (2). }

\lemmaproof{lm:F}
{Let $t\in\term(\func,\viplus)$ and
$\mu\in(\viplus\mapsto\rtype)$. Then
$\theta(t)\in\sem{\rules}{\type(t,\mu)}$ for all $\theta\in\vtcon(\mu)$.
}
{ The proof is done by induction on $h(t)$.

Basis. $h(t)=0$. Then $t\in\viplus$ and
$\type(t,\mu)=\mu(t)$. The lemma holds.

Induction. $h(t)>0$. Let $t=f(t_1,\cdots,t_n)$ and
$\rt_i=\type(t_i,\mu)$ for $i\in\{1,\cdots,n\}$ and
$\theta\in\vtcon(\mu)$.  By the induction hypothesis, we have
$\theta(t_i)\in\sem{\rules}{\rt_i}$ for all $1\leq{i}\leq{n}$. Let
$\tau\tdef f(\Vector{\tau}{i})$ be a type rule in $\rules$ and
$\kcal_i=\tvs(\rt_i,\tau_i)$. By Lemma~\ref{lm:D},
$\sem{\rules}{\rt_i}\subseteq\bigcup_{\Bbbk\in\kcal_i}\sem{\rules}{\Bbbk(\tau_i)}$. Thus,
$\sem{\rules}{\langle\Vector{\rt}{n}\rangle}\subseteq
\bigcup_{\Bbbk\in(\bigtvallub_{1\leq{i}\leq{n}}\kcal_i)}
\sem{\rules}{\Bbbk(\langle\Vector{\tau}{n}\rangle)}$ by
Lemma~\ref{P:2}, which implies $\theta(t)\in
\bigcup_{\Bbbk\in(\bigtvallub_{1\leq{i}\leq{n}}\kcal_i)}
\sem{\rules}{\Bbbk(\tau)}$ by the definition of
$\sem{\rules}{\cdot}$. This is true of each type rule for
$f/n$. Therefore, $\theta(t)\in\sem{\rules}{\type(t,\mu)}$.  }

\lemmaproof{lm:H}
{ Let $\scal\in\wp(\viplus\mapsto\rtype)$ and
                           $E\in\wp(\eqs)$. Then
$\mgu(\theta(E))\circ\theta\in\rcon({[\mup(E,\scal)]}_\req)$ for all
$\theta\in\rcon({[\scal]}_\req)$.
}
{ Let
\[\mu'=\lambda x\in\viplus.
    \left(\begin{array}{l}
    if~\exists t.(x=t)\in E\\
    then~\mu(x)~\tand~\type(t,\mu)\\
    else~\mu(x)
    \end{array}
    \right)
\]
 It
                            suffices to prove that
                            $\mgu(\sigma(E))\circ\sigma\in\vtcon(\mu')$
                            for all $\sigma\in\vtcon(\mu)$. By
                            Lemma~\ref{lm:F},
                            $\sigma(t)\in\sem{\rules}{\type(t,\mu)}$. We
                            have
                            $(\mgu(\sigma(E))\circ\sigma)(x)\in\sem{\rules}{\type(t,\mu)}$
                            for all $x$ and $t$ such that
                            $(x=t)\in{E}$.  Therefore,
                            $(\mgu(\sigma(E))\circ\sigma)\in\vtcon(\mu')$.
}

\theoremproof{lm:I}
{ For any
  ${[\scal_{1}]}_\req,{[\scal_{2}]}_\req\in
  \rsub$ and any $a_1,a_2\in\atom_{P}$,
  \begin{eqnarray*}
  {\cunify(a_1,\rcon({[\scal_{1}]}_\req),
  a_2,\rcon({[\scal_{2}]}_\req))}
 & \subseteq & \rcon
  (\munify(a_1,{[\scal_{1}]}_\req,a_2,{[\scal_{2}]}_\req))
  \end{eqnarray*}
} { We first prove a preliminary result on substitution and
 unification.
 Let $\eta,\theta\in\sub$ and $E,E_1,E_2\in\wp(\eqs)$ and assume that
 $\theta=mgu(\eta(E))\circ\eta\neq\fail$. Recall that $mgu(E_1\cup
 E_2)=mgu(E_1\cup eq(mgu(E_2)))$ and $mgu(\eta(E))=mgu(eq(\eta)\cup
 E)$~\cite{eder}. Then
\begin{eqnarray*} mgu(\theta(E))\circ\theta &=&
   mgu(mgu(\eta(E))\circ\eta(E))\circ\mgu(\eta(E))\circ\eta \\
 &=&
   mgu(mgu(\eta(E))(\eta(E)))\circ\mgu(\eta(E))\circ\eta\\
 &=& mgu(eq(mgu(\eta(E)))\cup eq(\eta)\cup E)\circ\mgu(\eta(E))\circ\eta\\
&=& mgu(eq(mgu(eq(\eta)\cup E))\cup eq(\eta)\cup E)\circ\mgu(\eta(E))\circ\eta\\
&=& mgu(eq(\eta)\cup E\cup eq(\eta)\cup E)\circ\mgu(\eta(E))\circ\eta\\
&=& mgu(eq(\eta)\cup E)\circ\mgu(\eta(E))\circ\eta\\
&=& mgu(\eta(E))\circ\mgu(\eta(E))\circ\eta\\
&=& mgu(\eta(E))\circ\eta\\
&=& \theta
\end{eqnarray*}
We are now ready to prove the theorem.  Let
$\theta_1\in\rcon({[\scal_{1}]}_\req)$,
$\theta_2\in\rcon({[\scal_{2}]}_\req)$ and $E_0 =
eq\circ\mgu(\Psi(a_1),a_2)$. Assume that
$\unify(a_1,\theta_1,a_2,\theta_2)\neq\fail$. It is equivalent to
prove
$\unify(a_1,\theta_1,a_2,\theta_2)\in\rcon(\munify(a_1,{[\scal_{1}]}_\req,a_2,{[\scal_{2}]}_\req))$.
By the definition of $\rcon$ and $\mrestrict$, if
$\zeta\in\rcon({[\scal]}_\req)$ then
$\zeta\in\rcon({[\mrestrict(\scal)]}_\req)$ for any substitution
$\zeta$ and any set of variable typings over $\viplus$. Thus, it
suffices to prove that $ \unify(a_1,\theta_1,a_2,\theta_2)\in
\rcon({[\mup(E_0,\mdown(E_0,\Psi(\scal_1)\biguplus\scal_2))]}_\req)$
by the definitions for $\munify$ and $\msolve$. Without loss of
generality, assume that $\Psi$ renames $\theta_1(a_1)$ apart from
$\theta_2(a_2)$. Let $\eta=\theta_2\cup\Psi(\theta_1)$ and
$\theta=\mgu(\eta(E_0))\circ\eta$. Then
\begin{eqnarray*}
\lefteqn{\unify(a_1,\theta_1,a_2,\theta_2) \in \rcon({[\mup(E_0,\mdown(E_0,\Psi(\scal_1)\biguplus\scal_2))]}_\req)} \\
 &\lequiv &
 \mgu((\Psi(\theta_1))(\Psi(a_1)),\theta_2(a_2))\circ\theta_2 \in \rcon({[\mup(E_0,\mdown(E_0,\Psi(\scal_1)\biguplus\scal_2))]}_\req) \\
 &\lequiv &
 \mgu(\eta(E_0))\circ\eta\in \rcon({[\mup(E_0,\mdown(E_0,\Psi(\scal_1)\biguplus\scal_2))]}_\req) \\
 &\lequiv &
 \theta\in \rcon({[\mup(E_0,\mdown(E_0,\Psi(\scal_1)\biguplus\scal_2))]}_\req)
\end{eqnarray*}
Thus, it remains to prove $\theta\in
\rcon({[\mup(E_0,\mdown(E_0,\Psi(\scal_1)\biguplus\scal_2))]}_\req)$. Since
$\eta\in\rcon(\Psi(\scal_1)\biguplus\scal_2)$ and
$\theta=\mgu(\eta(E_0))\circ\eta$, it holds that $\theta\in
\rcon({[\mdown(E_0,\Psi(\scal_1)\biguplus\scal_2)]}_\req)$ according
to Lemma~\ref{lm:G}. According to Lemma~\ref{lm:H}, we have
$\mgu(\theta(E))\circ\theta\in
\rcon({[\mup(E_0,\mdown(E_0,\Psi(\scal_1)\biguplus\scal_2))]}_\req)$. Note
that $\mgu(\theta(E))\circ\theta=\theta$. Thus, $\theta\in
\rcon({[\mup(E_0,\mdown(E_0,\Psi(\scal_1)\biguplus\scal_2))]}_\req)$.
}

\theoremproof{th:eq}
{ For any term $t$ in $\term(\func,\allvars)$ and any type
$\rt$ in $\rtype$, $t\in\sem{\rules}{\rt}$ iff
$\chi(t)\in\semb{\rules}{\rt}$.
}
{ We first prove  necessity. Assume that $t\in\sem{\rules}{\rt}$. We prove that
$\chi(t)\in\semb{\rules}{\rt}$ by induction on $\langle
h(t),h(\rt)\rangle$.

Basis. $h(t)=0$ and $h(\rt)=1$. We have that $t\in\allvars$ and that
$\rt=\ttop$ by the definition of $\sem{\rules}{\cdot}$. Thus,
$\chi(t)=\varrho\in\semb{\rules}{\rt}$.

Induction. Either $h(t)=0$ and $h(\rt)>1$ or $h(t)>0$ and $h(\rt)\geq
1$. Consider first the case where $h(t)=0$ and $h(\rt)>1$. Then
$t\in\allvars$ and either (i) $\rt=(\rt_1~\tor~\rt_2)$; or (ii)
$\rt=(\rt_1~\tand~\rt_2)$. We only prove the case (i) since the case
(ii) is dual to the case (i). Since $t\in\sem{\rules}{\rt}$, either
$t\in\sem{\rules}{\rt_1}$ or $t\in\sem{\rules}{\rt_2}$. So, we have
either $\chi(t)\in\semb{\rules}{\rt_1}$ or
$\chi(t)\in\semb{\rules}{\rt_2}$ by the induction hypothesis. Thus,
$\chi(t)\in\semb{\rules}{\rt}$.

Now consider the case $h(t)>0$ and $h(\rt)\geq 1$. Then
$t=f(t_1,\cdots,t_n)$. Either (i) $\rt=(\rt_1~\tor~\rt_2)$; (ii)
$\rt=(\rt_1~\tand~\rt_2)$; (iii) $\rt=\ttop$ or (iv)
$\rt=c(\rt_1,\cdots,\rt_m)$. The proof for that
$\chi(t)\in\semb{\rules}{\rt}$ in cases (i) and (ii) are the same as
in the previous paragraph. The case (iii) is vacuous. Consider the
case (iv). Since $t\in\sem{\rules}{\rt}$, by the definition of
$\sem{\rules}{\cdot}$, there is a type rule
$c(\beta_1,\cdots,\beta_m)\tdef f(\tau_1,\cdots,\tau_n)$ in $\rules$
such that $t_i\in\sem{\rules}{\Bbbk(\tau_i)}$ for all $1\leq i\leq n$
where
$\Bbbk=\{\beta_1\mapsto\rt_1,\cdots,\beta_m\mapsto\rt_m\}$. Observe
that $h(t_i)<h(t)$ and $h(\Bbbk(\tau_i))\leq h(\rt)$. By induction
hypothesis, $\chi(t_i)\in\semb{\rules}{\Bbbk(\tau_i)}$. By the
definition of $\semb{\rules}{\cdot}$, $t\in\semb{\rules}{\rt}$.

We now prove  sufficiency.  Assume that $\chi(t)\in\semb{\rules}{\rt}$. We
prove that $t\in\sem{\rules}{\rt}$ by induction on $\langle
h(t),h(\rt)\rangle$.

Basis. $h(t)=0$ and $h(\rt)=1$. Then $t\in\allvars$ and
$\chi(t)=\varrho$. We have that $\rt=\ttop$ by the definition of
$\semb{\rules}{\cdot}$. Thus, $t\in\sem{\rules}{\rt}$.

Induction. Either $h(t)=0$ and $h(\rt)>1$ or $h(t)>0$ and $h(\rt)\geq
1$. Consider first the case where $h(t)=0$ and $h(\rt)>1$. Then
$t\in\allvars$ and either (i) $\rt=(\rt_1~\tor~\rt_2)$; or (ii)
$\rt=(\rt_1~\tand~\rt_2)$. We only prove the case (i) since the case
(ii) is dual to the case (i). Since $\chi(t)\in\semb{\rules}{\rt}$, either
$\chi(t)\in\semb{\rules}{\rt_1}$ or $\chi(t)\in\semb{\rules}{\rt_2}$. So, we have
either $t\in\sem{\rules}{\rt_1}$ or
$t\in\sem{\rules}{\rt_2}$ by the induction hypothesis. Thus,
$t\in\sem{\rules}{\rt}$.

Now consider the case $h(t)>0$ and $h(\rt)\geq 1$. Then
$t=f(t_1,\cdots,t_n)$. Either (i) $\rt=(\rt_1~\tor~\rt_2)$; (ii)
$\rt=(\rt_1~\tand~\rt_2)$; (iii) $\rt=\ttop$ or (iv)
$\rt=c(\rt_1,\cdots,\rt_m)$. The proof for that
$t\in\sem{\rules}{\rt}$ in cases (i) and (ii) are the same as
in the previous paragraph. The case (iii) is vacuous. Consider the
case (iv). Since $\chi(t)\in\semb{\rules}{\rt}$, by the definition of
$\semb{\rules}{\cdot}$, there is a type rule
$c(\beta_1,\cdots,\beta_m)\tdef f(\tau_1,\cdots,\tau_n)$ in $\rules$
such that $\chi(t_i)\in\semb{\rules}{\Bbbk(\tau_i)}$ for all $1\leq i\leq n$
where
$\Bbbk=\{\beta_1\mapsto\rt_1,\cdots,\beta_m\mapsto\rt_m\}$. Observe
that $h(t_i)<h(t)$ and $h(\Bbbk(\tau_i))\leq h(\rt)$. By induction
hypothesis, $t_i\in\sem{\rules}{\Bbbk(\tau_i)}$. By the
definition of $\sem{\rules}{\cdot}$, $\chi(t)\in\sem{\rules}{\rt}$.
}

\coproof{co:eq}
{For any $\rt_1,\rt_2\in\rtype$,
$\sem{\rules}{\rt_1}\subseteq\sem{\rules}{\rt_2}$ iff
$\semb{\rules}{\rt_1}\subseteq\semb{\rules}{\rt_2}$.
}
{
Both sufficiency and necessity are proved by contradiction. 
We first consider  sufficiency. Assume that $\semb{\rules}{\rt_1}\subseteq
\semb{\rules}{\rt_2}$ but $\sem{\rules}{\rt_1}\not\subseteq
\sem{\rules}{\rt_2}$. Then there is a term $t$ such that $t\in
\sem{\rules}{\rt_1}$ and $t\not\in\sem{\rules}{\rt_2}$. By
Theorem~\ref{th:eq}, we have that $\chi(t)\in
\semb{\rules}{\rt_1}$. By the assumption, $\chi(t)\in
\semb{\rules}{\rt_2}$. By Theorem~\ref{th:eq}, $t\in
\sem{\rules}{\rt_2}$ that contradicts with that
$t\not\in\sem{\rules}{\rt_2}$.

We now prove necessity. Assume that $\sem{\rules}{\rt_1}\subseteq
\sem{\rules}{\rt_2}$ but $\semb{\rules}{\rt_1}\not\subseteq
\semb{\rules}{\rt_2}$. Then there is a term $t$ such that $t\in
\semb{\rules}{\rt_1}$ and $t\not\in\semb{\rules}{\rt_2}$. By
Theorem~\ref{th:eq}, there is a term $t'$ such that $\chi(t')=t$ and
 $t'\in \sem{\rules}{\rt_1}$. By the assumption,
$t'\in \sem{\rules}{\rt_2}$. By Theorem~\ref{th:eq}, $t\in
\semb{\rules}{\rt_2}$ that contradicts with that
$t\not\in\semb{\rules}{\rt_2}$.
}


\begin{thebibliography}{}

\bibitem[\protect\citeauthoryear{Aiken and Lakshman}{Aiken and
  Lakshman}{1994}]{AikenL94}
{\sc Aiken, A.} {\sc and} {\sc Lakshman, T.} 1994.
\newblock Directional type checking of logic programs.
\newblock In {\em Proceedings of the {F}irst {I}nternational {S}tatic
  {A}nalysis {S}ymposium}, {B.~Le~Charlier}, Ed. Lecture {N}otes in {C}omputer
  {S}cience, vol. 864. Springer, 43--60.

\bibitem[\protect\citeauthoryear{Barbuti and Giacobazzi}{Barbuti and
  Giacobazzi}{1992}]{BarbutiG92}
{\sc Barbuti, R.} {\sc and} {\sc Giacobazzi, R.} 1992.
\newblock A bottom-up polymorphic type inference in logic programming.
\newblock {\em Science of {C}omputer {P}rogramming\/}~{\em 19,\/}~3, 133--181.

\bibitem[\protect\citeauthoryear{Barbuti, Giacobazzi, and Levi}{Barbuti
  et~al\mbox{.}}{1993}]{BarbutiGL93}
{\sc Barbuti, R.}, {\sc Giacobazzi, R.}, {\sc and} {\sc Levi, G.} 1993.
\newblock A general framework for semantics-based bottom-up abstract
  interpretation of logic programs.
\newblock {\em ACM {T}ransactions on {P}rogramming {L}anguages and
  {S}ystems\/}~{\em 15,\/}~1, 133--181.

\bibitem[\protect\citeauthoryear{Boye and Mal\'{u}szynski}{Boye and
  Mal\'{u}szynski}{1996}]{BoyeM96}
{\sc Boye, J.} {\sc and} {\sc Mal\'{u}szynski, J.} 1996.
\newblock Two aspects of directional types.
\newblock In {\em Proceedings of the {T}welfth {I}nternational {C}onference on
  {L}ogic {P}rogramming}. The MIT Press, 747--761.

\bibitem[\protect\citeauthoryear{Bronsard, Lakshman, and Reddy}{Bronsard
  et~al\mbox{.}}{1992}]{BronsardJICSLP92}
{\sc Bronsard, F.}, {\sc Lakshman, T.~K.}, {\sc and} {\sc Reddy, U.~S.} 1992.
\newblock A framework of directionality for proving termination of logic
  programs.
\newblock In {\em Proceedings of the {J}oint {I}nternational {C}onference and
  {S}ymposium on {L}ogic {P}rogramming}, {K.~Apt}, Ed. The MIT Press, 321--335.

\bibitem[\protect\citeauthoryear{Bruynooghe}{Bruynooghe}{1991}]{Bruynooghe91}
{\sc Bruynooghe, M.} 1991.
\newblock A practical framework for the abstract interpretation of logic
  progams.
\newblock {\em {J}ournal of {L}ogic {P}rogramming\/}~{\em 10,\/}~2, 91--124.

\bibitem[\protect\citeauthoryear{Charatonik and Podelski}{Charatonik and
  Podelski}{1998}]{CharatonikP98}
{\sc Charatonik, W.} {\sc and} {\sc Podelski, A.} 1998.
\newblock Directional type inference for logic programs.
\newblock In {\em Proceedings of the {F}ifth {I}nternational {S}ymposium on
  {S}tatic {A}nalysis}, {G.~Levi}, Ed. Lecture {N}otes in {C}omputer {S}cience,
  vol. 1503. Springer, 278--294.

\bibitem[\protect\citeauthoryear{Codish and Demoen}{Codish and
  Demoen}{1994}]{CodishD94}
{\sc Codish, M.} {\sc and} {\sc Demoen, B.} 1994.
\newblock Deriving polymorphic type dependencies for logic programs using
  multiple incarnations of {Prop}.
\newblock In {\em Proceedings of the {F}irst {I}nternational {S}tatic
  {A}nalysis {S}ymposium}, {B.~Le~Charlier}, Ed. Lecture {N}otes in {C}omputer
  {S}cience, vol. 864. Springer, 281--297.

\bibitem[\protect\citeauthoryear{Codish and Lagoon}{Codish and
  Lagoon}{2000}]{aci_types_full}
{\sc Codish, M.} {\sc and} {\sc Lagoon, V.} 2000.
\newblock Type dependencies for logic programs using {ACI}-unification.
\newblock {\em Theoretical {C}omputer {S}cience\/}~{\em 238,\/}~1--2, 131--159.

\bibitem[\protect\citeauthoryear{Comon, Dauchet, Gilleron, Jacquemard, Lugiez,
  Tison, and Tommasi}{Comon et~al\mbox{.}}{2002}]{tata02}
{\sc Comon, H.}, {\sc Dauchet, M.}, {\sc Gilleron, R.}, {\sc Jacquemard, F.},
  {\sc Lugiez, D.}, {\sc Tison, S.}, {\sc and} {\sc Tommasi, M.} 2002.
\newblock Tree automata techniques and applications.
\newblock {http://www.grappa.univ-lille3.fr/tata}.

\bibitem[\protect\citeauthoryear{Cousot and Cousot}{Cousot and
  Cousot}{1977}]{Cousot:Cousot:77}
{\sc Cousot, P.} {\sc and} {\sc Cousot, R.} 1977.
\newblock Abstract interpretation: a unified framework for static analysis of
  programs by construction or approximation of fixpoints.
\newblock In {\em Principles of {P}rogramming {L}anguages}. The ACM Press,
  238--252.

\bibitem[\protect\citeauthoryear{Cousot and Cousot}{Cousot and
  Cousot}{1992}]{Cousot:JLP92}
{\sc Cousot, P.} {\sc and} {\sc Cousot, R.} 1992.
\newblock Abstract interpretation and application to logic programs.
\newblock {\em {J}ournal of {L}ogic {P}rogramming\/}~{\em 13,\/}~1--4,
  103--179.

\bibitem[\protect\citeauthoryear{Cousot and Cousot}{Cousot and
  Cousot}{1995}]{CousotCousot95}
{\sc Cousot, P.} {\sc and} {\sc Cousot, R.} 1995.
\newblock Formal language, grammar and set-constraint-based program analysis by
  abstract interpretation.
\newblock In {\em Proceedings of the {S}eventh {ACM} {C}onference on
  {F}unctional {P}rogramming {L}anguages and {C}omputer {A}rchitecture}. The
  ACM Press, 170--181.

\bibitem[\protect\citeauthoryear{Dart and Zobel}{Dart and
  Zobel}{1992a}]{Dart:Zobel:JLP92}
{\sc Dart, P.} {\sc and} {\sc Zobel, J.} 1992a.
\newblock Efficient runtime type checking of typed logic programs.
\newblock {\em {J}ournal of {L}ogic {P}rogramming\/}~{\em 14,\/}~1-2, 31--69.

\bibitem[\protect\citeauthoryear{Dart and Zobel}{Dart and
  Zobel}{1992b}]{DartZ92}
{\sc Dart, P.} {\sc and} {\sc Zobel, J.} 1992b.
\newblock A regular type language for logic programs.
\newblock In {\em Types in {L}ogic {P}rogramming}, {F.~Pfenning}, Ed. The MIT
  Press, 157--189.

\bibitem[\protect\citeauthoryear{Eder}{Eder}{1985}]{eder}
{\sc Eder, E.} 1985.
\newblock Properties of substitutions and unifications.
\newblock {\em Journal of {S}ymbolic {C}omputation\/}~{\em 1,\/}~1, 31--46.

\bibitem[\protect\citeauthoryear{Fages and Coquery}{Fages and
  Coquery}{2001}]{FagesC01}
{\sc Fages, F.} {\sc and} {\sc Coquery, E.} 2001.
\newblock Typing constraint logic programs.
\newblock {\em {T}heory and {P}ractice of {L}ogic {P}rogramming\/}~{\em
  1,\/}~6, 751--777.

\bibitem[\protect\citeauthoryear{Fr\"{u}hwirth, Shapiro, Vardi, and
  Yardeni}{Fr\"{u}hwirth et~al\mbox{.}}{1991}]{FruhwirthSVY:LICS91}
{\sc Fr\"{u}hwirth, T.}, {\sc Shapiro, E.}, {\sc Vardi, M.}, {\sc and} {\sc
  Yardeni, E.} 1991.
\newblock Logic programs as types for logic programs.
\newblock In {\em Proceedings of the {S}ixth {A}nnual {IEEE} {S}ymposium on
  {L}ogic in {C}omputer {S}cience}. The {IEEE} {C}omputer {S}ociety {P}ress,
  300--309.

\bibitem[\protect\citeauthoryear{Gallagher and de~Waal}{Gallagher and
  de~Waal}{1994}]{GallagherW94}
{\sc Gallagher, J.} {\sc and} {\sc de~Waal, D.} 1994.
\newblock Fast and precise regular approximations of logic programs.
\newblock In {\em Proceedings of the {E}leventh {I}nternational {C}onference on
  {L}ogic {P}rogramming}, {M.~Bruynooghe}, Ed. The MIT Press, 599--613.

\bibitem[\protect\citeauthoryear{Gallagher, Boulanger, and Saglam}{Gallagher
  et~al\mbox{.}}{1995}]{GallagherBS95}
{\sc Gallagher, J.~P.}, {\sc Boulanger, D.}, {\sc and} {\sc Saglam, H.} 1995.
\newblock Practical model-based static analysis for definite logic programs.
\newblock In {\em Proceedings of the {F}ifteenth {I}nternational {S}ymposium on
  {L}ogic {P}rogramming}, {J.~W. Lloyd}, Ed. The MIT Press, 351--368.

\bibitem[\protect\citeauthoryear{Gallagher and Puebla}{Gallagher and
  Puebla}{2002}]{GallagherP02}
{\sc Gallagher, J.~P.} {\sc and} {\sc Puebla, G.} 2002.
\newblock Abstract interpretation over non-deterministic finite tree automata
  for set-based analysis of logic programs.
\newblock In {\em Proceedings of the Fourth International Symposium on
  Practical Aspects of Declarative Languages}, {S.~Krishnamurthi} {and} {C.~R.
  Ramakrishnan}, Eds. Lecture {N}otes in {C}omputer {S}cience, vol. 2257.
  Springer, 243--261.

\bibitem[\protect\citeauthoryear{G\'{e}cseg and Steinby}{G\'{e}cseg and
  Steinby}{1984}]{GecsegS84}
{\sc G\'{e}cseg, F.} {\sc and} {\sc Steinby, M.} 1984.
\newblock {\em Tree Automata}.
\newblock Akad\'{e}miai Kiad\'{o}.

\bibitem[\protect\citeauthoryear{Heintze and Jaffar}{Heintze and
  Jaffar}{1990}]{HeintzeJ90}
{\sc Heintze, N.} {\sc and} {\sc Jaffar, J.} 1990.
\newblock A finite presentation theorem for approximating logic programs.
\newblock In {\em {P}rinciples of {P}rogramming {L}anguages}. The ACM Press,
  197--209.

\bibitem[\protect\citeauthoryear{Heintze and Jaffar}{Heintze and
  Jaffar}{1992}]{HeintzeJ92}
{\sc Heintze, N.} {\sc and} {\sc Jaffar, J.} 1992.
\newblock Semantic types for logic programs.
\newblock In {\em Types in {L}ogic {P}rogramming}, {F.~Pfenning}, Ed. The MIT
  Press, 141--155.

\bibitem[\protect\citeauthoryear{Hermenegildo, Warren, and Debray}{Hermenegildo
  et~al\mbox{.}}{1992}]{HermenegildoWD92}
{\sc Hermenegildo, M.}, {\sc Warren, R.}, {\sc and} {\sc Debray, S.} 1992.
\newblock Global flow analysis as a practical compilation tool.
\newblock {\em {J}ournal of {L}ogic {P}rogramming\/}~{\em 13,\/}~1--4,
  349--366.

\bibitem[\protect\citeauthoryear{Hermenegildo, Bueno, Puebla, and
  L\'{o}pez}{Hermenegildo et~al\mbox{.}}{1999}]{HermenegildoBPL99}
{\sc Hermenegildo, M.~V.}, {\sc Bueno, F.}, {\sc Puebla, G.}, {\sc and} {\sc
  L\'{o}pez, P.} 1999.
\newblock Program analysis, debugging, and optimization using the {C}iao system
  preprocessor.
\newblock In {\em Proceedings of the 1999 {I}nternational {C}onference on
  {L}ogic {P}rogramming}. The MIT Press, 52--65.

\bibitem[\protect\citeauthoryear{Hill and Lloyd}{Hill and Lloyd}{1994}]{Godel}
{\sc Hill, P.} {\sc and} {\sc Lloyd, J.} 1994.
\newblock {\em The {G}\"{o}del Programming Language}.
\newblock The MIT Press.

\bibitem[\protect\citeauthoryear{Hill and Spoto}{Hill and
  Spoto}{2002}]{HillS02}
{\sc Hill, P.~M.} {\sc and} {\sc Spoto, F.} 2002.
\newblock Generalising \emph{Def} and \emph{Pos} to type analysis.
\newblock {\em Journal of {L}ogic and {C}omputation\/}~{\em 12,\/}~3, 497--542.

\bibitem[\protect\citeauthoryear{Horiuchi and Kanamori}{Horiuchi and
  Kanamori}{1988}]{HoriuchiK87}
{\sc Horiuchi, K.} {\sc and} {\sc Kanamori, T.} 1988.
\newblock Polymorphic type inference in {P}rolog by abstract interpretation.
\newblock In {\em Proceedings of the {S}ixth {C}onference on {L}ogic
  {P}rogramming}, {K.~Furukawa}, {H.~Tanaka}, {and} {T.~Fujisaki}, Eds. Lecture
  {N}otes in {C}omputer {S}cience, vol. 315. Springer, 195--214.

\bibitem[\protect\citeauthoryear{Janssens and Bruynooghe}{Janssens and
  Bruynooghe}{1992}]{Janssens:JLP92}
{\sc Janssens, G.} {\sc and} {\sc Bruynooghe, M.} 1992.
\newblock Deriving descriptions of possible values of program variables by
  means of abstract interpretation.
\newblock {\em {J}ournal of {L}ogic {P}rogramming\/}~{\em 13,\/}~1--4,
  205--258.

\bibitem[\protect\citeauthoryear{Kahrs}{Kahrs}{1996}]{Kahrs}
{\sc Kahrs, S.} 1996.
\newblock Limits of {ML}-definability.
\newblock In {\em Proceedings of the {E}ighth {I}nternational {S}ymposium on
  {P}rogramming {L}anguages: {I}mplementation, {L}ogic and {P}rograms},
  {H.~Kuchen} {and} {S.~D. Swierstra}, Eds. Lecture Notes in Computer Science,
  vol. 1140. Springer, 17--31.

\bibitem[\protect\citeauthoryear{Kanamori and Horiuchi}{Kanamori and
  Horiuchi}{1985}]{Kanamori:Horiuchi:85}
{\sc Kanamori, T.} {\sc and} {\sc Horiuchi, K.} 1985.
\newblock Type inference in {Prolog} and its application.
\newblock In {\em Proceedings of the {N}inth {I}nternational {J}oint
  {C}onference on {A}rtificial {I}ntelligence}, {A.~Joshi}, Ed. Morgan
  Kaufmann, 704--707.

\bibitem[\protect\citeauthoryear{Kanamori and Kawamura}{Kanamori and
  Kawamura}{1993}]{KanamoriJLP93}
{\sc Kanamori, T.} {\sc and} {\sc Kawamura, T.} 1993.
\newblock Abstract interpretation based on {OLDT} resolution.
\newblock {\em {J}ournal of {L}ogic {P}rogramming\/}~{\em 15,\/}~1 \& 2, 1--30.

\bibitem[\protect\citeauthoryear{Lagoon and Stuckey}{Lagoon and
  Stuckey}{2001}]{LS01}
{\sc Lagoon, V.} {\sc and} {\sc Stuckey, P.~J.} 2001.
\newblock A framework for analysis of typed logic programs.
\newblock In {\em Proceedings of the {F}ifth {I}nternational {S}ymposium on
  {F}unctional and {L}ogic {P}rogramming}, {H.~Kuchen} {and} {K.~Ueda}, Eds.
  Lecture {N}otes in {C}omputer {S}cience, vol. 2024. Springer, 296--310.

\bibitem[\protect\citeauthoryear{Lloyd}{Lloyd}{1987}]{Lloyd:87}
{\sc Lloyd, J.} 1987.
\newblock {\em Foundations of Logic Programming}.
\newblock Springer-Verlag.

\bibitem[\protect\citeauthoryear{Lu}{Lu}{1995}]{Lu95}
{\sc Lu, L.} 1995.
\newblock Type analysis of logic programs in the presence of type definitions.
\newblock In {\em Proceedings of the 1995 {ACM SIGPLAN} {S}ymposium on
  {P}artial {E}valuation and {S}emantics-{B}ased {P}rogram {M}anipulation}. The
  ACM Press, 241--252.

\bibitem[\protect\citeauthoryear{Lu}{Lu}{1998}]{LuJLP98}
{\sc Lu, L.} 1998.
\newblock A polymorphic type analysis in logic programs by abstract
  interpretation.
\newblock {\em {J}ournal of {L}ogic {P}rogramming\/}~{\em 36,\/}~1, 1--54.

\bibitem[\protect\citeauthoryear{Lu}{Lu}{2003}]{Lu03path}
{\sc Lu, L.} 2003.
\newblock Path dependent analysis of logic programs.
\newblock {\em Higher-Order and Symbolic Computation\/}~{\em 16}, 341--377.

\bibitem[\protect\citeauthoryear{Lu and Cleary}{Lu and
  Cleary}{1998}]{LuC:empty}
{\sc Lu, L.} {\sc and} {\sc Cleary, J.} 1998.
\newblock An emptiness algorithm for regular types with set operators.
\newblock Technical report, Department of Computer Science, The University of
  Waikato.
\newblock http://xxx.lanl.gov/abs/cs.LO/9811015.

\bibitem[\protect\citeauthoryear{Marriott and S{\o}ndergaard}{Marriott and
  S{\o}ndergaard}{1989}]{MarriottS89}
{\sc Marriott, K.} {\sc and} {\sc S{\o}ndergaard, H.} 1989.
\newblock Semantics-based dataflow analysis of logic programs.
\newblock In {\em Information {P}rocessing 89, {P}roceedings of the {E}leventh
  {IFIP} {W}orld {C}omputer {C}ongress}, {G.~Ritter}, Ed. North-Holland,
  601--606.

\bibitem[\protect\citeauthoryear{Mishra}{Mishra}{1984}]{Mishra:84}
{\sc Mishra, P.} 1984.
\newblock Towards a theory of types in {P}rolog.
\newblock In {\em Proceedings of the {IEEE} {I}nternational {S}ymposium on
  {L}ogic {P}rogramming}. The IEEE Computer Society Press, 289--298.

\bibitem[\protect\citeauthoryear{Mycroft and O'Keefe}{Mycroft and
  O'Keefe}{1984}]{Mycroft:OKeefe:84}
{\sc Mycroft, A.} {\sc and} {\sc O'Keefe, R.} 1984.
\newblock A polymorphic type system for {P}rolog.
\newblock {\em Artificial {I}ntelligence\/}~{\em 23,\/}~3, 295--307.

\bibitem[\protect\citeauthoryear{Nilsson}{Nilsson}{1988}]{Nilsson:88}
{\sc Nilsson, U.} 1988.
\newblock Towards a framework for abstract interpretation of logic programs.
\newblock In {\em Proceedings of the {F}irst {I}nternational {W}orkshop on
  {P}rogramming {L}anguage {I}mplementation and {L}ogic {P}rogramming},
  {P.~Deransart}, {B.~Lorho}, {and} {J.~Ma{\l}uszynski}, Eds. Lecture {N}otes
  in {C}omputer {S}cience, vol. 348. Springer, 68--82.

\bibitem[\protect\citeauthoryear{Reddy}{Reddy}{1990}]{Reddy:NACLP90}
{\sc Reddy, U.} 1990.
\newblock Types for logic programs.
\newblock In {\em Proceedings of the 1990 {N}orth {A}merican {C}onference on
  {L}ogic {P}rogramming}. The MIT Press, 836--40.

\bibitem[\protect\citeauthoryear{Rychlikowski and Truderung}{Rychlikowski and
  Truderung}{2001}]{RychlikowskiT01}
{\sc Rychlikowski, P.} {\sc and} {\sc Truderung, T.} 2001.
\newblock Polymorphic directional types for logic programming.
\newblock In {\em Proceedings of the {Third ACM SIGPLAN} {I}nternational
  {C}onference on {P}rinciples and {P}ractice of {D}eclarative {P}rogramming}.
  The ACM Press, 61--72.

\bibitem[\protect\citeauthoryear{Saglam and Gallagher}{Saglam and
  Gallagher}{1995}]{1995-saglam}
{\sc Saglam, H.} {\sc and} {\sc Gallagher, J.} 1995.
\newblock Approximating constraint logic programs using polymorphic types and
  regular descriptions.
\newblock Technical report CSTR-95-017, Department of Computer Science,
  University of Bristol.

\bibitem[\protect\citeauthoryear{Smaus}{Smaus}{2001}]{Smaus01}
{\sc Smaus, J.-G.} 2001.
\newblock Analysis of polymorphically typed logic programs using
  {ACI}-unification.
\newblock In {\em Proceedings of the {E}ighth {I}nternational {C}onference on
  {L}ogic for {P}rogramming, {A}rtificial {I}ntelligence, and {R}easoning}.
  Lecture Notes in Artificial Intelligence, vol. 2250. Springer, 282--298.

\bibitem[\protect\citeauthoryear{Somogyi, Henderson, and Conway}{Somogyi
  et~al\mbox{.}}{1996}]{Mercury}
{\sc Somogyi, Z.}, {\sc Henderson, F.}, {\sc and} {\sc Conway, T.} 1996.
\newblock The execution algorithm of {M}ercury: An efficient purely declarative
  logic programming language.
\newblock {\em {J}ournal of {L}ogic {P}rogramming\/}~{\em 29,\/}~1--3, 19--64.

\bibitem[\protect\citeauthoryear{Van~Hentenryck, Cortesi, and
  Le~Charlier}{Van~Hentenryck et~al\mbox{.}}{1995}]{CortesiCH_JLP95}
{\sc Van~Hentenryck, P.}, {\sc Cortesi, A.}, {\sc and} {\sc Le~Charlier, B.}
  1995.
\newblock Type analysis of {Prolog} using type graphs.
\newblock {\em {J}ournal of {L}ogic {P}rogramming\/}~{\em 22,\/}~3, 179--208.

\bibitem[\protect\citeauthoryear{Warren}{Warren}{1992}]{Warren92}
{\sc Warren, D.~S.} 1992.
\newblock Memoing for logic programs.
\newblock {\em Communications of the {ACM}\/}~{\em 35,\/}~3, 93--111.

\bibitem[\protect\citeauthoryear{Yardeni, Fr\"{u}hwirth, and Shapiro}{Yardeni
  et~al\mbox{.}}{1991}]{YardeniFS:ICLP91}
{\sc Yardeni, E.}, {\sc Fr\"{u}hwirth, T.}, {\sc and} {\sc Shapiro, E.} 1991.
\newblock Polymorphically typed logic programs.
\newblock In {\em Proceedings of the {E}ighth {I}nternational {C}onference on
  {L}ogic {P}rogramming}, {K.~Furukawa}, Ed. The MIT Press, 379--93.

\bibitem[\protect\citeauthoryear{Yardeni and Shapiro}{Yardeni and
  Shapiro}{1991}]{Yardeni:Shapiro:91}
{\sc Yardeni, E.} {\sc and} {\sc Shapiro, E.} 1991.
\newblock A type system for logic programs.
\newblock {\em {J}ournal of {L}ogic {P}rogramming\/}~{\em 10,\/}~2, 125--153.

\bibitem[\protect\citeauthoryear{Zobel}{Zobel}{1987}]{Zobel:87}
{\sc Zobel, J.} 1987.
\newblock Derivation of polymorphic types for {P}rolog programs.
\newblock In {\em Proceedings of the {F}ourth {I}nternational {C}onference on
  {L}ogic {P}rogramming}, {J.~Lassez}, Ed. The MIT Press, 817--838.

\end{thebibliography}
\end{document}